\def\verPreprint{1}
\def\verPAPER{2}
\def\ver{1}

\def\lineNumbersEnabled{true}

\def\showLineNumbers{false}

\ifx\ver\verPreprint
\documentclass[a4paper,english,11pt]{article}
\usepackage[bindingoffset=0.5cm,left=2.5cm,right=2.5cm,top=2.5cm,bottom=2.5cm,footskip=1.0cm]{geometry}
\usepackage{hepnames,bm,multirow,amssymb,xspace,hyperref,lineno,amsmath,bm,multirow,authblk,graphicx,newclude,rotating,cite}
\fi
\ifx\ver\verPAPER
\documentclass[review,sort&compress]{elsarticle}
\usepackage{hepnames,bm,multirow,amssymb,xspace,hyperref}
\fi

\ifx\showLineNumbers\lineNumbersEnabled
\usepackage{lineno}
\newenvironment{linenowrapper}{\begin{linenomath*}}{\end{linenomath*}}
\modulolinenumbers[5]
\else
\newenvironment{linenowrapper}{}{} 
\fi

\usepackage{soul}
\usepackage[dvipsnames]{xcolor}

\newcommand{\virt}{\ensuremath{^{\ast}}}
\renewcommand{\Plepton}{\ensuremath{\ell}}
\newcommand{\ellStar}{\ensuremath{\ell\virt}}
\newcommand{\nuStar}{\ensuremath{\Pnu\virt}}
\newcommand{\ellMinus}{\ensuremath{\Plepton^{-}}\xspace}
\newcommand{\ellPlus}{\ensuremath{\Plepton^{+}}\xspace}
\newcommand{\ellnu}{\ensuremath{\Plepton\Pnu}\xspace}
\newcommand{\ellnuStar}{\ensuremath{\Plepton\virt\Pnu\virt}\xspace}
\newcommand{\ellMinusnu}{\ensuremath{\Plepton^{-}\APnu}\xspace}
\newcommand{\ellPlusnu}{\ensuremath{\Plepton^{+}\Pnu}\xspace}
\newcommand{\ellnuellnuStar}{\ensuremath{\Plepton\Pnu\Plepton\virt\Pnu\virt}\xspace}
\newcommand{\ellnuellStar}{\ensuremath{\Plepton\Pnu\Plepton\virt}\xspace}
\newcommand{\dihiggs}{\ensuremath{\PHiggs\PHiggs}}
\renewcommand{\Pbottom}{\ensuremath{\Pqb}}
\renewcommand{\APbottom}{\ensuremath{\APqb}}
\renewcommand{\Ptop}{\ensuremath{\Pqt}}
\renewcommand{\APtop}{\ensuremath{\APqt}}
\newcommand{\ttbar}{\ensuremath{\Ptop\APtop}}
\newcommand{\ggHH}{\ensuremath{\Pgluon\Pgluon\PHiggs\PHiggs}}
\newcommand{\qqHH}{\ensuremath{\textrm{qq}\PHiggs\PHiggs}}
\newcommand{\VHH}{\ensuremath{\textrm{V}\PHiggs\PHiggs}}
\newcommand{\ttHH}{\ensuremath{\Ptop\APtop\PHiggs\PHiggs}}
\newcommand{\lambdaHHH}{\ensuremath{\lambda}}
\newcommand{\yt}{\ensuremath{y_{\Ptop}}}
\newcommand{\vecy}{\ensuremath{\bm{p}}\xspace}
\newcommand{\vecyhat}{\ensuremath{\bm{\hat{p}}}\xspace}
\newcommand{\xhat}{\ensuremath{\hat{x}}\xspace}
\newcommand{\vecp}{\ensuremath{\bm{p}}\xspace}
\newcommand{\vecphat}{\ensuremath{\bm{\hat{p}}}\xspace}
\newcommand{\pT}{\ensuremath{p_{\textrm{T}}}\xspace}
\newcommand{\pThat}{\ensuremath{\hat{p}_{\textrm{T}}}\xspace}

\newcommand{\thetahat}{\ensuremath{\hat{\theta}}\xspace}
\newcommand{\phihat}{\ensuremath{\hat{\phi}}\xspace}
\newcommand{\betahat}{\ensuremath{\hat{\beta}}\xspace}

\newcommand{\Ehat}{\ensuremath{\hat{E}}\xspace}

\newcommand{\pX}{\ensuremath{p_{\textrm{x}}}\xspace}
\newcommand{\pXhat}{\ensuremath{\hat{p}_{\textrm{x}}}\xspace}
\newcommand{\pY}{\ensuremath{p_{\textrm{y}}}\xspace}
\newcommand{\pYhat}{\ensuremath{\hat{p}_{\textrm{y}}}\xspace}

\newcommand{\pZhat}{\ensuremath{\hat{p}_{\textrm{z}}}\xspace}

\newcommand{\MET}{\ensuremath{p_{\textrm{T}}^{\textrm{\kern0.10em miss}}}\xspace}
\newcommand{\vecMET}{\ensuremath{\bm{p}_{\textrm{T}}^{\textrm{\kern0.10em miss}}}\xspace}
\newcommand{\METx}{\ensuremath{p_{\textrm{x}}^{\textrm{\kern0.10em miss}}}\xspace}
\newcommand{\METxhat}{\ensuremath{\hat{p}_{\textrm{x}}^{\textrm{\kern0.10em miss}}}\xspace}
\newcommand{\METy}{\ensuremath{p_{\textrm{y}}^{\textrm{\kern0.10em miss}}}\xspace}
\newcommand{\METyhat}{\ensuremath{\hat{p}_{\textrm{y}}^{\textrm{\kern0.10em miss}}}\xspace}
\newcommand{\vece}{\ensuremath{\bm{e}}\xspace}
\newcommand{\vecehat}{\ensuremath{\bm{\hat{e}}}\xspace}
\newcommand{\defL}{\ensuremath{\equiv}\xspace}
\newcommand{\defR}{\ensuremath{\equiv}\xspace}
\newcommand{\X}{\ensuremath{\textrm{X}}}

\newcommand{\GeV}{\ensuremath{\textrm{GeV}}\xspace}
\newcommand{\TeV}{\ensuremath{\textrm{TeV}}\xspace}

\newcommand{\mbb}{\ensuremath{m_{\Pbottom\Pbottom}}}
\newcommand{\cf}{cf.\xspace}
\newcommand{\ie}{i.e.\xspace}
\newcommand{\eg}{e.g.\xspace}
\newcommand{\pb}{\ensuremath{\textrm{~pb}}\xspace}
\newcommand{\fb}{\ensuremath{\textrm{~fb}}\xspace}

\newcommand{\kt}{\ensuremath{k_{\textrm{T}}}\xspace}
\newcommand{\abs}[1]{\left\lvert #1 \right\rvert}
\def\TReg{\textsuperscript{\textregistered}}
\usepackage{array}
\newcolumntype{C}[1]{>{\centering\arraybackslash}p{#1}}

\begin{document}

\ifx\ver\verPAPER
\begin{frontmatter}
\fi

\title{Application of the matrix element method to Higgs boson pair production in the channel $\dihiggs \to \Pbottom\APbottom\PW\PW^{*}$ at the LHC}


\ifx\ver\verPreprint
\author[1]{Karl Ehat\"aht}
\author[1]{Christian Veelken}
\affil[1]{National Institute for Chemical Physics and Biophysics, 10143 Tallinn, Estonia}
\fi
\ifx\ver\verPAPER
\author[tallinn]{Karl Ehat\"aht}
\ead{karl.ehataht@cern.ch}
\author[tallinn]{Christian Veelken}
\ead{christian.veelken@cern.ch}
\address[tallinn]{National Institute for Chemical Physics and Biophysics, 10143 Tallinn, Estonia}
\fi

\ifx\ver\verPreprint
\maketitle
\fi

\begin{abstract}
We apply the matrix element method (MEM) to the search for non-resonant Higgs boson pair ($\dihiggs$) production in the channel $\dihiggs \to \Pbottom\APbottom\PW\PW^{*}$ at the LHC
and study the separation between the $\dihiggs$ signal and the large irreducible background, which arises from the production of top quark pairs ($\ttbar$).
Our study focuses on events containing two leptons (electrons or muons) in the final state.
The separation between signal and background is studied for experimental conditions characteristic for the ATLAS and CMS experiments during LHC Run $2$,
using the $\textsc{DELPHES}$ fast-simulation package.
We find that the $\ttbar$ background can be reduced to a level of $0.26\%$ for a signal efficiency of $35\%$.
\end{abstract}

\ifx\ver\verPAPER
\end{frontmatter}
\fi

\ifx\ver\verPreprint
\clearpage
\fi

\ifx\showLineNumbers\lineNumbersEnabled
\linenumbers
\fi

\section{Introduction}
\label{sec:introduction}

The discovery of a Higgs ($\PHiggs$) boson by the ATLAS and CMS experiments~\cite{Higgs-Discovery_ATLAS,Higgs-Discovery_CMS}
represents a major step towards our understanding of electroweak symmetry breaking (EWSB),
as well as of the mechanism that generates the masses of quarks and leptons, 
the particles that constitute the `` ordinary'' matter in our universe.
In a combined analysis of the data recorded by ATLAS and CMS during LHC Run $1$, 
the mass of the $\PHiggs$ boson has been measured to be $125.09 \pm 0.24$~\GeV~\cite{HIG-14-042}.
Recent analyses of data collected during LHC Run $2$ corroborate this value~\cite{ATLAS:2020coj,CMS:2020xrn}.
The Standard Model (SM) of particle physics makes precise predictions for all properties of the $\PHiggs$ boson, given its mass. 
The predictions have been probed by measurements of its spin and CP quantum numbers~\cite{HIG-14-018,Aad:2015mxa,ATLAS:2020evk,CMS:2021nnc},
of its couplings to gauge bosons and to down-type fermions~\cite{HIG-15-002},
and of its total decay width, including decays to invisible particles~\cite{CMS:2018yfx,CMS:2019ekd,Aad:2015pla,ATLAS:2018jym}.
So far, all measured properties of the discovered particle are consistent with the expectation for a SM $\PHiggs$ boson 
within the uncertainties of these measurements.
Evidence for its coupling to up-type fermions, at a strength compatible with the SM expectation, has been observed recently~\cite{Aaboud:2018urx,HIG-17-035}.

The SM predicts $\PHiggs$ boson self-interactions via trilinear and quartic couplings. 
Measurements of the $\PHiggs$ boson self-interactions will allow to determine the potential of the Higgs field,
thereby ultimately either confirming or falsifying that the Brout-Englert-Higgs mechanism of the SM is responsible for EWSB.
The measurement of the quartic coupling is not possible at the LHC~\cite{deFlorian:2019app}, 
even with the $3000$~fb$^{-1}$ of data foreseen to be recorded at $\sqrt{s}=14$~\TeV center-of-mass energy during the upcoming HL-LHC data-taking period~\cite{HL-LHC-TDR},
as the cross section of the corresponding process, triple $\PHiggs$ boson production, is much too small, 
on the level of $5 \cdot 10^{-2}$~\fb~\cite{Plehn:2005nk,Binoth:2006ym}.

\begin{figure}
\setlength{\unitlength}{1mm}
\begin{center}
\begin{tabular}{ccc}
\raisebox{-.45\height}{\resizebox{0.43\textwidth}{!}{\includegraphics[bb=0 0 214 147]{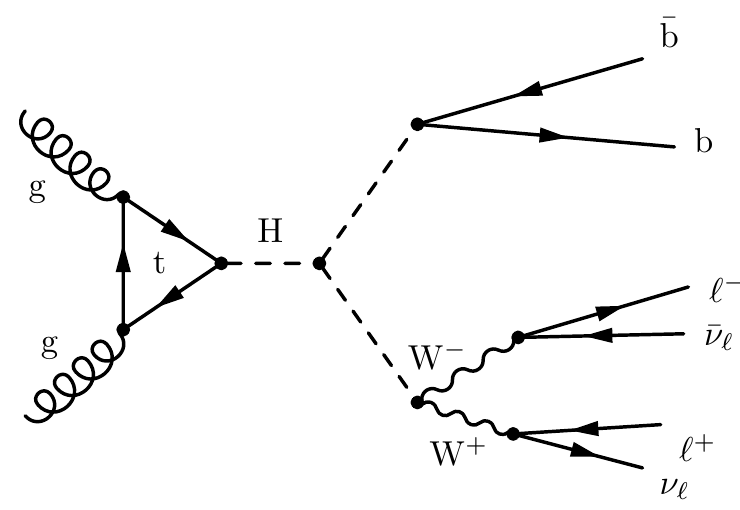}}} &
\qquad + &
\raisebox{-.45\height}{\resizebox{0.43\textwidth}{!}{\includegraphics[bb=0 0 214 147]{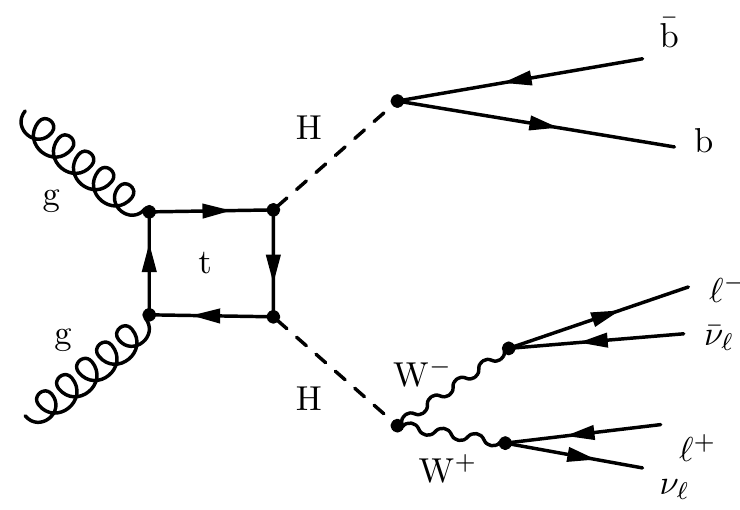}}}
\end{tabular}
\end{center}
\caption{
  Leading order (LO) Feynman diagrams for the $\ggHH$ production process in $\Pp\Pp$ collisions at the LHC,
  with subsequent decay of the $\PHiggs$ boson pair via $\PHiggs\PHiggs \to \Pbottom\APbottom\PW\PW\virt \to \Pbottom\APbottom \, \ellPlusnu\ellMinusnu$.
  The asterisk ($\virt$) denotes an off-shell $\PW$ boson.
}
\label{fig:ggHH_FeynmanDiagram}
\end{figure}

The trilinear coupling ($\lambdaHHH$) can be determined at the LHC, by measuring the rate for $\PHiggs$ boson pair production ($\dihiggs$). 
In analogy to the production of single $\PHiggs$ bosons, 
four different processes are relevant for $\dihiggs$ production at the LHC: 
gluon fusion ($\ggHH$), vector boson fusion ($\qqHH$), the associated production with a $\PW$ or $\PZ$ boson ($\VHH$),
and associated production of the $\PHiggs$ boson pair with a pair of top quarks ($\ttHH$).
The total $\dihiggs$ production rate is dominated by the $\ggHH$ process.
Its cross section has been computed at next-to-next-to-leading order (NNLO) in perturbative quantum chromodynamics (pQCD),
with resummation of soft gluon contributions at next-to-next-to-leading logarithmic accuracy.
Including corrections for finite top quark mass effects, computed at next-to-leading order (NLO),
the SM cross section for the $\ggHH$ process amounts to $31.05^{+1.40}_{-1.98}$~\fb at $\sqrt{s} = 13$~\TeV center-of-mass energy~\cite{Grazzini:2018hh}.
The cross section is rather small, as the production of $\PHiggs$ boson pairs through gluon fusion is a loop induced process,
and is further reduced by the negative interference of two competing production mechanisms.
The leading order (LO) Feynman diagrams for the two competing production mechanisms are shown in Fig.~\ref{fig:ggHH_FeynmanDiagram}.
The right diagram, referred to as the ``box'' diagram, does actually not depend on the trilinear $\PHiggs$ boson self-coupling $\lambdaHHH$.
The diagram that provides the sensitivity to $\lambdaHHH$ is the ``triangle'' diagram shown on the left.
Both diagrams depend on the coupling of the $\PHiggs$ boson to the top quark, denoted by the symbol $\yt$,
which is measured with an uncertainty of order $10\%$ at present~\cite{Aaboud:2018urx,HIG-17-035}.
The cross sections for the $\qqHH$, $\VHH$, and $\ttHH$ process are more than one order of magnitude smaller~\cite{Baglio:2012np}.
As the sensitivity of experimental analyses at the LHC is limited by the small signal rate at present,
we will focus on the $\ggHH$ production process in this paper.

The $\dihiggs$ production rate may be enhanced significantly in case an as yet unknown resonance decays to pairs of $\PHiggs$ bosons.
Such resonances are predicted in models with two Higgs doublets~\cite{Craig:2013hca,Nhung:2013lpa}, composite $\PHiggs$ boson models~\cite{Grober:2010yv,Contino:2010mh}, 
Higgs portal models~\cite{Englert:2011yb,No:2013wsa}, and models involving extra dimensions~\cite{Randall:1999ee}.
In the absence of new resonances decaying into $\PHiggs$ boson pairs,
the $\dihiggs$ production rate may be enhanced by deviations of the couplings $\lambdaHHH$ and $\yt$ from the SM expectation for these couplings
and by the contribution of new particles to the loops 
that are present in the triangle and box diagrams shown in Fig.~\ref{fig:ggHH_FeynmanDiagram}.
The effect of contributions from new particles to these loops can adequately be described by anomalous $\PHiggs$ boson couplings
in an effective field theory (EFT) approach~\cite{Buchalla:2015wfa, Goertz:2014qta}.
The production of $\PHiggs$ boson pairs in the absence of new resonances is referred to as non-resonant $\dihiggs$ production,
the case that we focus on in this paper.


The ATLAS and CMS collaborations have searched for non-resonant $\dihiggs$ production in the decay channels 
$\dihiggs \to \Pbottom\APbottom\Pbottom\APbottom$, $\Pbottom\APbottom\Pgt\Pgt$, $\Pbottom\APbottom\PW\PW\virt$, $\Pbottom\APbottom\Pgamma\Pgamma$
using the data recorded during LHC Runs $1$ and $2$~\cite{HIG-13-032,HIG-15-013,HIG-17-006,HIG-17-030,Sirunyan:2020xok,Aad:2015xja,Aaboud:2018knk,Aaboud:2018ftw,Aaboud:2018sfw,Aaboud:2018zhh}.
ATLAS has further performed searches in the decay channels $\dihiggs \to \Pgamma\Pgamma\PW\PW\virt$ and $\PW\PW\virt\PW\PW\virt$~\cite{Aad:2015xja,Aaboud:2018ewm,Aaboud:2018ksn}.
The asterisk ($\virt$) denotes $\PW$ bosons that are off-shell.
Phenomenological studies of non-resonant $\dihiggs$ production are presented in 
Refs.~\cite{Baur:2002rb,Baur:2002qd,Baur:2003gpa,Baur:2003gp,Dolan:2012rv,Papaefstathiou:2012qe,Baglio:2012np,deLima:2014dta,Wardrope:2014kya,Behr:2015oqq,Li:2015yia,Adhikary:2017jtu}.
No evidence for a signal has been found in the LHC data so far.
The present analyses are able to probe the existence of a SM-like $\dihiggs$ signal produced with a cross section of order $10$ times the SM production rate.

The decay channel providing the highest sensitivity for an SM-like $\dihiggs$ signal 
is the $\Pbottom\APbottom\Pgt\Pgt$ channel in case of ATLAS~\cite{Aaboud:2018sfw} and the $\Pbottom\APbottom\Pgamma\Pgamma$ channel in case of CMS~\cite{Sirunyan:2020xok}.
Both channels provide a favorable signal-to-background ratio and are limited mainly by statistical uncertainties at present,
resulting from the limited amount of data that has been recorded so far, compared to the small SM $\ggHH$ production cross section.
The channels $\Pbottom\APbottom\Pbottom\APbottom$ and $\Pbottom\APbottom\PW\PW\virt$ provide a significantly larger signal rate,
but suffer from sizeable backgrounds,
arising from QCD multijet production in case of the $\Pbottom\APbottom\Pbottom\APbottom$ channel 
and from top quark pair ($\ttbar$) production in case of the $\Pbottom\APbottom\PW\PW\virt$ channel.
In this paper, we focus on the $\Pbottom\APbottom\PW\PW\virt$ channel,
and in particular on events in which both $\PW$ bosons decay to leptons (electrons or muons).
The latter are denoted by the symbol $\Plepton$.
 
\begin{figure}
\setlength{\unitlength}{1mm}
\begin{center}
\begin{tabular}{ccc}
\raisebox{-.45\height}{\resizebox{0.4\textwidth}{!}{\includegraphics[bb=0 0 214 172]{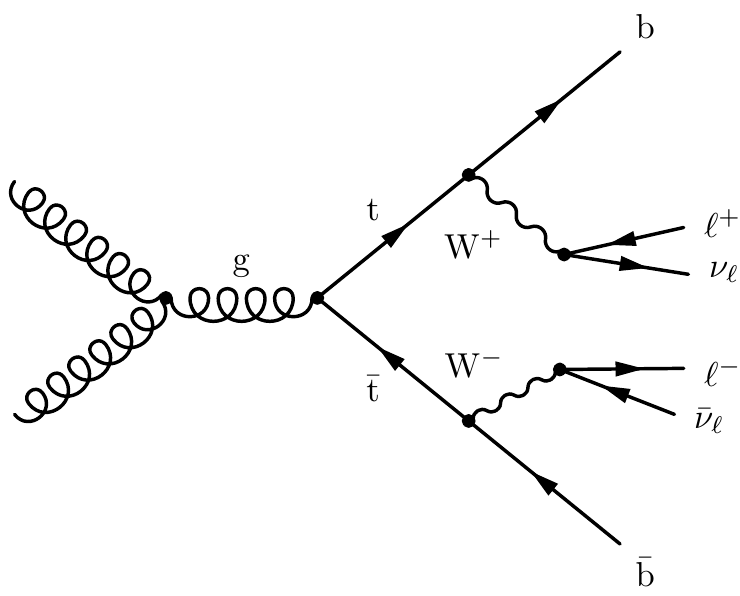}}} &
\qquad + &
\raisebox{-.45\height}{\resizebox{0.4\textwidth}{!}{\includegraphics[bb=0 0 214 172]{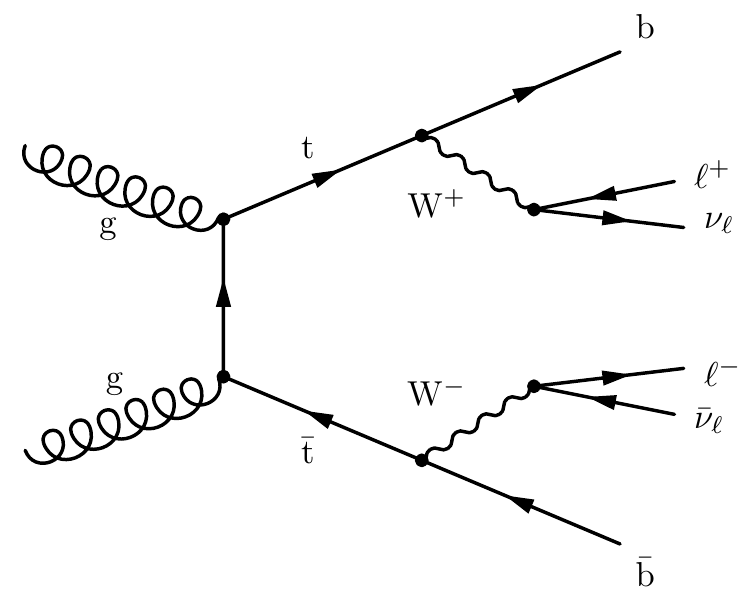}}}
\end{tabular}
\end{center}
\caption{
  LO Feynman diagrams for $\ttbar$ production in $\Pp\Pp$ collisions at the LHC,
  with subsequent decay of the top quark pair via $\ttbar \to \Pbottom\PW\APbottom\PW \to \Pbottom\ellPlusnu \, \APbottom\ellMinusnu$.
}
\label{fig:ttbar_FeynmanDiagram}
\end{figure}

The separation of the $\dihiggs$ signal from the large $\ttbar$ background constitutes the main experimental challenge in the $\Pbottom\APbottom\PW\PW\virt$ channel.
For a top quark mass of $m_{\Ptop} = 172.8$~\GeV~\cite{PDG},
the cross section for $\ttbar$ production amounts to $825.9^{+46.1}_{-50.5}$~\pb at $\sqrt{s} = 13$~\TeV center-of-mass energy~\cite{Czakon:2011xx}.
The $\ttbar$ background is irreducible in this channel, as it produces the exact same multiplicity of charged leptons, neutrinos, and $\Pbottom$-jets as the $\dihiggs$ signal.
The LO Feynman diagrams for $\ttbar$ production are shown in Fig.~\ref{fig:ttbar_FeynmanDiagram}.
The main handle to separate the $\dihiggs$ signal from the $\ttbar$ background is the difference in event kinematics,
that is, the differences in the distributions of the energies and angles of the charged leptons, of the $\Pbottom$-jets, and of the missing transverse momentum reconstructed in the event.

The present CMS analysis~\cite{HIG-17-006} utilizes machine-learning methods, based on a Deep Neural Network~\cite{ANN,chollet2015keras},
to separate the $\dihiggs$ signal from the $\ttbar$ background, while the current ATLAS analysis~\cite{Aaboud:2018zhh} employs a sequence of hard cuts for this purpose.
In this paper, we propose an alternative multivariate method for the separation of the $\dihiggs$ signal from the $\ttbar$ background,
the matrix element method (MEM)~\cite{Kondo:1988yd,Kondo:1991dw}.

The paper is structured as follows:
In Section~\ref{sec:mem} we describe the MEM and its application to the $\Pbottom\APbottom\PW\PW\virt$ channel.
The application of the MEM requires the computation of multi-dimensional integrals.
The evaluation of the integrals is performed numerically and demands a significant amount of computing time, in the order of a few seconds per event.
In order to make the integrals suitable for numeric integration, analytic transformations need to be performed.
The most relevant of these transformations will be described in Section~\ref{sec:mem} and further details will be given in the appendix.
The performance of the MEM in separating the $\dihiggs$ signal from the $\ttbar$ background is studied in Section~\ref{sec:performance}.
The performance is studied on Monte-Carlo truth and on detector level,
for experimental conditions that are characteristic for the ATLAS and CMS experiments during LHC Run $2$.
The latter are simulated using the $\textsc{DELPHES}$ fast-simulation framework~\cite{deFavereau:2013fsa}.
Section~\ref{sec:performance} also presents a study of the effect of using matrix elements of leading order when applying the MEM to the $\Pbottom\APbottom\PW\PW\virt$ channel
and discusses the computing-time requirements of the MEM.
We conclude the paper with a summary in Section~\ref{sec:summary}.

\section{The Matrix element method}
\label{sec:mem}

The MEM computes probability densities (PDs) $w_{i}(\vecy)$
for obtaining the measured observables $\vecy$, assuming that the event has been produced by the process $i$.
The PDs $w_{i}(\vecy)$ are interpreted as quantifying the compatibility of the measured observables $\vecy$
with the signal ($i=0$) and background ($i=1$) hypothesis.
In the analysis of $\PHiggs\PHiggs$ production in the decay channel 
$\PHiggs\PHiggs \to \Pbottom\APbottom\PW\PW\virt \to \Pbottom\APbottom \, \ellPlusnu\ellMinusnu$
the observables $\vecy$ refer to 
the measured momenta of the two $\Pbottom$-jets, of the two charged leptons, and of the measured missing transverse momentum ($\vecMET$) in the event.
The vector $\vecMET$ represents the measured value of the vectorial sum of the two neutrino momenta in the plane transverse to the beam axis.
We use symbols with a hat to denote the true values of energies and momenta.
Bold letters denote vector quantities.
The vector $\vecyhat$ denotes the true values of the $\Pbottom$-jet and charged lepton momenta and the true values of the momenta of the two neutrinos produced in the $\PW$ boson decays.

As already mentioned, $\PHiggs\PHiggs \to \Pbottom\APbottom\PW\PW\virt \to \Pbottom\APbottom \, \ellPlusnu\ellMinusnu$ signal events
contain the same number of $\Pbottom$-jets, charged leptons, and neutrinos as the dominant background,
arising from $\ttbar \to \Pbottom\PW\APbottom\PW \to \Pbottom\ellPlusnu \, \APbottom\ellMinusnu$.
The separation of the $\PHiggs\PHiggs$ signal from the irreducible $\ttbar$ background is based on the difference in event kinematics,
causing the PD $w_{0}(\vecy)$ to be in general higher when evaluated on signal events
and lower when evaluated on background events, and vice versa for the PD $w_{1}(\vecy)$.
Given the PDs $w_{0}(\vecy)$ and $w_{1}(\vecy)$ for the signal and background hypotheses,
the Neyman-Pearson lemma~\cite{Neyman:1933wgr} postulates that the likelihood ratio (LR):
\begin{linenowrapper}
\begin{equation}
P(\vecy) = \frac{w_{0}(\vecy)}{w_{0}(\vecy) + w_{1}(\vecy)}
\label{eq:memLR}
\end{equation}
\end{linenowrapper}
provides the optimal separation of the $\PHiggs\PHiggs$ signal from the irreducible $\ttbar$ background.

Different nomenclatures and conventions for the MEM exist in the literature.
In this note, we follow the nomenclature and conventions introduced in Ref.~\cite{Volobouev:2011vb}.
The PDs $w_{i}(\vecy)$ are given by the integral:
\begin{linenowrapper}
\begin{align}
w_{i}(\vecy) = & \frac{\Omega(\vecy)}{\sigma_{i} \cdot \mathcal{A}_{i}} \, \int \, d\xhat_{a} \, d\xhat_{b} \,
  d\Phi_{n} \, \frac{f(\xhat_{a}, Q) \, f(\xhat_{b}, Q)}{2 \, \xhat_{a} \, \xhat_{b} \, s} \, (2\pi)^{4} \,
  \delta\left( \xhat_{a} \, \Ehat_{a} + \xhat_{b} \, \Ehat_{b} - \sum_{k}^{n} \Ehat_{(k)} \right) \nonumber \\
 & \quad \cdot \, \delta^{3}\left( \xhat_{a} \, \vecphat^{a} + \xhat_{b} \, \vecphat^{b} - \sum_{k}^{n} \vecphat^{(k)}\right) \, 
  \vert \mathcal{M}_{i}(\vecyhat) \vert^{2} \, W(\vecy|\vecyhat) \, \epsilon_{i}(\vecyhat) \, .
\label{eq:mem1}
\end{align}
\end{linenowrapper}
The symbol $\vert \mathcal{M}_{i}(\vecyhat) \vert^{2}$ denotes the squared modulus of the matrix element (ME),
averaged over helicity states,
for either the signal ($i=0$) or for the background ($i=1$) hypothesis.
We use ME generated at LO accuracy with the program $\textsc{MadGraph\_aMCatNLO}$ $2.3.3$~\cite{MadGraph_aMCatNLO} for the signal as well as for the background hypothesis.
The ME for the signal hypothesis is generated using the infinite top quark mass approximation~\cite{Grinstein:2007iv}.
In this approximation, the top quark degrees of freedom are integrated out,
replacing the top quark loops in the triangle and box diagrams shown in Fig.~\ref{fig:ggHH_FeynmanDiagram} by point-like effective couplings of, respectively, 
two gluons to one $\PHiggs$ boson and two gluons to two $\PHiggs$ bosons.
The point-like effective couplings are specified via a Universal $\textsc{FeynRules}$ Output (UFO) model file~\cite{Degrande:2011ua,Hespel:2014sla} given to $\textsc{MadGraph\_aMCatNLO}$.
The usage of the infinite top quark mass approximation is necessary, because at present the program $\textsc{MadGraph\_aMCatNLO}$ does not support the generation of ME code for processes involving loops.

The symbols $\Ehat_{a}$ and $\Ehat_{b}$ ($\vecphat^{a}$ and $\vecphat^{b}$) denote the energies (momenta) of the two colliding protons,
$\sqrt{s}$ their center-of-mass energy,
$\xhat_{a}$ and $\xhat_{b}$ the Bjorken scaling variables~\cite{Bjorkenx},
and $f(\xhat_{a}, Q)$ and $f(\xhat_{b}, Q)$ the corresponding parton distribution functions (PDFs)~\cite{LHAPDF}.
Besides their dependence on the Bjorken scaling variables, 
the PDFs depend on a scale parameter $Q$.
We use the \textrm{MSTW} 2008 LO PDF set~\cite{MSTW} to evaluate $f(\xhat_{a}, Q)$ and $f(\xhat_{b}, Q)$
and set the scale $Q$ to half the mass of the $\PHiggs\PHiggs$ system and to twice the value of the top quark mass
when computing the PD $w_{0}(\vecy)$ and $w_{1}(\vecy)$ for the signal and background hypothesis, respectively.
We denote by $n$ the number of particles in the final state,
and by $\vecp^{(k)}$ ($\vecphat^{(k)}$) the measured (true) momentum of the $k$-th final state particle. 
The $\delta$-functions $\delta( \xhat_{a} \, \Ehat_{a} + \xhat_{b} \, \Ehat_{b} - \sum_{k}^{n} \Ehat_{(k)})$
and $\delta^{3}( \xhat_{a} \, \vecphat^{a} + \xhat_{b} \, \vecphat^{b} - \sum_{k}^{n} \vecphat^{(k)})$ 
impose conservation of energy and momentum.

The functions $W(\vecy|\vecyhat)$ are referred to as ``transfer functions'' (TF) in the literature.
They represent the PD to observe the measured values $\vecy$, given the true values $\vecyhat$.
The function $\Omega(\vecy)$ is referred to as ``indicator function'' in the literature~\cite{Fiedler:2010sg,Volobouev:2011vb}.
It attains the value $1$ in case the event represented by the measured observables $\vecy$ passes the event selection criteria and otherwise attains the value $0$.
The efficiency for an event originating at the phase-space (PS) point
$\vecyhat$ to pass the event selection, \ie to end up with measured
observables $\vecy$ for which $\Omega(\vecy) = 1$,
is denoted by $\epsilon_{i}(\vecyhat)$. 
Finally, the symbol $\mathcal{A}_{i}$ denotes the acceptance of the event selection, 
that is, the percentage of events which pass the event selection criteria,
while $\sigma_{i}$ denotes the cross section of process $i$.
The subscript $i$ of the symbols $\sigma_{i}$, $\mathcal{A}_{i}$, and $\epsilon_{i}$ 
emphasize that the cross section, acceptance, and efficiency differ between the signal and the background hypothesis.
Division of the right-hand-side (RHS) of Eq.~(\ref{eq:mem1}) by the product $\sigma_{i} \cdot \mathcal{A}_{i}$
ensures that $w_{i}(\vecy)$ has the correct normalization required for a probability density, 
\ie $\int \, d\vecy \, w_{i}(\vecy) = 1$,
provided that the TF satisfy the normalization condition
$\int \, d\vecy \, \Omega(\vecy) \, W(\vecy|\vecyhat) = 1$
for every $\vecyhat$.

The symbol $d\Phi_{n} = \prod_{k}^{n} \, \frac{d^{3}\vecphat^{(k)}}{(2\pi)^{3} \, 2 \Ehat_{(k)}}$ 
represents the differential $n$-particle PS element.
For the $\PHiggs\PHiggs$ signal 
as well as for the $\ttbar$ background hypothesis, $n=6$.  
We express the PS element $d\Phi_{6}$ in terms of the energies $\Ehat_{(k)}$, the polar angles $\thetahat_{(k)}$, and the azimuthal angles $\phihat_{(k)}$ 
of the two $\Pbottom$ quarks, of the two charged leptons, and of the two neutrinos:
\begin{linenowrapper}
\begin{eqnarray}
d\Phi_{6} 
 & = &\prod_{k}^{6} \, \frac{d^{3}\vecphat^{(k)}}{(2 \, \pi)^{3} \, 2 \, \Ehat_{(k)}} 
  = \frac{1}{2^{24} \, \pi^{18}} \, \prod_{k}^{6} \, 
\frac{d^{3}\vecphat^{(k)}}{\Ehat_{(k)}} \nonumber \\
 & = & \frac{1}{2^{24} \, \pi^{18}} \, \prod_{k}^{6} \, 
\frac{d\Ehat_{(k)} \, d\thetahat_{(k)} \, d\phihat_{(k)} \, |\vecphat^{(k)}| \, \Ehat_{(k)} \, \sin\thetahat_{(k)}}{\Ehat_{(k)}} \nonumber \\
 & = & \frac{1}{2^{24} \, \pi^{18}} \, \prod_{k}^{6} \, 
d\Ehat_{(k)} \, d\thetahat_{(k)} \, d\phihat_{(k)} \, \betahat_{(k)} \Ehat_{(k)} \, \sin\thetahat_{(k)} \, .
\label{eq:PS_inPolarCoordinates}
\end{eqnarray}
\end{linenowrapper}
All energies $\Ehat_{(k)}$ as well as the angles $\thetahat_{(k)}$ and $\phihat_{(k)}$ refer to the laboratory (detector) frame.
The velocity $\betahat_{(k)}$ of particle $k$,
given by $\betahat_{(k)} \equiv \frac{|\vecphat^{(k)}|}{\Ehat_{(k)}}$,
has been used to simplify the expression for $d\Phi_{6}$ in the last step.
Note that the velocity $\betahat_{(k)}$ is a function of energy $\Ehat_{(k)}$ and hence cannot be treated as constant when evaluating the integral over $d\Ehat_{(k)}$.
Similarly, the magnitude of the momentum $|\vecphat^{(k)}|$ is a function of the energy $\Ehat_{(k)}$.
In the following, we use the identities $|\vecphat^{(k)}| = \sqrt{\Ehat_{(k)}^{2} - m_{(k)}^{2}}$ 
and $\betahat_{(k)} = \frac{\sqrt{\Ehat_{(k)}^{2} - m_{(k)}^{2}}}{\Ehat_{(k)}}$ to make the dependency on the energy $\Ehat_{(k)}$ explicit.

The form of Eq.~(\ref{eq:PS_inPolarCoordinates}) is useful, 
as it allows to trivially perform the integration over the 
angles $\thetahat_{(k)}$ and $\phihat_{(k)}$ for the two $\Pbottom$ quarks and for the two charged leptons,
taking advantage of the fact that the directions of quarks (jets) and charged leptons can be measured with negligible experimental resolution.
With the further assumption that also the energy of charged leptons can be measured with negligible experimental resolution,
the integration over $d\Ehat_{\Plepton^{+}}$ and $d\Ehat_{\Plepton^{-}}$ can be carried out trivially too.
We shall only consider events that pass the event selection criteria, \ie for which the indicator function $\Omega(\vecy)$ is equal to $1$.
For simplicity, we neglect the effect of the efficiency $\epsilon_{i}(\vecyhat)$ and of the acceptance $\mathcal{A}_{i}$.
With these assumptions and upon inserting the expressions for the TF given by Eqs.~(\ref{eq:TF_ell}),~(\ref{eq:TF_b}), and~(\ref{eq:TF_f}) in the appendix
into Eq.~(\ref{eq:PS_inPolarCoordinates}), we obtain:
\begin{linenowrapper}
\begin{align}
w_{i}(\vecy) 
 = & \frac{1}{2^{21} \, \pi^{14} \, \sigma_{i} \, E_{\Plepton^{+}} \, E_{\Plepton^{-}} \, E_{\Pbottom} \, E_{\APbottom}} \, \int \, d\xhat_{a} \, d\xhat_{b} \,
d\Ehat_{\Pbottom} \, d\Ehat_{\APbottom} \, \frac{d^{3}\vecphat_{\Pnu}}{\Ehat_{\Pnu}} \, \frac{d^{3}\vecphat_{\APnu}}{\Ehat_{\APnu}} \,
\frac{f(\xhat_{a}, Q) \, f(\xhat_{b}, Q)}{\xhat_{a} \, \xhat_{b} \, s} \nonumber \\
 & \quad \cdot \, \delta\left( \xhat_{a} \, \Ehat_{a} + \xhat_{b} \, \Ehat_{b} - \sum_{k}^{6} \Ehat_{(k)}\right) \,
\delta^{3}\left( \xhat_{a} \, \vecphat^{a} + \xhat_{b} \, \vecphat^{b} - \sum_{k}^{6} \vecphat^{(k)}\right) \nonumber \\
 & \quad \cdot \, \vert \mathcal{M}_{i}(\vecphat) \vert^{2}\, \frac{\betahat_{\Pbottom} \, \Ehat_{\Pbottom}}{\beta_{\Pbottom} \, E_{\Pbottom}} \, W(E_{\Pbottom}|\Ehat_{\Pbottom}) \, 
\frac{\betahat_{\APbottom} \, \Ehat_{\APbottom}}{\beta_{\APbottom} \, E_{\APbottom}} \, W(E_{\APbottom}|\Ehat_{\APbottom}) \, .
\label{eq:mem2}
\end{align}
\end{linenowrapper}
The terms $\frac{\betahat_{\Pbottom} \, \Ehat_{\Pbottom}}{\beta_{\Pbottom} \, E_{\Pbottom}}$ and $\frac{\betahat_{\APbottom} \, \Ehat_{\APbottom}}{\beta_{\APbottom} \, E_{\APbottom}}$ 
arise because the integration over the PS elements $d^{3}\vecphat$ of the $\Pbottom$ and $\APbottom$ quarks yields a factor $\betahat \, \Ehat^{2} \, \sin\thetahat$,
while the normalization of the TF yields a factor $\frac{1}{\beta \, E^{2} \, \sin\theta}$, \cf Eq.~(\ref{eq:TF_f}).
The terms $\sin\thetahat$ and $\frac{1}{\sin\theta}$ cancel, due to the presence of the $\delta$-function $\delta(\theta - \thetahat)$ in the integrand, \cf Eq.~(\ref{eq:TF_b}).
No similar terms arise for the charged leptons, as the TF for charged leptons demand $\betahat = \beta$, $\Ehat = E$, and $\thetahat = \theta$, \cf Eq.~(\ref{eq:TF_ell}).

We simplify the four-dimensional $\delta$-function 
$\delta( \xhat_{a} \, \Ehat_{a} + \xhat_{b} \, \Ehat_{b} - \sum_{k}^{6} \Ehat_{(k)}) \cdot \delta^{3}( \xhat_{a} \, \vecphat^{a} + \xhat_{b} \, \vecphat^{b} - \sum_{k}^{6} \vecphat^{(k)})$
by assuming the momentum vectors of the colliding protons to be aligned in direction parallel and anti-parallel to the beam axis 
and neglecting the small transverse momenta of the partons within the protons as well as parton masses.
With this assumption, we can eliminate the energy and longitudinal momentum components of the $\delta$-function 
and solve for the Bjorken scaling variables $\xhat_{a}$ and $\xhat_{b}$ as function of the energies and longitudinal momenta of the particles in the final state.
This yields:
\begin{linenowrapper}
\begin{equation}
\xhat_{a} = \frac{1}{\sqrt{s}} \, \sum_{k}^{6} \left( \Ehat_{(k)} + \pZhat^{(k)} \right) \quad \mbox{ and } \quad
\xhat_{b} = \frac{1}{\sqrt{s}} \, \sum_{k}^{6} \left( \Ehat_{(k)} - \pZhat^{(k)} \right) \, .
\label{eq:Bjorkenx}
\end{equation}
\end{linenowrapper}

For the purpose of eliminating the transverse momentum components of the four-dimensional $\delta$-function,
we follow the approach of Ref.~\cite{SVfitMEM}.
The approach is based on introducing the ``hadronic recoil'', denoted by the symbol $\rho$, as a means to account for QCD radiation,
which causes additional jets to be produced besides the two $\Pbottom$-jets that originate from the decay of the $\PHiggs$ boson (in signal events) 
or from the decay of the two top quarks (in background events).
As detailed in Ref.~\cite{Alwall:2010cq}, significant amounts of QCD radiation, in particular initial-state radiation (ISR),
are a typical feature of most signal and background processes at the LHC.
The longitudinal momentum of the additional jets produced by QCD radiation alters the relations for $\xhat_{a}$ and $\xhat_{b}$ somewhat,
compared to the values given by Eq.~(\ref{eq:Bjorkenx}).
We expect the effect of QCD radiation on the energy and longitudinal momentum components to be small and thus neglect it.
The effect on the transverse momentum balance is important, however,
as QCD radiation distorts the kinematic relations that would be expected to hold in the absence of such radiation.
As a consequence, the $\delta$-functions that ensure the conservation of momentum in the transverse plane need to be modified. 
Their modified form reads: 
$\delta( \pXhat^{\rho} + \sum_{k}^{6} \pXhat^{(k)})$ and $\delta( \pYhat^{\rho} + \sum_{k}^{6} \pYhat^{(k)})$,
where $\pXhat^{\rho}$ and $\pYhat^{\rho}$ denote the true value of the momentum of the hadronic recoil $\rho$ in $x$ and $y$ direction, respectively.
They imply the relations:
\begin{linenowrapper}
\begin{equation}
\pXhat^{\rho} = - \left( \pXhat^{\Pbottom} + \pXhat^{\APbottom} + \pXhat^{\Plepton^{+}} + \pXhat^{\Pnu} + \pXhat^{\Plepton^{-}} + \pXhat^{\APnu} \right) \, \mbox{ and } \,
\pYhat^{\rho} = - \left( \pYhat^{\Pbottom} + \pYhat^{\APbottom} + \pYhat^{\Plepton^{+}} + \pYhat^{\Pnu} + \pYhat^{\Plepton^{-}} + \pYhat^{\APnu} \right) \, .
\label{eq:hadRecoil_true}
\end{equation}
\end{linenowrapper}
The corresponding relations for the measured momenta read:
\begin{linenowrapper}
\begin{equation}
\pX^{\rho} = - \left( \pX^{\Pbottom} + \pX^{\APbottom} + \pX^{\Plepton^{+}} + \pX^{\Plepton^{-}} + \METx \right) \, \mbox{ and } \,
\pY^{\rho} = - \left( \pY^{\Pbottom} + \pY^{\APbottom} + \pY^{\Plepton^{+}} + \pY^{\Plepton^{-}} + \METy \right) \, .
\label{eq:hadRecoil}
\end{equation}
\end{linenowrapper}
We use Eq.~(\ref{eq:hadRecoil}) to compute the measured values of $\pX^{\rho}$ and $\pY^{\rho}$, 
given the measured momenta of the two $\Pbottom$-jets, of the two charged leptons, and of the measured $\vecMET$.
The experimental resolution on $\pX^{\rho}$ and $\pY^{\rho}$ is accounted for by introducing a TF for the hadronic recoil into the integrand of Eq.~(\ref{eq:mem2}).
We assume that the resolution on the transverse momentum components of $\rho$ follows a two-dimensional normal distribution:
\begin{linenowrapper}
\begin{equation}
W_{\rho}( \pX^{\rho},\pY^{\rho} | \pXhat^{\rho},\pYhat^{\rho} ) = 
 \frac{1}{2\pi \, \sqrt{\vert V \vert}} \, \exp \left( -\frac{1}{2}
 \left( \begin{array}{c} \pX^{\rho} - \pXhat^{\rho} \\ \pY^{\rho} - \pYhat^{\rho} \end{array} \right)^{T}
  \cdot V^{-1} \cdot
   \left( \begin{array}{c} \pX^{\rho} - \pXhat^{\rho} \\ \pY^{\rho} - \pYhat^{\rho} \end{array} \right)
 \right) \, ,
\label{eq:TF_hadRecoil}
\end{equation}
\end{linenowrapper}
where the matrix $V$ quantifies the resolution on the hadronic recoil in the transverse plane.

The CMS collaboration computes the matrix $V$ on an event-by-event basis, using an algorithm referred to as the ``$\MET$-significance'' algorithm~\cite{JME-10-009}.
Alternatively, one could determine an average resolution $\sigma_{\rho}$ for a sample of $\dihiggs$ signal and $\ttbar$ background events using the Monte Carlo simulation
and take the matrix $V$ to be $V = \sigma_{\rho}^{2} \cdot I_{2}$, where $I_{2}$ denotes the identity matrix of size $2$.
We follow the procedure detailed in Ref.~\cite{SVfitMEM} and replace the $\delta$-functions 
$\delta( \pXhat^{\rho} + \sum_{k}^{6} \pXhat^{(k)})$ and $\delta( \pYhat^{\rho} + \sum_{k}^{6} \pYhat^{(k)})$,
which ensure the momentum conservation in the transverse plane,
with the TF for the hadronic recoil, given by Eq.~(\ref{eq:TF_hadRecoil}).

A remaining issue is that we use LO ME $\mathcal{M}_{i}(\vecphat)$ for the $\PHiggs\PHiggs$ signal and for the $\ttbar$ background in Eq.~(\ref{eq:mem2}).
The LO ME for the signal (background) requires that the $\PHiggs\PHiggs$ ($\ttbar$) system has zero $\pT$, 
a condition that only holds in case the hadronic recoil has zero $\pT$.
As previously discussed, the case that the hadronic recoil has negligible $\pT$ is rare at the LHC, due to the abundance of QCD radiation.
The issue that the LO ME is only well-defined for events with zero ISR
is resolved by evaluating the ME $\mathcal{M}_{i}(\vecyhat)$ in a frame in which the $\PHiggs\PHiggs$ ($\ttbar$) system has zero $\pT$, 
to which we refer as the zero-transverse-momentum (ZTM) frame.
The Lorentz transformation of the energy $\Ehat_{(k)}$ and momenta $\vecphat^{(k)}$ in Eq.~(\ref{eq:mem2})
from the laboratory to the ZTM frame is performed using the vector $\left(-\frac{\pXhat^{\rho}}{\pThat^{\rho}},-\frac{\pYhat^{\rho}}{\pThat^{\rho}},0\right)$ as the boost vector.
The values of $\pXhat^{\rho}$, $\pYhat^{\rho}$, and $\pThat^{\rho}$ are computed using Eq.~(\ref{eq:hadRecoil_true}).
The momentum components $\pXhat^{\Plepton^{+}}$, $\pXhat^{\Plepton^{-}}$, $\pYhat^{\Plepton^{+}}$, and $\pYhat^{\Plepton^{-}}$ are set to their measured values,
while the components $\pXhat^{\Pbottom}$, $\pXhat^{\APbottom}$, $\pXhat^{\Pnu}$, $\pXhat^{\APnu}$, $\pYhat^{\Pbottom}$, $\pYhat^{\APbottom}$, $\pYhat^{\Pnu}$, and $\pYhat^{\APnu}$
are recomputed as function of the integration variables $\Ehat_{\Pbottom}$, $\Ehat_{\APbottom}$, $\vecphat_{\Pnu}$, and $\vecphat_{\APnu}$ when evaluating Eq.~(\ref{eq:hadRecoil_true}).

Eliminating the energy and longitudinal momentum components of the four-dimensional $\delta$-function 
$\delta( \xhat_{a} \, \Ehat_{a} + \xhat_{b} \, \Ehat_{b} - \sum_{k}^{6} \Ehat_{(k)}) \cdot \delta( \xhat_{a} \, \pZhat^{a} + \xhat_{b} \, \pZhat^{b} - \sum_{k}^{6} \pZhat^{(k)}) \cdot \delta( \pXhat^{\rho} + \sum_{k}^{6} \pXhat^{(k)}) \cdot \delta( \pYhat^{\rho} + \sum_{k}^{6} \pYhat^{(k)})$
by means of Eq.~(\ref{eq:Bjorkenx})
and replacing its transverse momentum components by the TF $W_{\rho}( \pX^{\rho},\pY^{\rho} | \pXhat^{\rho},\pYhat^{\rho} )$ for the hadronic recoil $\rho$,
the expression for the PD $w_{i}(\vecy)$ in Eq.~(\ref{eq:mem2}) becomes:
\begin{linenowrapper}
\begin{align}
w_{i}(\vecy) 
 = & \frac{1}{2^{21} \, \pi^{14} \, \sigma_{i} \, E_{\Plepton^{+}} \, E_{\Plepton^{-}} \, E_{\Pbottom} \, E_{\APbottom}} \, \int \, 
d\Ehat_{\Pbottom} \, d\Ehat_{\APbottom} \, d\Ehat_{\Pnu} \, d\thetahat_{\Pnu} \, d\phihat_{\Pnu} \, d\Ehat_{\APnu} \, d\thetahat_{\APnu} \, d\phihat_{\APnu} \nonumber \\
 & \quad \cdot \, \betahat_{\Pnu} \, \Ehat_{\Pnu} \, \sin\thetahat_{\Pnu} \, 
\betahat_{\APnu} \, \Ehat_{\APnu} \, \sin\thetahat_{\APnu} \, 
\frac{f(\xhat_{a}, Q) \, f(\xhat_{b}, Q)}{\xhat_{a} \, \xhat_{b} \, s} \nonumber \\
 & \quad \cdot \, \vert \mathcal{M}_{i}(\vecphat) \vert^{2} \, 
\frac{\betahat_{\Pbottom} \, \Ehat_{\Pbottom}}{\beta_{\Pbottom} \, E_{\Pbottom}} \, W(E_{\Pbottom}|\Ehat_{\Pbottom}) \, 
\frac{\betahat_{\APbottom} \, \Ehat_{\APbottom}}{\beta_{\APbottom} \, E_{\APbottom}} \, W(E_{\APbottom}|\Ehat_{\APbottom}) \, W_{\rho}( \pX^{\rho},\pY^{\rho} | \pXhat^{\rho},\pYhat^{\rho} ) \, .
\label{eq:mem3}
\end{align}
\end{linenowrapper}
The expression in Eq.~(\ref{eq:mem3}) concludes our discussion of analytic transformations of the expressions for the PD $w_{i}(\vecy)$ 
that are common to the signal as well as to the background hypothesis.

A few more analytic transformations need to be performed to handle the presence of Breit-Wigner (BW) propagators in the ME $\mathcal{M}_{i}(\vecphat)$,
as the presence of these propagators represent an obstacle for the numeric integration of Eq.~(\ref{eq:mem3}).
The effect of the BW propagators is that only narrow slices in the $6$-particle PS yield sizeable contributions to the integral,
namely the regions where the $6$ final state particles satisfy certain mass constraints.
The mass constraints arise from the presence of on-shell $\PHiggs$ bosons, $\PW$ bosons, and top quarks in the decay chains
$\PHiggs\PHiggs \to \Pbottom\APbottom\PW\PW\virt \to \Pbottom\APbottom \, \ellPlusnu\ellMinusnu$ and
$\ttbar \to \Pbottom\PW\APbottom\PW \to \Pbottom\ellPlusnu \, \APbottom\ellMinusnu$.
Their presence renders the numeric integration inefficient, unless the mass constraints are treated analytically. 
We use the narrow-width approximation (NWA)~\cite{NWA} to handle the mass constraints and replace the BW propagators by $\delta$-functions.
The NWA has the effect of restricting the numerical integration to the narrow slices in the $6$-particle PS where the mass constraints are satisfied
and the ME $\mathcal{M}_{i}(\vecphat)$ yields a sizeable contribution to the integral.
The analytic transformations that are needed to handle the BW propagators differ for the signal and for the background hypothesis,
reflecting the presence of different resonances in the respective decay chains.
The transformations that are specific to the signal hypothesis are detailed in Section~\ref{sec:mem_signal},
while those specific to the background hypothesis are presented in Section~\ref{sec:mem_background}.

Finally, the numeric integration is performed using the VAMP algorithm~\cite{VAMP}, a variant of the popular VEGAS algorithm~\cite{VEGAS},
which has been optimized for the case of integrating multimodal functions that typically appear in the integration of ME over regions in PS.
We use $2500$ evaluations of the integrand when computing the PD $w_{0}(\vecy)$ for the signal hypothesis 
and $25000$ evaluations of the integrand for the computation of the PD $w_{1}(\vecy)$ for the background hypothesis.
The number of evaluations has been chosen such that the computation of $w_{0}(\vecy)$ and $w_{1}(\vecy)$ take approximately the same time
and the computation of the likelihood ratio $P(\vecy)$ takes about one minute per event,
using a single core of a $2.30$~GHz Intel\TReg~Xeon\TReg~E5-2695V3 processor.

\subsection{Analytic transformations specific to the signal hypothesis}
\label{sec:mem_signal}

When evaluating the integrand in Eq.~(\ref{eq:mem3}) for the signal hypothesis,
only those points in the $6$-particle PS provide a sizeable contribution to the value of the integral $w_{0}(\vecy)$
which satisfy the following conditions:
\begin{itemize}
\item The mass of the $2$-particle system comprised of the two $\Pbottom$ quarks equals $m_{\PHiggs} = 125.1$~\GeV~\cite{HIG-14-042}.
\item The mass of the $2$-particle system comprised of the charged lepton and of the neutrino, which originate from the decay of the on-shell $\PW$ boson, equals $m_{\PW} = 80.4$~\GeV~\cite{PDG}.
\item The mass of the $4$-particle system comprised of the two charged leptons and of the two neutrinos equals $m_{\PHiggs}$.
\end{itemize}

We formally introduce these mass constraints by inserting three $\delta$-functions $\delta\left( g(x) \right)$ into the integrand of Eq.~(\ref{eq:mem3}).
The procedure is explained in Section~\ref{sec:appendix_mass_constraints} of the appendix.
More specifically, we insert
one $\delta$-function of the type $g(\Ehat_{\APbottom})$ given by Eq.~(\ref{eq:bEn_Hbb1}), 
one of the type $g(\Ehat_{\Pnu})$ given by Eq.~(\ref{eq:nuEn_Wlnu1}), and one of the type $g(\Ehat_{\nuStar})$ given by Eq.~(\ref{eq:nuEn_Hww1}) into the integrand of Eq.~(\ref{eq:mem3}).
We denote the charged lepton and the neutrino originating from the decay of the off-shell $\PW$ boson, which can be either the $\PW^{+}$ or the $\PW^{-}$, by an asterisk.
The charged lepton and the neutrino that are referred to without asterisks are subject to the $\PW$ mass constraint.

After solving for the $\delta$-functions analytically, as detailed in Sections~\ref{sec:appendix_bEn_Hbb}, ~\ref{sec:appendix_nuEn_Wlnu}, and~\ref{sec:appendix_nuEn_Hww} of the appendix,
the resulting expression for the PD $w_{0}(\vecy)$ of the signal hypothesis reads:
\begin{linenowrapper}
\begin{align}
w_{0}(\vecy) 
 = & \frac{(m_{\PHiggs} \, \Gamma_{\PHiggs})^{2} \, m_{\PW} \, \Gamma_{\PW}}{2^{23} \, \pi^{14} \, \sigma_{0} \, s \, 
  E_{\Plepton}^{2} \, E_{\ellStar} \, \beta_{\Pbottom} \, E_{\Pbottom}^{2} \, \beta_{\APbottom} \, E_{\APbottom}^{2}} \, \int \,
d\Ehat_{\Pbottom} \, d\thetahat_{\Pnu} \, d\phihat_{\Pnu} \, d\thetahat_{\nuStar} \, d\phihat_{\nuStar} \nonumber \\
 & \quad \cdot \, \pThat^{\Pnu} \, \pThat^{\nuStar} \, 
\frac{f(\xhat_{a}, Q) \, f(\xhat_{b}, Q)}{\xhat_{a} \, \xhat_{b}} \nonumber \\
 & \quad \cdot \, \vert \mathcal{M}_{0}(\vecphat) \vert^{2} \, 
\betahat_{\Pbottom} \, W(E_{\Pbottom}|\Ehat_{\Pbottom}) \, 
\betahat_{\APbottom} \, \Ehat_{\APbottom} \, W(E_{\APbottom}|\Ehat_{\APbottom}) \,
W_{\rho}( \pX^{\rho},\pY^{\rho} | \pXhat^{\rho},\pYhat^{\rho} ) \nonumber \\
 & \quad \cdot \, \left[ \left\lvert 1 - \frac{\betahat_{\Pbottom}}{\betahat_{\APbottom}} \, \cos\sphericalangle(\vece_{\Pbottom},\vece_{\APbottom}) \right\rvert \right. \nonumber \\
 & \qquad \cdot \, \left. \sin^{2}\left(\frac{\sphericalangle(\vece_{\Plepton},\vecehat_{\Pnu})}{2}\right) \,
\Ehat_{\ellnuellStar} \left( 1 - \betahat_{\ellnuellStar} \, \cos\sphericalangle(\vecehat_{\ellnuellStar},\vecehat_{\nuStar}) \right) \right]^{-1} \, ,
\label{eq:mem_signal}
\end{align}
\end{linenowrapper}
with:
\begin{linenowrapper}
\begin{eqnarray}
\Ehat_{\APbottom} & = & \frac{a \, \Delta_{m_{\PHiggs}} + |b| \, \sqrt{\Delta_{m_{\PHiggs}}^{2} - (a^{2} - b^{2}) \, m_{\Pbottom}^{2}}}{a^{2} - b^{2}} \nonumber \\
\Ehat_{\Pnu} & = & \frac{m_{\PW}^{2}}{4 \, \Ehat_{\Plepton} \, \sin^{2}\left(\frac{\sphericalangle(\vece_{\Plepton},\vecehat_{\Pnu})}{2}\right)} \nonumber \\
\Ehat_{\nuStar} & = & \frac{m_{\PHiggs}^{2} - m_{\ellnuellStar}^{2}}{2 \, \Ehat_{\ellnuellStar} \, 
 \left( 1 - \betahat_{\ellnuellStar} \, \cos\sphericalangle(\vecehat_{\ellnuellStar},\vecehat_{\nuStar}) \right)} \, ,
\label{eq:mem_signal_defs}
\end{eqnarray}
\end{linenowrapper}
where:
\begin{linenowrapper}
\begin{eqnarray}
\Delta_{m_{\PHiggs}} & = & \frac{m_{\PHiggs}^{2}}{2} - m_{\Pbottom}^{2} \nonumber \\
a & = & \Ehat_{\Pbottom} \nonumber \\
b & = & \betahat_{\Pbottom} \, \Ehat_{\Pbottom} \, \cos\sphericalangle(\vece_{\Pbottom},\vece_{\APbottom}) \, .
\end{eqnarray}
\end{linenowrapper}
The integral on the RHS of Eq.~(\ref{eq:mem_signal}) is ready to be evaluated by numeric integration. 
The integral extends over the $5$ variables
 $\Ehat_{\Pbottom}$, $\thetahat_{\Pnu}$, $\phihat_{\Pnu}$, $\thetahat_{\nuStar}$, and $\phihat_{\nuStar}$.
The symbol $\vece_{k}$ refers to a unit vector in direction of particle $k$,
and the symbol $\sphericalangle(\vecehat_{k},\vecehat_{k'})$ denotes the angle between the directions of particles $k$ and $k'$.
This notation includes the case that the ``particles'' $k$ and $k'$ are systems of multiple particles,
\eg $\vecehat_{\ellnuellStar}$ denotes the direction of the momentum vector of the $3$-particle system composed of
the charged lepton and the neutrino produced in the decay of the on-shell $\PW$ boson and of the charged lepton produced in the decay of the off-shell $\PW$ boson.
The hat in the symbol $\vecehat_{\ellnuellStar}$ indicates that this direction refers to the true momenta of the neutrinos,
which are computed as function of the integration variables $\thetahat_{\Pnu}$, $\phihat_{\Pnu}$, $\thetahat_{\nuStar}$, $\phihat_{\nuStar}$ and Eq.~(\ref{eq:mem_signal_defs}).
For the charged leptons and the $\Pbottom$-jets, the true direction is equal to the measured direction,
as the direction of charged leptons and jets is measured with negligible experimental resolution.

There is one further aspect, which needs to be taken into account when computing the compatibility of a given event with the signal hypothesis,
and that is that there exists a fourfold ambiguity in associating the two measured $\Pbottom$-jets to the $\Pbottom$ and $\APbottom$ quarks 
and in associating the two measured charged leptons to the on-shell and off-shell $\PW$ bosons.
We deal with the fourfold ambiguity by evaluating the integral $w_{0}(\vecy)$ given by Eq.~(\ref{eq:mem_signal}) four times,
once for each of the four possible associations of measured $\Pbottom$-jets to the $\Pbottom$ and $\APbottom$ quarks and of the measured charged leptons to the on-shell and off-shell $\PW$ bosons,
and using the average of these four values when evaluating the LR in Eq.~(\ref{eq:memLR}).

\subsection{Analytic transformations specific to the background hypothesis}
\label{sec:mem_background}

In $\ttbar \to \Pbottom\PW\APbottom\PW \to \Pbottom\ellPlusnu \, \APbottom\ellMinusnu$ background events,
both $\PW$ bosons are on-shell. Sizeable contributions to the value of the integral $w_{1}(\vecy)$ are obtained only
for those points $\vecphat$ in the $6$-particle PS for which:
\begin{itemize}
\item The masses of the $\ellPlusnu$ as well as of the $\ellMinusnu$ system are equal to $m_{\PW} = 80.4$~\GeV~\cite{PDG}.
\item The masses of the $\Pbottom\ellPlusnu$ and $\APbottom\ellMinusnu$ systems are equal to the top quark mass of $m_{\Ptop} = 172.8$~\GeV~\cite{PDG}.
\end{itemize}

We account for these mass constraints by inserting four $\delta$-functions $\delta\left( g(x) \right)$ into the integrand of Eq.~(\ref{eq:mem3}):
two $\delta$-functions of the type $g(\Ehat_{\Pnu})$, given by Eq.~(\ref{eq:nuEn_Wlnu1}),
and two $\delta$-functions of the type $g(\Ehat_{\Pbottom})$, given by Eq.~(\ref{eq:bEn_top1}).
We denote the second $\delta$-function of the type given by Eq.~(\ref{eq:nuEn_Wlnu1}) by the symbol $g(\Ehat_{\APnu})$
and the second $\delta$-function of the type given by Eq.~(\ref{eq:bEn_top1}) by the symbol $g(\Ehat_{\APbottom})$
to indicate that they refer to the anti-neutrino and to the anti-bottom quark, which both are produced in the decay of the anti-top quark.

After solving for the $\delta$-functions analytically, following Sections~\ref{sec:appendix_nuEn_Wlnu} and~\ref{sec:appendix_bEn_top} of the appendix,
we obtain the following expression for the integral $w_{1}(\vecy)$ for the background hypothesis:
\begin{linenowrapper}
\begin{align}
w_{1}(\vecy) 
 = & \frac{(m_{\Ptop} \, \Gamma_{\Ptop})^{2} \, (m_{\PW} \, \Gamma_{\PW})^{2}}{2^{25} \, \pi^{14} \, \sigma_{1} \, s \, 
  E_{\Plepton^{+}}^{2} \, E_{\Plepton^{-}}^{2} \, \beta_{\Pbottom} \, E_{\Pbottom}^{2} \, \beta_{\APbottom} \, E_{\APbottom}^{2}} \, \int \,
d\thetahat_{\Pnu} \, d\phihat_{\Pnu} \, d\thetahat_{\APnu} \, d\phihat_{\APnu}  \nonumber \\
 & \quad \cdot \, \pThat^{\Pnu} \, \pThat^{\APnu} \,
\frac{f(\xhat_{a}, Q) \, f(\xhat_{b}, Q)}{\xhat_{a} \, \xhat_{b}} \nonumber \\
 & \quad \cdot \, \vert \mathcal{M}_{1}(\vecphat) \vert^{2} \, 
\betahat_{\Pbottom} \, \Ehat_{\Pbottom} \, W(E_{\Pbottom}|\Ehat_{\Pbottom}) \, 
\betahat_{\APbottom} \, \Ehat_{\APbottom} \, W(E_{\APbottom}|\Ehat_{\APbottom}) \,
W_{\rho}( \pX^{\rho},\pY^{\rho} | \pXhat^{\rho},\pYhat^{\rho} ) \nonumber \\
 & \quad \cdot \, \left[ \sin^{2}\left(\frac{\sphericalangle(\vece_{\ellPlus},\vecehat_{\Pnu})}{2}\right) \, 
\sin^{2}\left(\frac{\sphericalangle(\vece_{\ellMinus},\vecehat_{\APnu})}{2}\right) \right. \nonumber \\
 & \qquad \cdot \, \left. \Ehat_{\ellPlusnu} \, \left\lvert 1 - \frac{\betahat_{\ellPlusnu}}{\betahat_{\Pbottom}} \, \cos\sphericalangle(\vecehat_{\ellPlusnu},\vece_{\Pbottom}) \right\rvert \,
\Ehat_{\ellMinusnu} \, \left\lvert 1 - \frac{\betahat_{\ellMinusnu}}{\betahat_{\APbottom}} \, \cos\sphericalangle(\vecehat_{\ellMinusnu},\vece_{\APbottom}) \right\rvert \right]^{-1} \, ,
\label{eq:mem_background}
\end{align}
\end{linenowrapper}
with:
\begin{linenowrapper}
\begin{eqnarray}
\Ehat_{\Pbottom} & = & \frac{a_{\Ptop} \, \Delta_{m_{\Ptop}}
 + |b_{\Ptop}| \, \sqrt{\Delta_{m_{\Ptop}}^{2} - (a_{\Ptop}^{2} - b_{\Ptop}^{2}) \, m_{\Pbottom}^{2}}}{a_{\Ptop}^{2} - b_{\Ptop}^{2}} \nonumber \\
\Ehat_{\APbottom} & = & \frac{a_{\APtop} \, \Delta_{m_{\Ptop}}
 + |b_{\APtop}| \, \sqrt{\Delta_{m_{\Ptop}}^{2} - (a_{\APtop}^{2} - b_{\APtop}^{2}) \, m_{\Pbottom}^{2}}}{a_{\APtop}^{2} - b_{\APtop}^{2}} \nonumber \\
\Ehat_{\Pnu} & = & \frac{m_{\PW}^{2}}{4 \, \Ehat_{\ellPlus} \, \sin^{2}\left(\frac{\sphericalangle(\vece_{\ellPlus},\vecehat_{\Pnu})}{2}\right)} \nonumber \\
\Ehat_{\APnu} & = & \frac{m_{\PW}^{2}}{4 \, \Ehat_{\ellMinus} \, \sin^{2}\left(\frac{\sphericalangle(\vece_{\ellMinus},\vecehat_{\APnu})}{2}\right)} \, ,
\label{eq:mem_background_defs}
\end{eqnarray}
\end{linenowrapper}
where:
\begin{linenowrapper}
\begin{eqnarray}
\Delta_{m_{\Ptop}} & = & \frac{m_{\Ptop}^{2} - m_{\Pbottom}^{2} - m_{\PW}^{2}}{2} \nonumber \\
a_{\Ptop} & = & \Ehat_{\ellPlusnu} \nonumber \\
b_{\Ptop} & = & \sqrt{\Ehat_{\ellPlusnu}^{2} - m_{\PW}^{2}} \, \cos\sphericalangle(\vecehat_{\ellPlusnu},\vece_{\Pbottom}) \nonumber \\
a_{\APtop} & = & \Ehat_{\ellMinusnu} \nonumber \\
b_{\APtop} & = & \sqrt{\Ehat_{\ellMinusnu}^{2} - m_{\PW}^{2}} \, \cos\sphericalangle(\vecehat_{\ellMinusnu},\vece_{\APbottom}) \, .
\end{eqnarray}
\end{linenowrapper}
The integral given by Eq.~(\ref{eq:mem_background}) extends over $4$ remaining variables,
which are integrated numerically: $\thetahat_{\Pnu}$, $\phihat_{\Pnu}$, $\thetahat_{\APnu}$, and $\phihat_{\APnu}$.
The symbol $\ellPlusnu$ ($\ellMinusnu$) refers to the true direction of the $\PW^{+}$ ($\PW^{-}$) boson,
which are computed by summing the momenta of the lepton of positive (negative) charge and of the neutrino (anti-neutrino).
The neutrino and anti-neutrino momenta are computed as function of the integration variables $\thetahat_{\Pnu}$, $\phihat_{\Pnu}$, $\thetahat_{\APnu}$, $\phihat_{\APnu}$ and Eq.~(\ref{eq:mem_background_defs}).

When evaluating the compatibility of a given event with the background hypothesis,
there exists a twofold ambiguity in associating the two measured $\Pbottom$-jets to the $\Pbottom$ and $\APbottom$ quarks.
We deal with this ambiguity by evaluating the integral $w_{1}(\vecy)$ given by Eq.~(\ref{eq:mem_signal}) two times,
corresponding to the two possible associations of the measured $\Pbottom$-jets to the $\Pbottom$ and $\APbottom$ quarks.
In contrast to the signal hypothesis,
there is no ambiguity in associating the two measured leptons to the two $\PW$ bosons,
as in $\ttbar$ background events both $\PW$ bosons are on-shell,
and the measurement of the lepton charge allows for a unique association of each charged lepton to either the $\PW^{+}$ or the $\PW^{-}$ boson.

\section{Performance}
\label{sec:performance}

We study the separation of the $\dihiggs$ signal from the $\ttbar$ background,
achieved by the LR $P(\vecy)$ given by Eq.~(\ref{eq:memLR}),
using samples of signal and background events produced by Monte Carlo (MC) simulation.
The samples are simulated at LO and at NLO accuracy in pQCD
and are analyzed at MC-truth as well as at detector level.
The former corresponds to the case of an ideal experimental resolution, 
while the latter aims to simulate the experimental conditions characteristic for the ATLAS and CMS experiments during LHC Run $2$.
The LO and NLO $\dihiggs$ signal samples each contain about three hundred thousand events
and the LO and NLO $\ttbar$ background samples each contain about five million events.
All samples simulated at LO accuracy in pQCD are produced with the program $\textsc{MadGraph\_aMCatNLO}$ $2.2.2$,
while the samples simulated at NLO accuracy in pQCD are produced using the program $\textsc{POWHEG}$ $v2$~\cite{POWHEG1,POWHEG2,POWHEG3,POWHEGTTBAR1,POWHEGTTBAR2,POWHEGHH1,POWHEGHH2}.
The \textrm{NNPDF3.0} LO set of PDF is used for the simulation of the LO samples and the \textrm{NNPDF3.0} NLO set for the NLO samples~\cite{NNPDF1,NNPDF2,NNPDF3}.
Parton shower and hadronization processes are modeled using the program $\textsc{PYTHIA}$ $v8.2$~\cite{Sjostrand:2014zea} with the tune \textrm{CP5}~\cite{Sirunyan:2019dfx}.
All events are generated for proton-proton collisions at $\sqrt{s} = 13$~\TeV center-of-mass energy.
Events in which the electrons or muons originate from $\Pgt$ lepton decays,
\ie from the decay chains $\PW^{+} \to \Pgt^{+}\Pnu_{\Pgt} \to \Plepton^{+}\Pnu_{\Plepton}\APnu_{\Pgt}\Pnu_{\Pgt}$ or 
$\PW^{-} \to \Pgt^{-}\APnu_{\Pgt} \to \Plepton^{-}\APnu_{\Plepton}\Pnu_{\Pgt}\APnu_{\Pgt}$, are discarded.
Detector effects are simulated using the program $\textsc{DELPHES}$ $v3.5.0$~\cite{deFavereau:2013fsa} with the card for the CMS detector.
On average forty inelastic proton-proton interactions (pileup) are added to each simulated event
in order to simulate the data-taking conditions during Run $2$ of the LHC.

Jets are reconstructed using the anti-$\kt$ algorithm~\cite{Cacciari:2008gp, Cacciari:2011ma} with a distance parameter of $0.4$,
using the detector-level particle-flow objects created by $\textsc{DELPHES}$ as input.
We refer to these jets as detector-level jets.
Their energy is corrected for pileup effects using the method described in Refs.~\cite{Cacciari:2008gn, Cacciari:2007fd}
and is calibrated as function of jet $\pT$ and $\eta$, where $\eta = -\ln\tan(\theta/2)$ denotes the pseudorapidity of the jet.
The calibration is performed such that the energy of the jets that are tagged, at detector level, as originating from the hadronization of a bottom quark
on average matches the energy of the bottom quarks that result from $\PHiggs$ boson or top quark decays at the parton level.
We refer to detector-level jets that pass the $\Pbottom$-tagging criteria as $\Pbottom$-jets.
By calibrating detector-level jets to the energy of the bottom quarks at the parton level,
the calibration procedure corrects the jet energy for out-of-cone effects and for the energy carried away, on average, by the neutrinos produced in heavy-flavor decays.

The simulated $\dihiggs$ signal and $\ttbar$ background events considered in this section are required to pass event selection criteria
similar to the analysis of $\dihiggs$ production performed, in the channel $\dihiggs \to \Pbottom\APbottom\PW\PW\virt$, by the CMS collaboration during LHC Run $2$~\cite{HIG-17-006}.
The events are required to contain two electrons or muons and two $\Pbottom$-jets.
The leptons must be within the region $\abs{\eta} < 2.5$ if they are electrons and $\abs{\eta} < 2.4$ if they are muons, and are required to be isolated.
Their isolation is computed by summing the $\pT$ of detector-level particle-flow objects that are within a cone of size
$\delta R = \sqrt{(\delta\eta)^{2} + (\delta\phi)^{2}} = 0.5$ around the lepton direction, excluding the lepton itself.
The sum is corrected for the contribution of particles from pileup using the method described in Refs.~\cite{Cacciari:2008gn, Cacciari:2007fd}.
Electrons and muons are considered isolated if the pileup-corrected sum amounts to less than $0.10$ times the $\pT$ of the lepton.
The lepton of higher $\pT$ is required to have $\pT > 25$~\GeV and the lepton of lower $\pT$ must have $\pT > 15$~\GeV.
These $\pT$ thresholds are motivated by trigger requirements.
The $\Pbottom$-jets are required to satisfy the conditions $\pT > 25$~\GeV and $\abs{\eta} < 2.4$ and to be both tagged as $\Pbottom$-jets at detector level.
The $\Pbottom$-tagging criteria implemented in the $\textsc{DELPHES}$ card for the CMS detector
corresponds to the medium working-point of the ``combined secondary vertex'' $\Pbottom$-tagging algorithm published in Ref.~\cite{CMS:2012feb}.
The algorithm identifies jets originating from the hadronization of a bottom quark with an efficiency of approximately $70\%$,
for a misidentification rate for light-quark and gluon jets of about $1.5\%$~\cite{CMS:2012feb}.
Events containing more than two $\Pbottom$-tagged jets of $\pT > 25$~\GeV and $\abs{\eta} < 2.4$ are vetoed.
The latter condition rejects a small fraction of events, amounting to $7.7\%$ of the $\dihiggs$ signal and $5.1\%$ of the $\ttbar$ background,
and avoids ambiguities in choosing the correct pair of $\Pbottom$-jets 
when computing the PDs $w_{0}(\vecy)$ and $w_{1}(\vecy)$ according to Eqs.~(\ref{eq:mem_signal}) and~(\ref{eq:mem_background}).
The selection criteria are applied to generator-level leptons and jets when analyzing simulated events at MC-truth level
and to detector-level leptons and jets when analyzing simulated events at the detector level.
In case the selection criteria are applied at MC-truth level,
no isolation requirements are applied to the leptons,
the conditions $\pT > 25$~\GeV and $\abs{\eta} < 2.4$ of the jet selection are applied at the parton level, 
to the bottom quarks that are produced in the $\PHiggs$ boson or top quark decays, and no detector-level $\Pbottom$-tagging criteria are applied.

\begin{figure}
\ifx\ver\verPreprint
\setlength{\unitlength}{1mm}
\begin{center}
\begin{picture}(160,67)(0,0)
\put(-1.0, 1.0){\mbox{\includegraphics*[height=66mm]
 {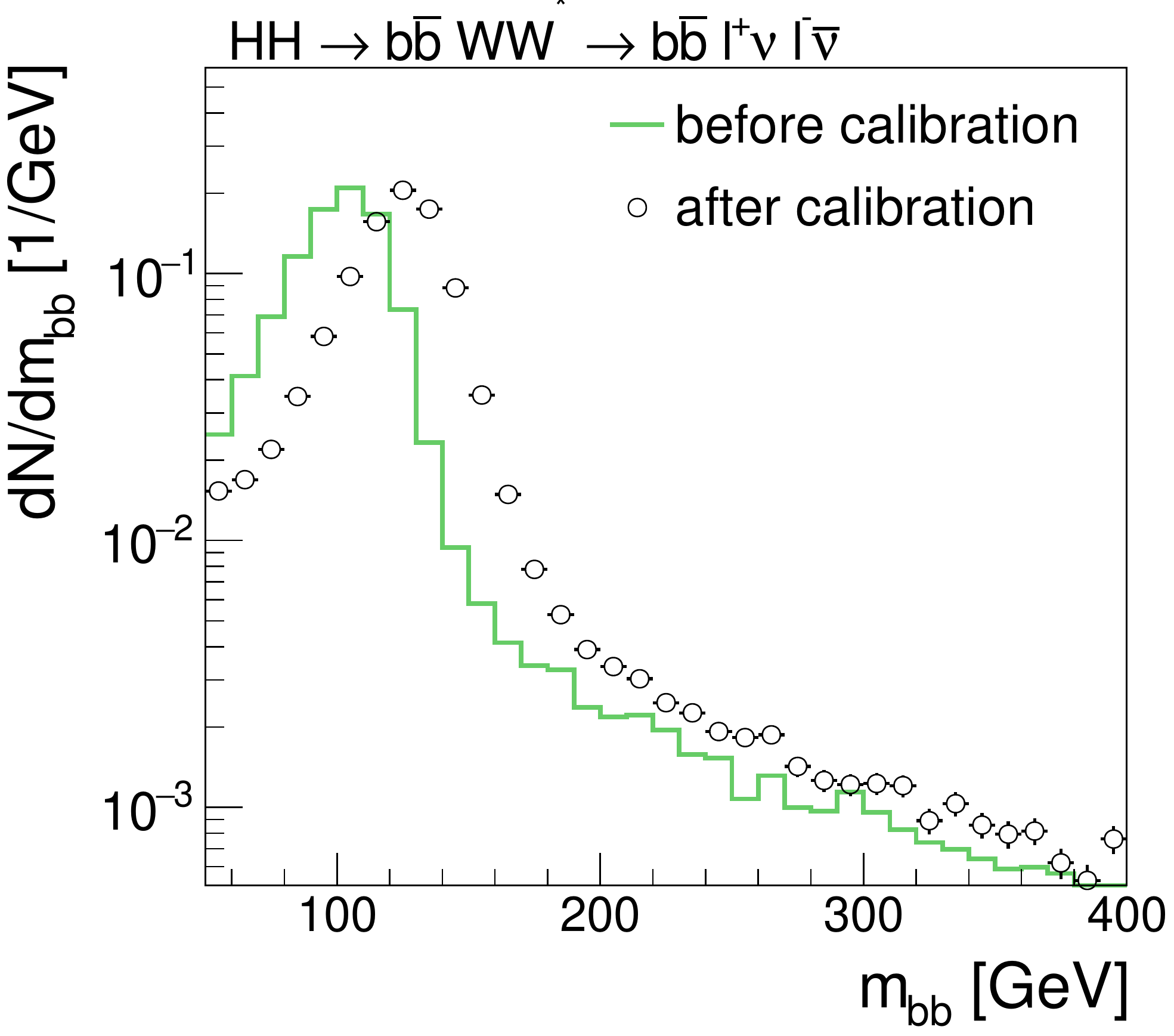}}}
\put(80.0, 0.0){\mbox{\includegraphics*[height=67mm]
 {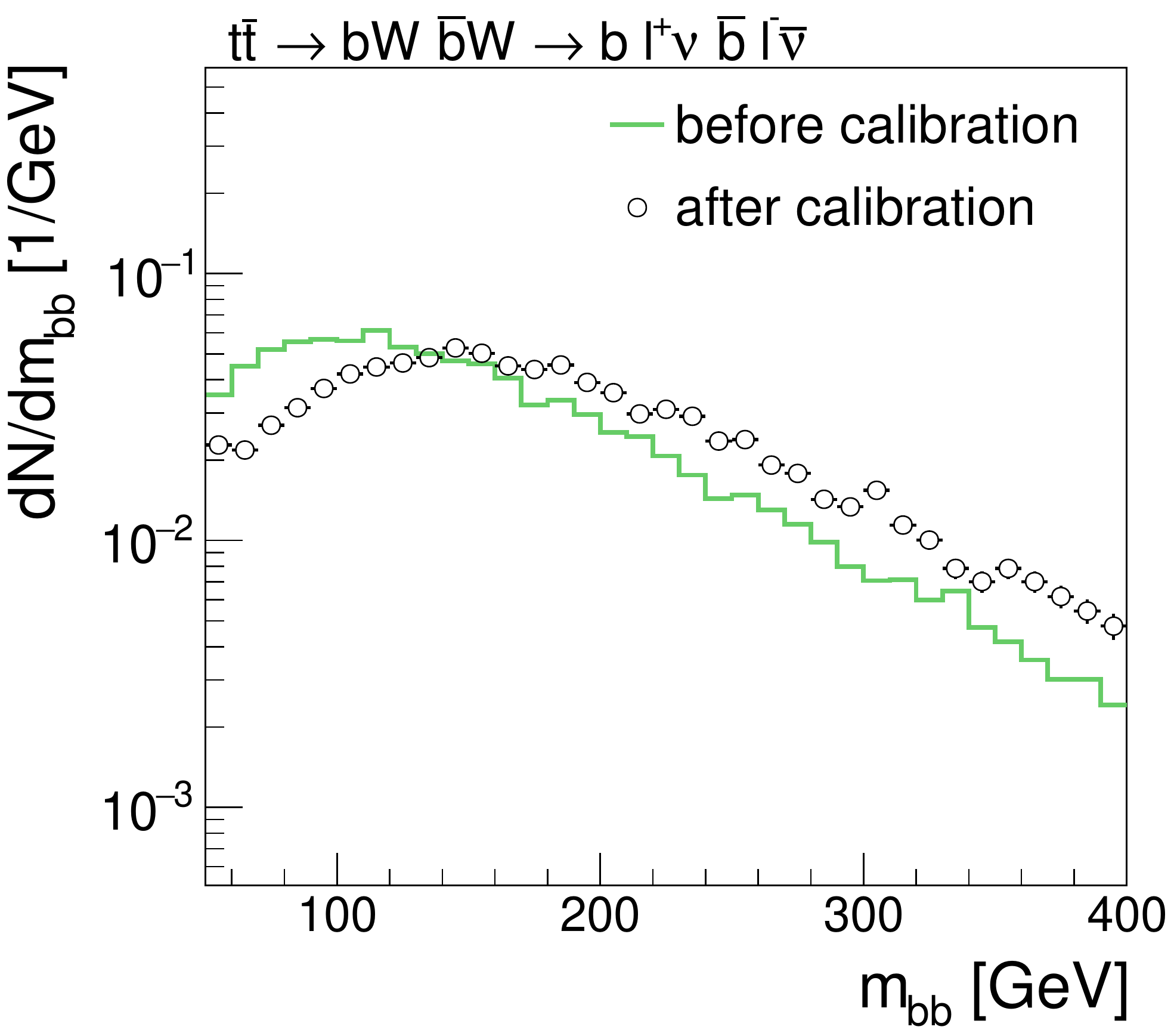}}}
\end{picture}
\end{center}
\fi
\ifx\ver\verPAPER
\centering
\includegraphics[width=0.48\textwidth]{plots/mbb_calibrated_vs_uncalibrated_signal.pdf}
\includegraphics[width=0.48\textwidth]{plots/mbb_calibrated_vs_uncalibrated_background.pdf}
\fi
\caption{
  Distribution in $\mbb$, the mass of the two detector-level jets that are tagged as $\Pbottom$-jets,
  in $\dihiggs$ signal (left) and $\ttbar$ background (right) events before and after the jet energy calibration is applied.
}
\label{fig:mbb}
\end{figure}

Fig.~\ref{fig:mbb} shows the distribution in $\mbb$, the mass of the two $\Pbottom$-tagged jets at detector level, 
in $\dihiggs$ signal and $\ttbar$ background events that pass the selection criteria described in the previous paragraph.
Only events in which both detector-level jets are matched, within a cone of size $\delta R = 0.3$, 
to bottom quarks that originate from either a $\PHiggs$ boson or from top quark decays, are shown in the figure.
According to the $\textsc{DELPHES}$ simulation, 
$95.0\%$ of $\dihiggs$ and $96.4\%$ of $\ttbar$ events that pass the selection criteria described in the previous paragraph fulfill this matching condition,
\ie in $5.0\%$ of selected $\dihiggs$ and $3.6\%$ of selected $\ttbar$ events one of the bottom quarks is not reconstructed as $\Pbottom$-jet at detector level
and a light quark or gluon jet is misidentified as $\Pbottom$-jet instead.
The figure shows that the jet calibration shifts the peak of the $\mbb$ distribution by about $20\%$.
After calibration, the $\mbb$ distribution in $\dihiggs$ signal events peaks close to $125$~\GeV.
The calibration also reduces the relative width, defined as the root mean square divided by the mean, of the $\mbb$ distribution in $\dihiggs$ signal events by about $20\%$.

In order to compute the PDs $w_{0}(\vecy)$ and $w_{1}(\vecy)$ according to Eqs.~(\ref{eq:mem_signal}) and~(\ref{eq:mem_background}),
we need to determine the TFs for the energy of $\Pbottom$-jets and for the transverse momentum components of the hadronic recoil
such that the TFs match the experimental resolution in the $\textsc{DELPHES}$ simulation.
We model the experimental resolution on the energy of $\Pbottom$-jets using a normal distribution:
\begin{linenowrapper}
\begin{equation}
W(E|\Ehat) = \frac{1}{\sqrt{2 \pi \sigma_{\Pbottom}^{2}}} \, e^{-\frac{(\pT - \pThat)^{2}}{2 \, \sigma_{\Pbottom}^{2}}} \, ,
\label{eq:resolution_b}
\end{equation}
\end{linenowrapper}
where $\pT = E \cdot \sin\theta$, $\pThat = \Ehat \cdot \sin\theta$, and $\theta$ refers to the polar angle of the jet.
The standard deviation $\sigma_{\Pbottom}$ depends on the jet energy and $\theta$.
We make the ansatz $\sigma_{\Pbottom} = k \cdot \sqrt{\Ehat \cdot \sin\theta}$ and determine the constant of proportionality $k$ such that it fits
the resolution on the energy of $\Pbottom$-jets in the $\textsc{DELPHES}$ simulation, yielding $k = 100\%$.
Our model for the jet energy resolution agrees with the resolution measured by the CMS collaboration during LHC Run $2$, shown in Fig.~3 of Ref.~\cite{JME-18-001}~\footnote{
  Our assumption that the polar angle $\theta$ of the jet is measured with negligible experimental resolution (\cf Section~\ref{sec:appendix_TF} of the appendix)
  is justified by Fig.~5 of Ref.~\cite{JME-18-001}, which shows that the resolution on $\theta$ amounts to about $0.02$ radians for jets of $\pT = 25$~\GeV and decreases for jets of higher $\pT$.}.
The hadronic recoil $\rho$ is not directly available in the $\textsc{DELPHES}$ simulation.
To determine the resolution on $\rho$, we compute the transverse momentum components of the hadronic recoil 
as function of the transverse momenta of the two leptons, the two $\Pbottom$-jets, and $\vecMET$,
using Eq.~(\ref{eq:hadRecoil_true}), with the substitutions $\pXhat^{\Pnu} + \pXhat^{\APnu} = \METxhat$ and $\pYhat^{\Pnu} + \pYhat^{\APnu} = \METyhat$, for the computation at MC-truth level 
and Eq.~(\ref{eq:hadRecoil}) for the computation at detector level.
The resolution on $\pX^{\rho}$ and $\pY^{\rho}$ in the $\textsc{DELPHES}$ simulation amounts to $32$~\GeV for the $\dihiggs$ signal and to $30$~\GeV for the $\ttbar$ background.
The resolution on the energy of $\Pbottom$-jets is small compared to the resolution on the hadronic recoil.
The resolution on the latter is thus similar to the resolution on $\vecMET$.
This similarity allows us to compare the resolutions on $\pX^{\rho}$ and $\pY^{\rho}$ in the $\textsc{DELPHES}$ simulation to the resolution on $\vecMET$
published by the ATLAS collaboration for simulated $\ttbar$ events during LHC Run $2$~\footnote{
  The CMS collaboration has not published the $\vecMET$ resolution during LHC Run $2$ specifically for $\ttbar$ events.},
which is shown in Fig.~9 of Ref.~\cite{ATLAS:2018txj} and amounts to $25$-$30$~\GeV.
We assume that the resolutions on $\pX^{\rho}$ and $\pY^{\rho}$ are uncorrelated and amount to the same for signal and background events.
Rounding the numbers for the resolution on $\pX^{\rho}$ and $\pY^{\rho}$ to one significant digit, we use:
\begin{linenowrapper}
\begin{equation}
V = \sigma_{\rho}^{2} \cdot I_{2} 
\label{eq:resolution_rho}
\end{equation}
\end{linenowrapper}
with $\sigma_{\rho} = 30$~\GeV for when computing the PDs $w_{0}(\vecy)$ and $w_{1}(\vecy)$ for $\dihiggs$ signal and $\ttbar$ background events.

\begin{figure}
\ifx\ver\verPreprint
\setlength{\unitlength}{1mm}
\begin{center}
\begin{picture}(160,144)(0,0)
\put(-1.0, 78.0){\mbox{\includegraphics*[height=66mm]
 {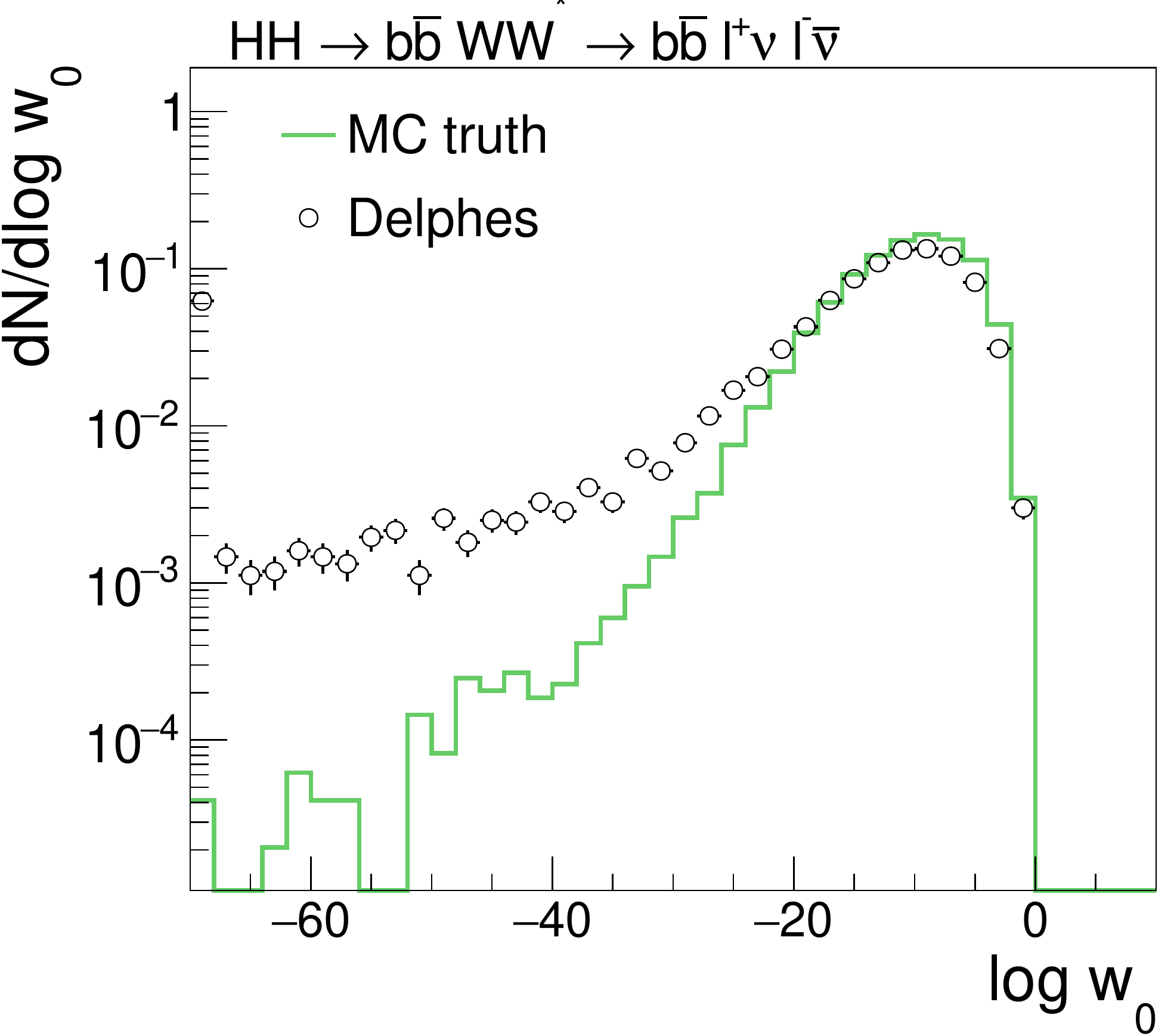}}}
\put(80.0, 78.0){\mbox{\includegraphics*[height=66mm]
 {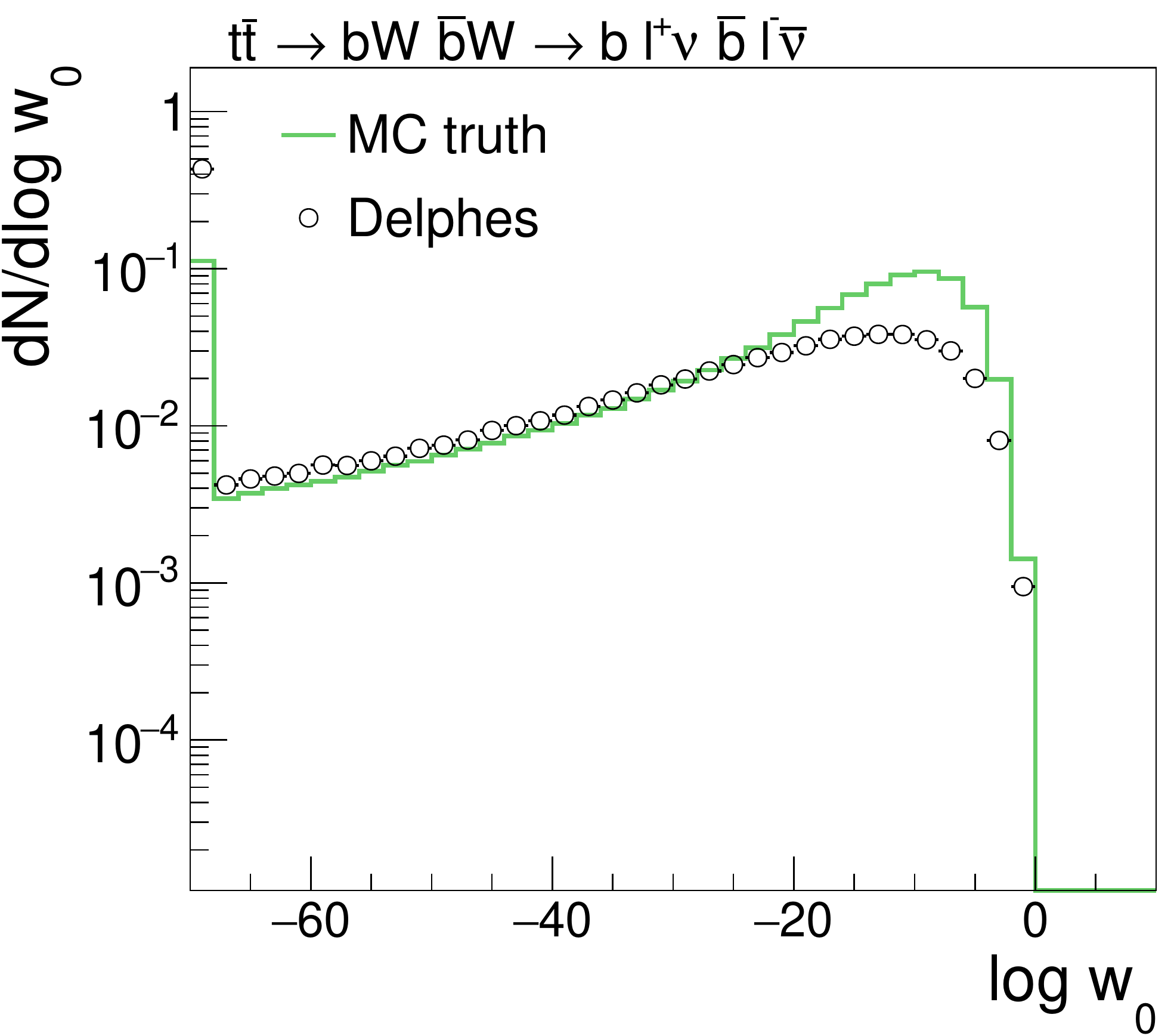}}}
\put(-1.0, 0.0){\mbox{\includegraphics*[height=66mm]
 {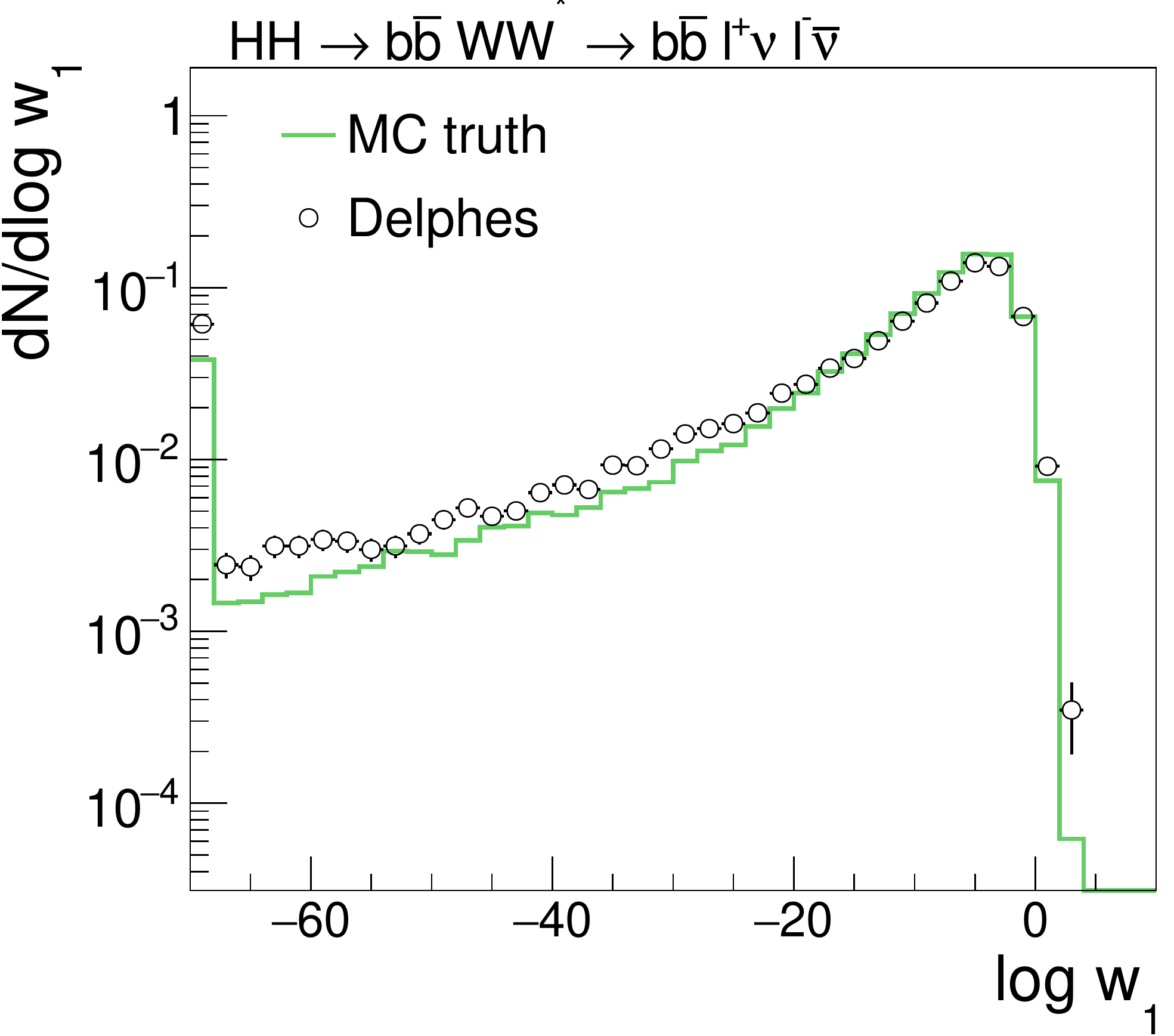}}}
\put(80.0, 0.0){\mbox{\includegraphics*[height=66mm]
 {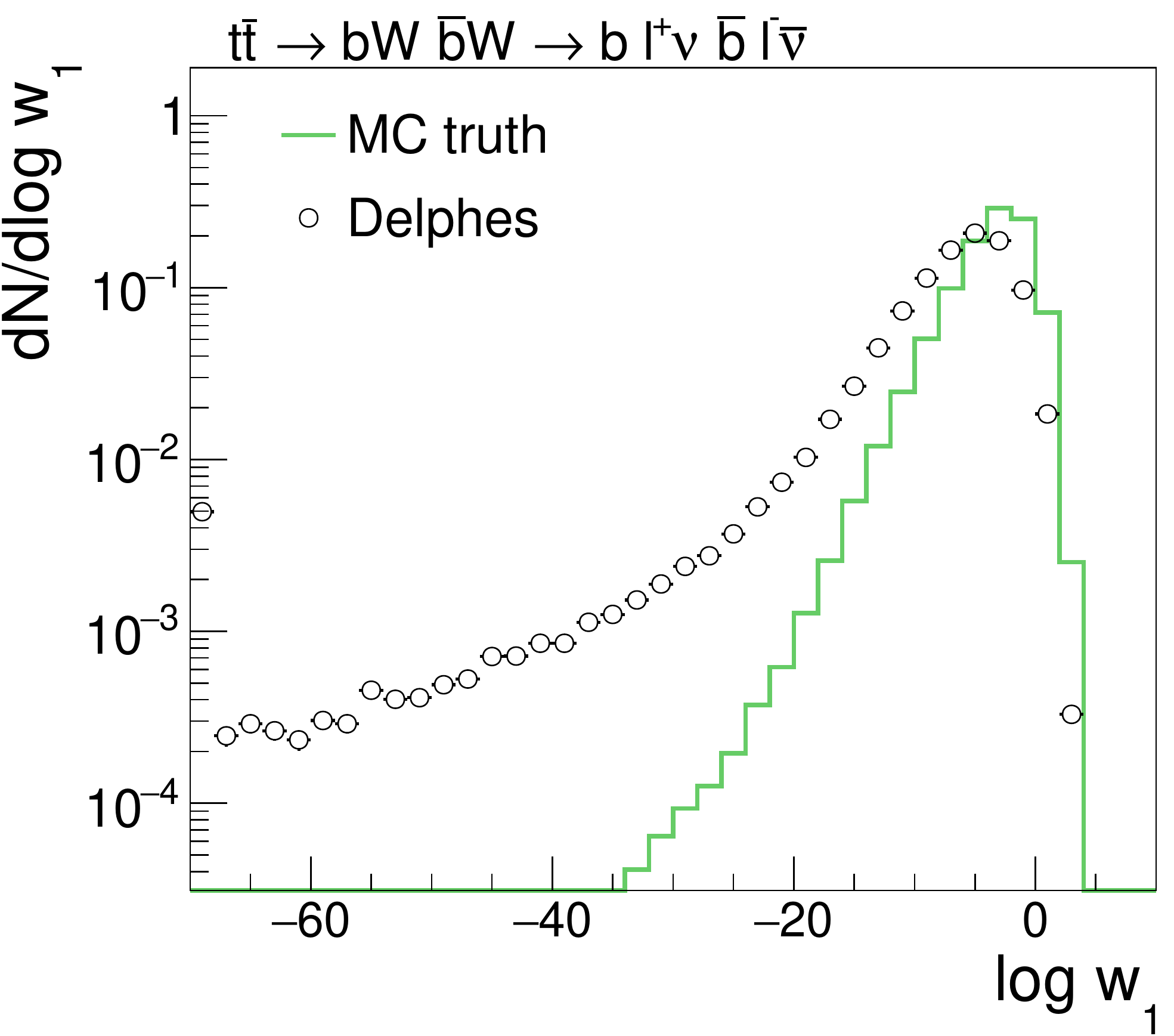}}}
\end{picture}
\end{center}
\fi
\ifx\ver\verPAPER
\centering
\includegraphics[width=0.48\textwidth]{plots/makePlotsForPaper_delphes_vs_mctruth_probS_signal.pdf}
\includegraphics[width=0.48\textwidth]{plots/makePlotsForPaper_delphes_vs_mctruth_probS_background.pdf}
\hspace{0.04\textwidth}
\includegraphics[width=0.48\textwidth]{plots/makePlotsForPaper_delphes_vs_mctruth_probB_signal.pdf}
\includegraphics[width=0.48\textwidth]{plots/makePlotsForPaper_delphes_vs_mctruth_probB_background.pdf}
\fi
\caption{
  Distributions in the PDs $w_{0}(\vecy)$ (upper) and $w_{1}(\vecy)$ (lower), computed according to Eqs.~(\ref{eq:mem_signal}) and~(\ref{eq:mem_background}),
  for $\dihiggs$ signal (left) and $\ttbar$ background (right) events.
  The PDs are computed at MC-truth and at detector level.
}
\label{fig:probS_and_probB}
\end{figure}

We can now proceed to compute the PDs $w_{0}(\vecy)$ and $w_{1}(\vecy)$.
Distributions in $w_{0}(\vecy)$ and $w_{1}(\vecy)$ for $\dihiggs$ signal and $\ttbar$ background events are shown in Fig.~\ref{fig:probS_and_probB}.
The horizontal axis is drawn in logarithmic scale to better visualize small values of the PDs.
The PDs are computed at MC-truth and at detector level.
When computing the PDs at MC-truth level, 
we set the ``measured'' momenta of electrons and muons to their generator-level values, the ``measured'' momenta of the $\Pbottom$-jets to the momenta of the corresponding parton-level bottom quarks,
and the ``measured'' transverse momentum components of the hadronic recoil to their true values $\pXhat^{\rho}$ and $\pYhat^{\rho}$.
The latter are computed according to Eq.~(\ref{eq:hadRecoil_true}).
We also demand that both $\Pbottom$-jets are matched, within a cone of size $\delta R = 0.3$,
to bottom quarks that originate from either a $\PHiggs$ boson or from top quark decays when we compute the PDs at MC-truth level.
The same TFs, described in the previous paragraph, are used when computing the PDs $w_{0}(\vecy)$ and $w_{1}(\vecy)$ at MC-truth and at detector level.
The distributions in the PDs for the ``correct'' hypothesis ($w_{0}(\vecy)$ for signal and $w_{1}(\vecy)$ for background events)
peak close to one and fall rapidly towards smaller values, while the distributions in the PDs for the ``wrong'' hypothesis
($w_{1}(\vecy)$ for signal and $w_{0}(\vecy)$ for background events)
exhibit more pronounced tails towards small values.
Interestingly, the distributions in the PDs for the wrong hypothesis change only by a small amount between MC-truth and detector level.
The main effect of the experimental resolutions on the energy of $\Pbottom$-jets and on the transverse momentum of the hadronic recoil
as well as of the misidentification of light quark or gluon jets as $\Pbottom$-jets is to increase the tail towards small values for the distributions in the PDs for the correct hypothesis.

\begin{figure}
\ifx\ver\verPreprint
\setlength{\unitlength}{1mm}
\begin{center}
\begin{picture}(160,66)(0,0)
\put(-1.0, 0.0){\mbox{\includegraphics*[height=66mm]
 {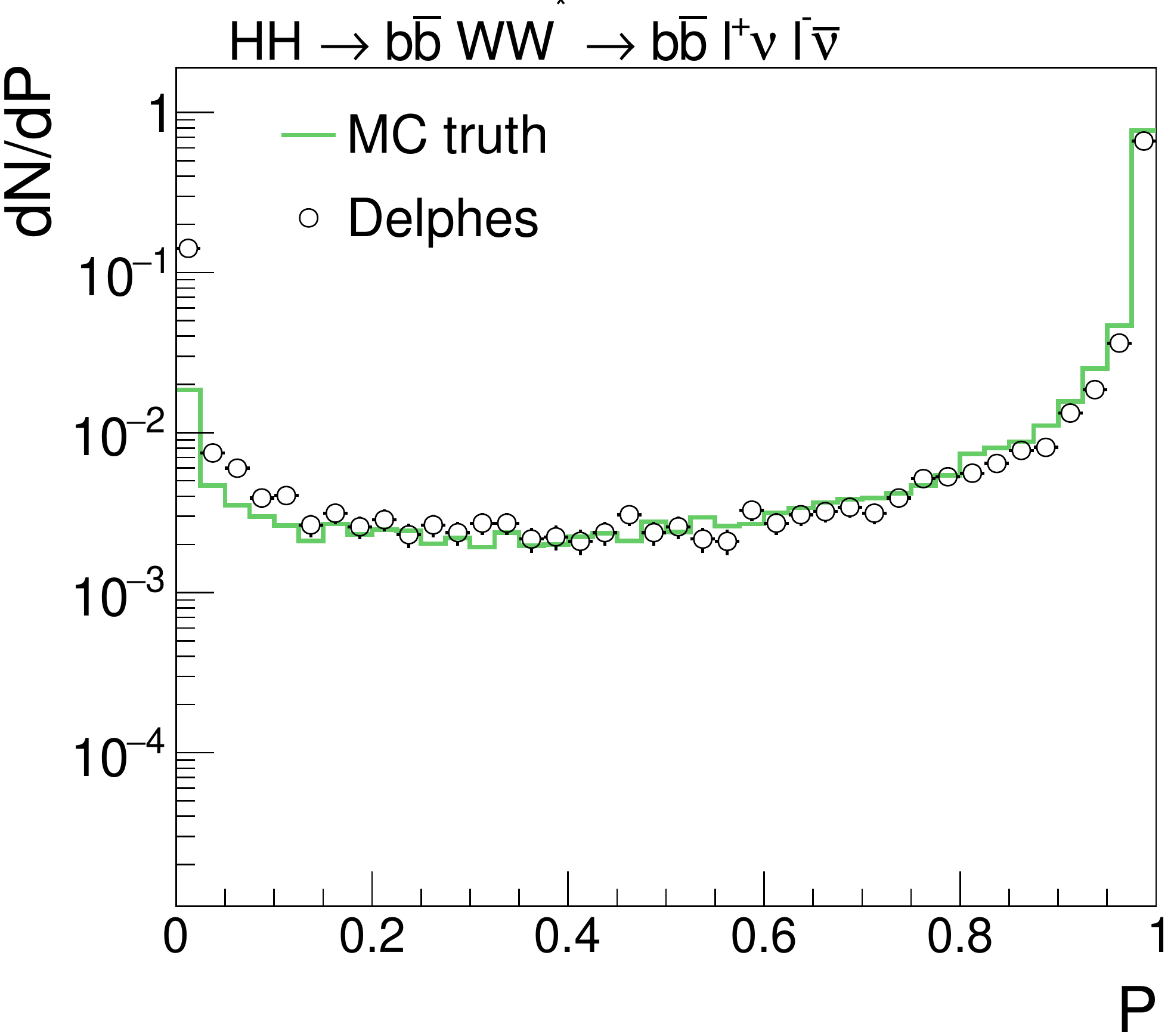}}}
\put(80.0, 0.0){\mbox{\includegraphics*[height=66mm]
 {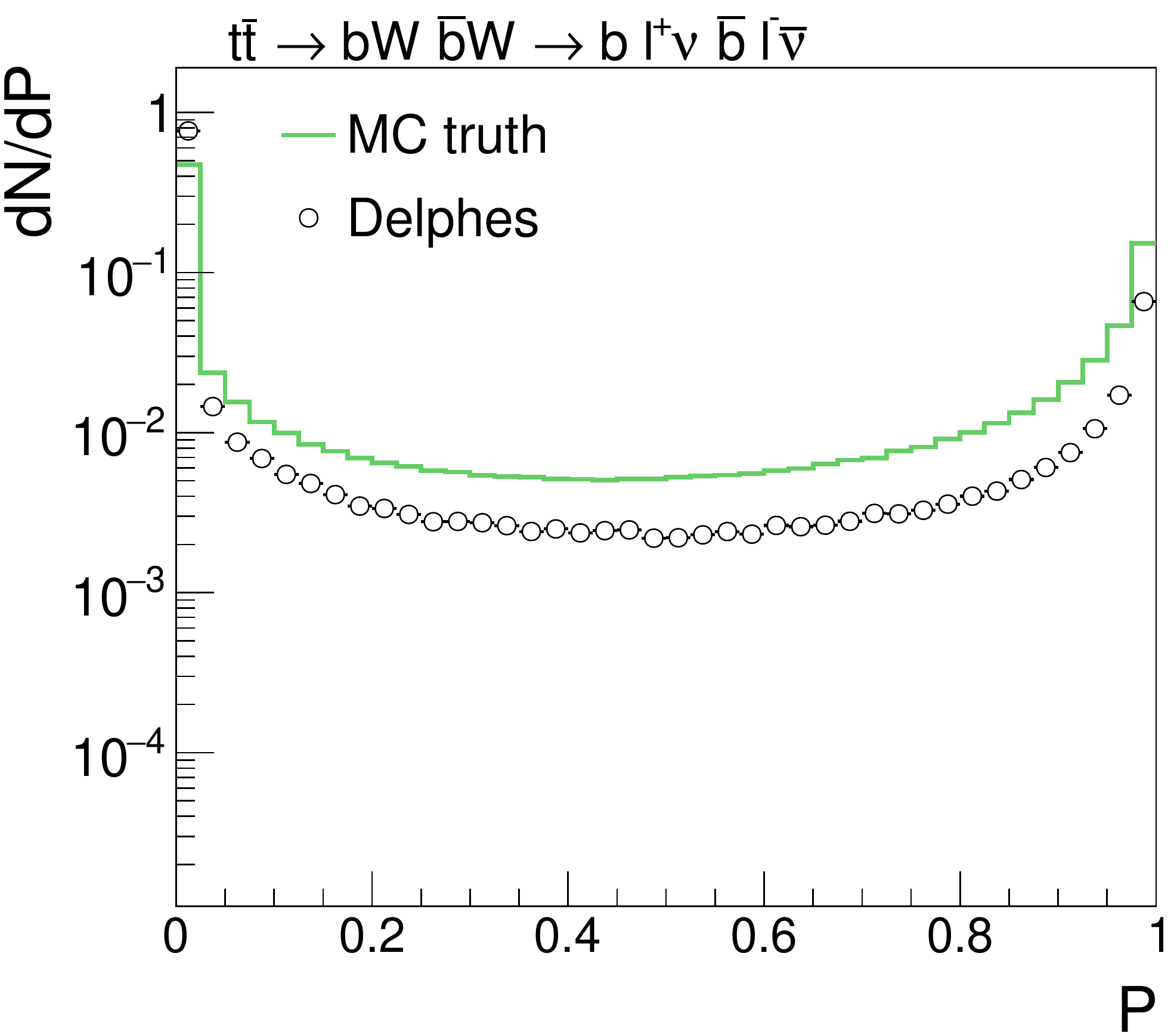}}}
\end{picture}
\end{center}
\fi
\ifx\ver\verPAPER
\centering
\includegraphics[width=0.48\textwidth]{plots/makePlotsForPaper_delphes_vs_mctruth_memLR_signal.pdf}
\includegraphics[width=0.48\textwidth]{plots/makePlotsForPaper_delphes_vs_mctruth_memLR_background.pdf}
\fi
\caption{
  Distributions in the LR $P(\vecy)$, computed according to Eq.~(\ref{eq:memLR}),
  for $\dihiggs$ signal (left) and $\ttbar$ background (right) events.
  The LR $P(\vecy)$ is computed using the PDs $w_{0}(\vecy)$ and $w_{1}(\vecy)$ shown in Fig.~\ref{fig:probS_and_probB} as input
  and is computed at MC-truth and at detector level.
}
\label{fig:memLR}
\end{figure}

\begin{figure}
\ifx\ver\verPreprint
\setlength{\unitlength}{1mm}
\begin{center}
\begin{picture}(160,67)(0,0)
\put(39.5, 0.0){\mbox{\includegraphics*[height=67mm]
 {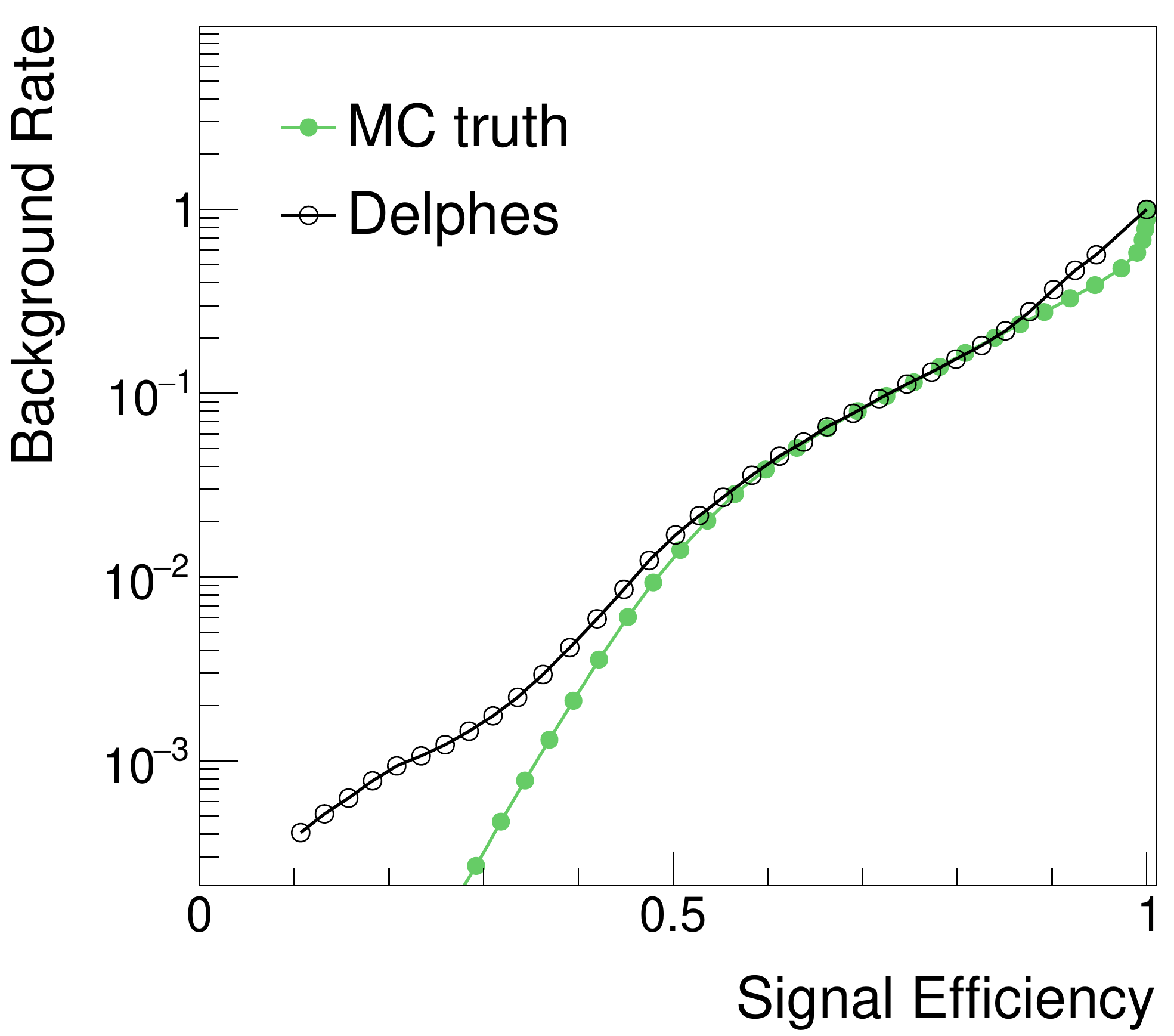}}}
\end{picture}
\end{center}
\fi
\ifx\ver\verPAPER
\centering
\includegraphics[width=0.57\textwidth]{plots/makePlotsForPaper_delphes_vs_mctruth_ROC.pdf}
\fi
\caption{
  Graphs of background rate versus signal efficiency (``ROC curve''), at MC-truth and at detector level,
  obtained by varying the threshold of a cut applied on the distributions in the LR $P(\vecy)$ shown in Fig.~\ref{fig:memLR}.
}
\label{fig:ROC}
\end{figure}

The corresponding distributions in the LR $P(\vecy)$, computed according to Eq.~(\ref{eq:memLR}), are shown in Fig.~\ref{fig:memLR}.
Signal events are characterized by high values of $P(\vecy)$, while background events typically have low values.
The secondary peaks in the leftmost (rightmost) bin of the distribution for the $\dihiggs$ signal ($\ttbar$ background)
are due to events in which the event kinematics are atypical for signal (background) events, 
resulting in the PD for the wrong hypothesis $w_{1}(\vecy)$ ($w_{0}(\vecy)$) to be higher than the PD for the correct hypothesis $w_{0}(\vecy)$ ($w_{1}(\vecy)$).
About $4\%$ of signal ($10\%$ of background) events populate the leftmost (rightmost) bin of the distribution in case the LR $P(\vecy)$ is computed at MC-truth level.
In case the LR is computed at detector level, the fraction of $\dihiggs$ signal ($\ttbar$ background) events 
that populate the leftmost (rightmost) bin increases to $14\%$ (decreases to $7\%$).
The ``receiver-operating-characteristic'' (ROC) curves~\cite{ROCcurve} that correspond to these distributions are shown in Fig.~\ref{fig:ROC}.
The ROC curve quantifies the separation between the $\dihiggs$ signal and the $\ttbar$ background
and is obtained by varying the threshold of a cut on the LR $P(\vecy)$ and plotting the fractions of signal and background events passing the cut.
For a signal efficiency of $35\%$, the $\ttbar$ background is reduced by about three orders of magnitude, to a level of $0.09\%$, in case the LR is computed at MC-truth level.
In case the LR is computed at detector level, the $\ttbar$ background is reduced to a level of $0.26\%$.
The degradation in separation power that occurs at detector level is mainly due to 
signal events in which one of the bottom quarks originating from the $\PHiggs$ boson decay is not reconstructed as $\Pbottom$-jet at detector level
and a light quark or gluon jet is misidentified as $\Pbottom$-jet instead.
If this happens, the mass of the two detector-level jets that are reconstructed as $\Pbottom$-jets are often incompatible with $m_{\PHiggs}$.
The presence of a BW propagator in the ME $\mathcal{M}_{0}(\vecphat)$ for the signal hypothesis,
which enforces that the mass of the pair of $\Pbottom$-jets equals $m_{\PHiggs}$,
then introduces large ``pulls'' in the TF $W(E|\Ehat)$ for the $\Pbottom$-jet energy, which diminish the value of the integrand.

To better gauge the level of separation of the $\dihiggs$ signal from the $\ttbar$ background presented in Fig.~\ref{fig:ROC},
we compute signal efficiencies and background rates that one would obtain by cutting on the mass, $\mbb$, of the $\Pbottom$-jet pair, shown in Fig.~\ref{fig:mbb}, for comparison.
The observable $\mbb$ is presumably one of the most powerful single observables to separate the $\dihiggs$ signal from the $\ttbar$ background.
Fitting the peak of the distribution in the mass of the $\Pbottom$-jet pair in $\dihiggs$ signal events, obtained after the jet energy calibration is applied,
with a normal distribution yields a mean of $123$~\GeV and a standard deviation of $20$~\GeV.
Requiring events to have a value of $\mbb$ within $1$ ($2$) standard deviations around the mean
selects $78\%$ ($89\%$) of the $\dihiggs$ signal and $27\%$ ($43\%$) of the $\ttbar$ background.
Compared to the cut on $\mbb$,
the LR $P(\vecy)$ allows for a significantly higher reduction in the rate of $\ttbar$ background
by exploiting the full difference in event kinematics between the $\dihiggs$ signal and the $\ttbar$ background.

We remark that the misidentification of hadrons as leptons is not simulated in $\textsc{DELPHES}$ and hence not accounted for in the detector-level ROC curve shown in Fig.~\ref{fig:ROC}.
Based on the analysis of $\dihiggs$ production performed in the decay channel $\Pbottom\APbottom\PW\PW\virt$ by the CMS collaboration during LHC Run $2$~\cite{HIG-17-006},
which found the background arising from the misidentification of hadrons as leptons to be negligible,
we expect the misidentification of hadrons as leptons to have at most a small effect on the ROC curve.

We conclude this section on the performance of the MEM with a study of the effect of using ME of LO when computing the weights $w_{0}(\vecy)$ and $w_{1}(\vecy)$ 
by means of Eqs.~(\ref{eq:mem_signal}) and~(\ref{eq:mem_background}) and with a discussion of the computing-time requirements of the MEM.

\begin{figure}
\ifx\ver\verPreprint
\setlength{\unitlength}{1mm}
\begin{center}
\begin{picture}(160,78)(0,0)
\put(0.0, 0.0){\mbox{\includegraphics*[height=78mm]
 {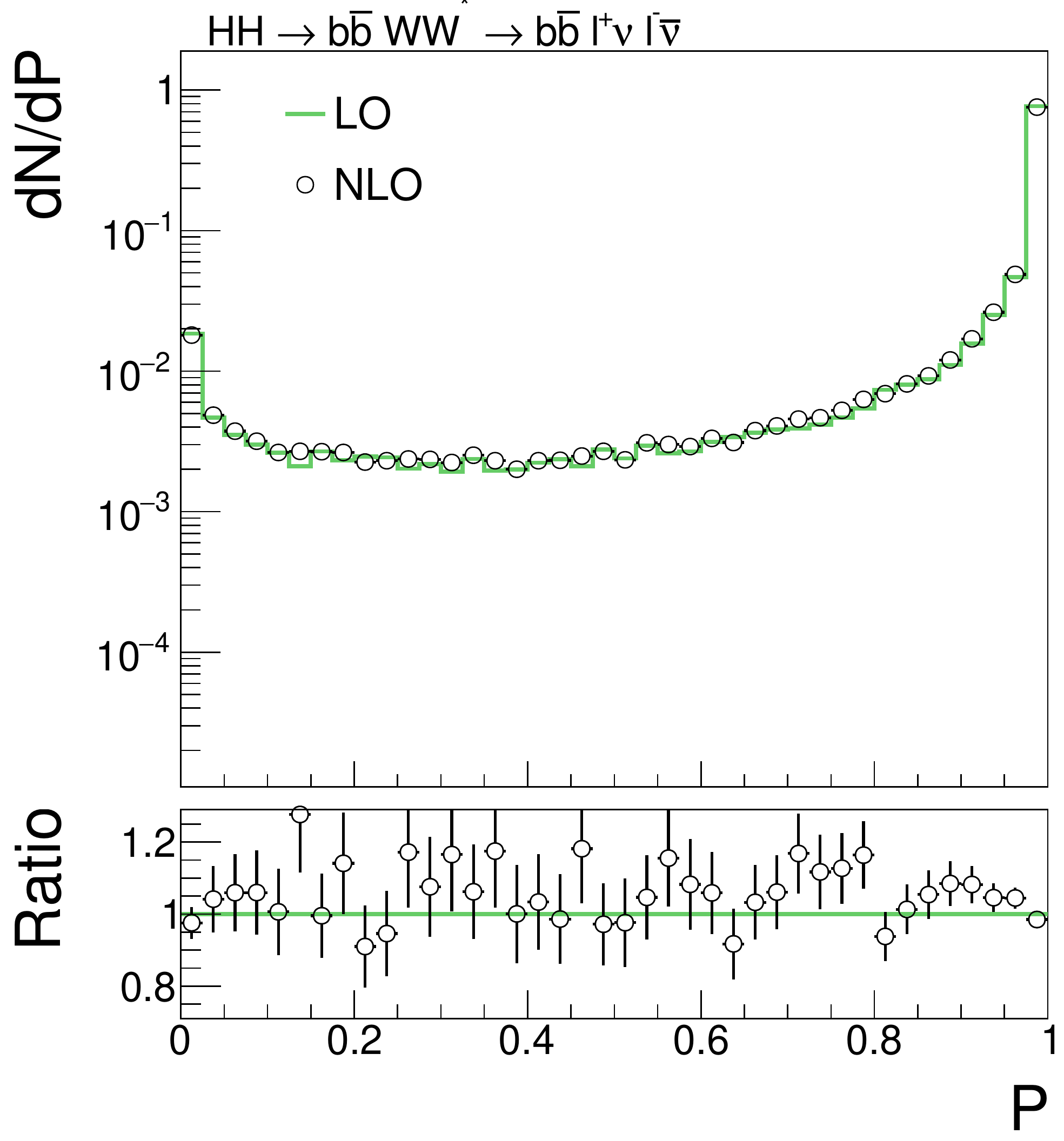}}}
\put(81.0, 0.0){\mbox{\includegraphics*[height=78mm]
 {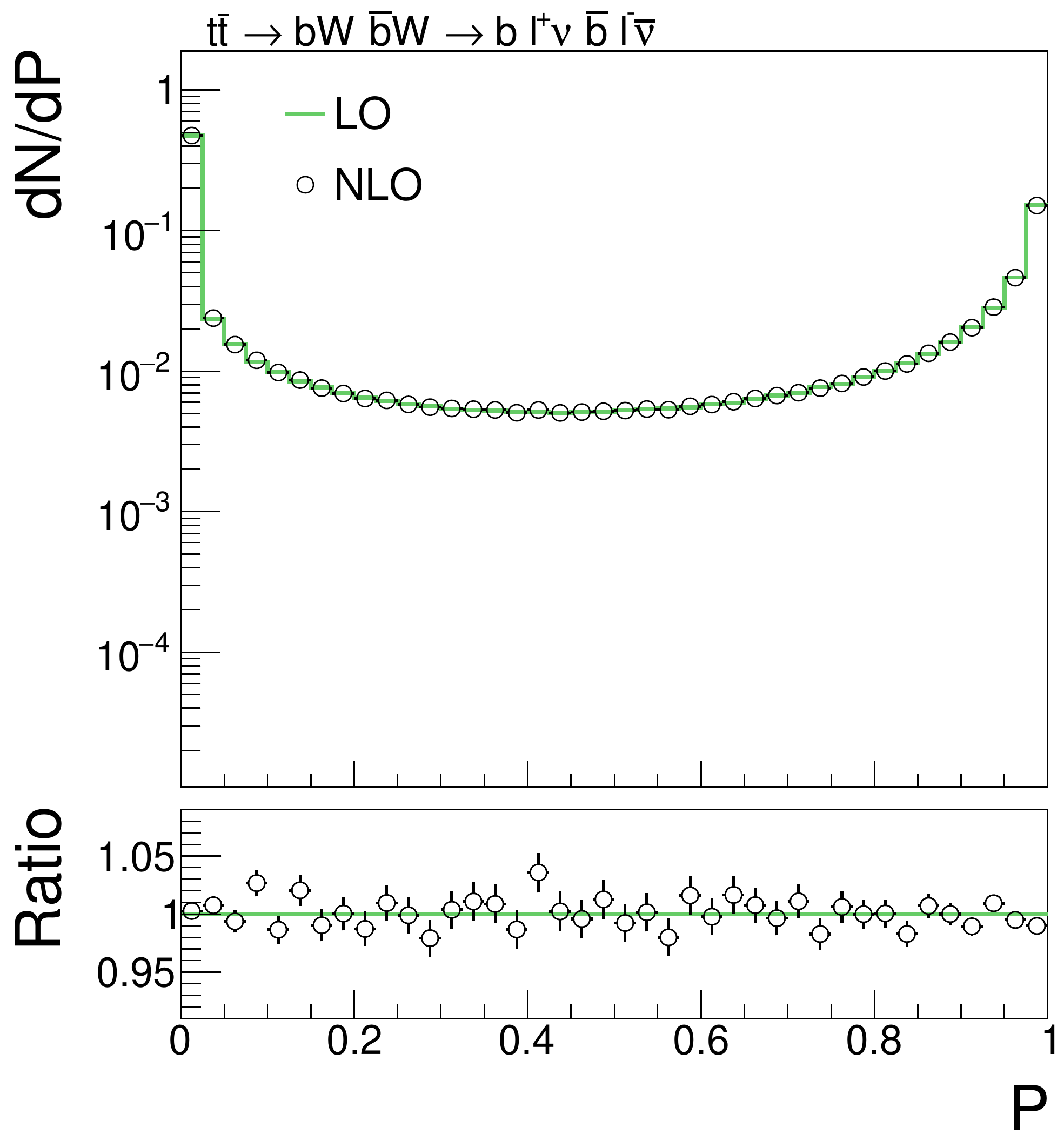}}}
\end{picture}
\end{center}
\fi
\ifx\ver\verPAPER
\centering
\includegraphics[width=0.48\textwidth]{plots/hh_bbwwMEM_dilepton_lo_vs_nlo_memLR_signal.pdf}
\includegraphics[width=0.48\textwidth]{plots/hh_bbwwMEM_dilepton_lo_vs_nlo_memLR_background.pdf}
\fi
\caption{
  Distribution in the LR $P(\vecy)$ 
  for $\dihiggs$ signal (left) and $\ttbar$ background (right) events
  simulated at LO and at NLO accuracy in pQCD.
  The likelihood ratios are computed at MC-truth level.
}
\label{fig:memLR_LO_vs_NLO}
\end{figure}

\begin{figure}
\ifx\ver\verPreprint
\setlength{\unitlength}{1mm}
\begin{center}
\begin{picture}(160,67)(0,0)
\put(39.5, 0.0){\mbox{\includegraphics*[height=67mm]
 {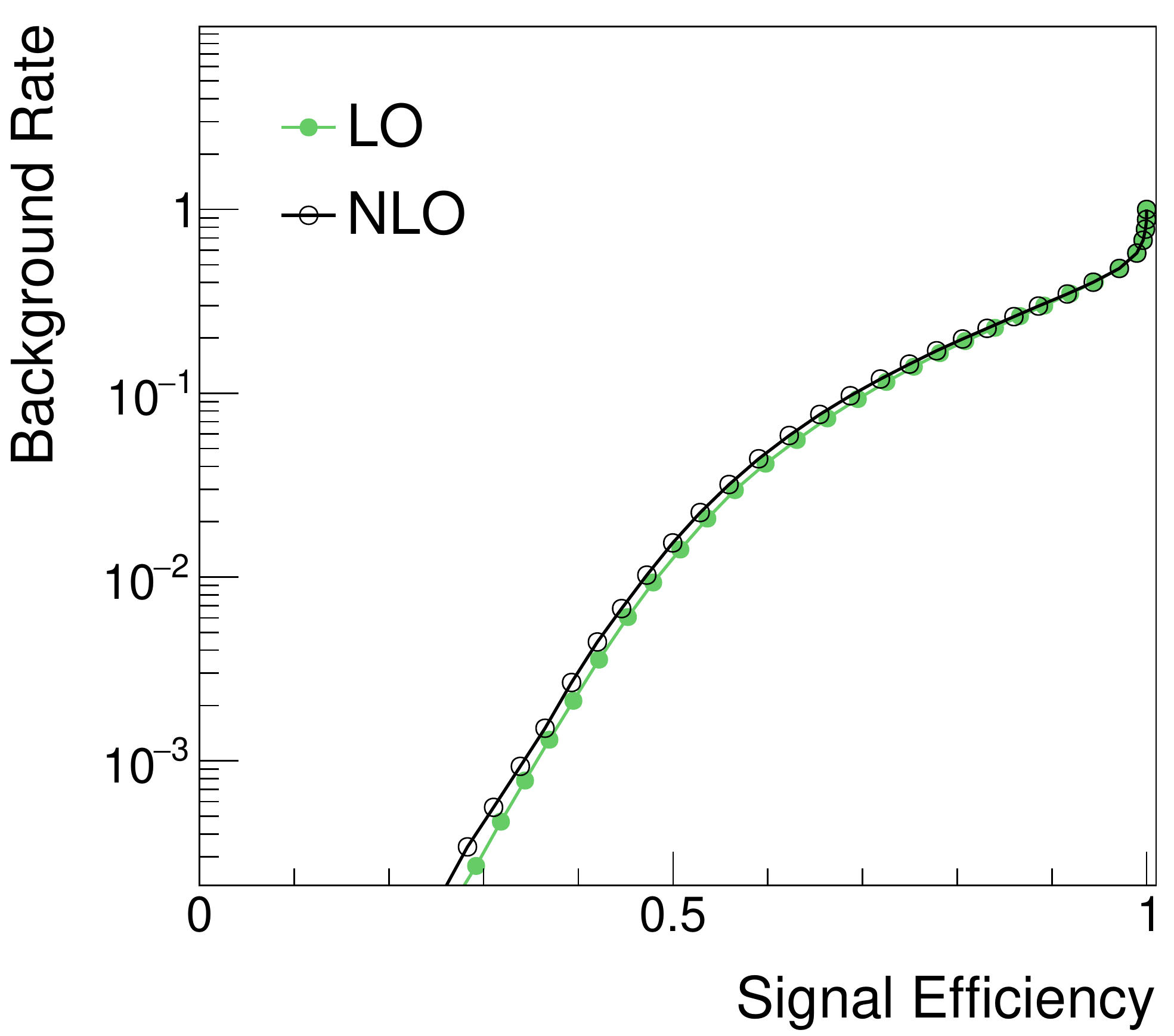}}}
\end{picture}
\end{center}
\fi
\ifx\ver\verPAPER
\centering
\includegraphics[width=0.57\textwidth]{plots/hh_bbwwMEM_dilepton_lo_vs_nlo_ROC.pdf}
\fi
\caption{
  Separation between the $\dihiggs$ signal and the $\ttbar$ background 
  for events simulated at LO and at NLO accuracy in pQCD.
  The graphs of background rate versus signal efficiency shown in the figure
  are obtained by applying a cut on the distributions in the likelihood ratios $P(\vecy)$ shown in Fig.~\ref{fig:memLR_LO_vs_NLO}.
}
\label{fig:ROC_LO_vs_NLO}
\end{figure}

Unfortunately, we cannot compare the performance of the MEM
for the case of using ME generated at LO versus ME generated at NLO in Eqs.~(\ref{eq:mem_signal}) and~(\ref{eq:mem_background}) directly,
because the program $\textsc{MadGraph\_aMCatNLO}$ does not support the generation of code for NLO ME at present
and also because the usage of NLO ME in the MEM would increase the computing-time requirements by $1$-$2$ orders of magnitude.
Instead, we use ME generated at LO accuracy in Eqs.~(\ref{eq:mem_signal}) and~(\ref{eq:mem_background}) 
and compare the resulting performance in separating the $\dihiggs$ signal from the $\ttbar$ background
for MC samples simulated at LO and at NLO accuracy in pQCD.
The NLO samples are expected to provide the more accurate modeling of real data and the LO samples are taken as a (more or less precise) approximation.
We take the difference in performance achieved by the MEM on the MC samples simulated at LO and at NLO accuracy
as an estimate for the loss in discrimination power that results from our choice of using LO ME and ignoring the effects of higher orders in the MEM.
Distributions in the LR $P(\vecy)$ computed for $\dihiggs$ signal and $\ttbar$ background events simulated at LO and at NLO accuracy in pQCD 
are shown in Fig.~\ref{fig:memLR_LO_vs_NLO}. The events are analyzed at MC-truth level.
The corresponding ROC curve is presented in Fig.~\ref{fig:ROC_LO_vs_NLO}.
The usage of LO ME causes a moderate loss in the separation of the $\dihiggs$ signal from the $\ttbar$ background,
amounting to a few percent loss in signal efficiency (for the same background rate).
We conclude from these figures that the usage of LO ME represents a viable approximation.

The computing time required to evaluate the integrals given by Eqs.~(\ref{eq:mem_signal}) and~(\ref{eq:mem_background})
may represent a challenge in practical applications of the MEM.
Experimental analyses will usually need to evaluate these integrals
multiple times for each event in order to assess the effect of systematic uncertainties.
Taken together with the large cross section for $\ttbar$ production at the LHC,
the integrals in Eqs.~(\ref{eq:mem_signal}) and~(\ref{eq:mem_background}) may need to be computed in the order of $100$ million times.
Even with several thousands of computing jobs running in parallel,
as it is nowadays commonplace for experimental data analyses performed at the LHC,
the computation still requires a few weeks of nonstop computing time.
Several possibilities to speed up the numeric integrations, which take most of the computing time in practical applications of the MEM,
have been explored in the literature.
One alternative is to use vector integrands to evaluate the likelihood ratio for all systematic uncertainties simultaneously~\cite{CUBA},
taking advantage of the fact that the systematic uncertainties typically constitute small changes with respect to the nominal value.
Another alternative is to take advantage of the parallelizability of multidimensional integration and perform the integration on graphics processing units (GPUs).
Speedup factors of order $100$, compared to using a single core of a general-purpose central processing unit (CPU) 
such as the $2.30$~GHz Intel\TReg~Xeon\TReg~E5-2695V3 processor that we used for the studies presented in this paper,
are reported in the literature for performing numeric integrations on GPUs~\cite{Hagiwara:2009aq,Hagiwara:2009cy,Kanzaki:2010ym,Hagiwara:2013oka,Schouten:2014yza,Grasseau:2015vfa}.

\section{Summary}
\label{sec:summary}

We presented an application of the matrix element method 
to the search for non-resonant $\dihiggs$ production in the channel $\dihiggs \to \Pbottom\APbottom\PW\PW\virt$ at LHC,
focusing on events in which the two $\PW$ bosons decay to a pair of electrons or muons.
According to the Neyman-Pearson lemma,
the likelihood ratio $P(\vecy)$ given by Eq.~(\ref{eq:memLR}) provides the optimal separation of the $\dihiggs$ signal from the dominant irreducible $\ttbar$ background.
We have studied the separation of the $\dihiggs$ signal from the $\ttbar$ background at Monte-Carlo truth and at detector level.
The latter has been simulated using the $\textsc{DELPHES}$ fast-simulation framework.
For experimental conditions characteristic for the ATLAS and CMS experiments during LHC Run $2$,
we find that the $\ttbar$ background can be reduced to a level of $0.26\%$ for a signal efficiency of $35\%$.
We regard the potential of the matrix element method for enhancing the sensitivity of the analysis of $\dihiggs$ production in the channel $\dihiggs \to \Pbottom\APbottom\PW\PW\virt$
as promising and we hope this paper will motivate the ATLAS and CMS collaborations to employ the method in a full analysis.

\section*{Acknowledgements}

This work has been supported by the Estonian Research Council grant PRG445.

\section{Appendix}
\label{sec:appendix}

In this section, we derive a few useful relations
that allow us to simplify the expression for the probability density $w_{i}(\vecy)$ starting from Eq.~(\ref{eq:mem1}).
We begin by deriving relations for the TF of charged leptons and of $\Pbottom$-jets, which we present in Section~\ref{sec:appendix_TF}.
In Section~\ref{sec:appendix_mass_constraints}, 
we will derive relations corresponding to various mass constraints.
The constraints arise from the presence of BW propagators 
in the ME $\mathcal{M}_{i}(\vecyhat)$ for the signal ($i=0$) and for the background ($i=1$) hypothesis.
The effect of the BW propagators is that only those points $\vecphat$ in the $6$-particle phase space contribute to the value of the integral in Eq.~(\ref{eq:mem1})
for which certain systems of final state particles satisfy certain mass conditions.
The relations derived in Sections~\ref{sec:appendix_TF} and~\ref{sec:appendix_mass_constraints}
are used to transform Eq.~(\ref{eq:mem1}) 
into Eq.~(\ref{eq:mem_signal}) for the $\dihiggs$ signal hypothesis and 
into Eq.~(\ref{eq:mem_background}) for the $\ttbar$ background hypothesis, respectively.

\subsection{Relations for transfer functions}
\label{sec:appendix_TF}

We assume that the directions of electrons, muons, and $\Pbottom$-jets
as well as the energies of electrons and muons are measured with negligible experimental resolution.
Our assumption implies that the TF for electrons and muons is given by:
\begin{linenowrapper}
\begin{equation}
W_{\Plepton}(\vecp|\vecphat) = f(E,\theta,\phi) \, \delta(E - \Ehat) \cdot \delta(\theta - \thetahat) \cdot \delta(\phi - \phihat) \, ,
\label{eq:TF_ell}
\end{equation}
\end{linenowrapper}
while the TF for $\Pbottom$-jets is given by:
\begin{linenowrapper}
\begin{equation}
W(\vecp|\vecphat) = f(E,\theta,\phi) \, W(E|\Ehat) \cdot \delta(\theta - \thetahat) \cdot \delta(\phi - \phihat) \, ,
\label{eq:TF_b}
\end{equation}
\end{linenowrapper}
where $E$ denotes the energy, $\theta$ the polar angle, and $\phi$ the azimuthal angle of the electron, muon, or $\Pbottom$-jet.
The function $W(E|\Ehat)$ quantifies the experimental resolution with which the energy of $\Pbottom$-jets is measured.
We choose the function $W(E|\Ehat)$ such that it satisfies the following normalization condition:
\begin{linenowrapper}
\begin{equation*}
\int \, dE \, W(E|\Ehat) \equiv 1.
\end{equation*}
\end{linenowrapper}
The function $f(E,\theta,\phi)$ ensures that the TF satisfy the normalization condition 
\begin{linenowrapper}
\begin{equation*}
\int \, d^{3}\vecp \, \Omega(\vecp) \, W(\vecp|\vecphat) = 1 \, .
\end{equation*}
\end{linenowrapper}
We only consider those events, which pass the event selection criteria, \ie for which $\Omega(\vecp)$ is equal to one.
With $d^{3}\vecp = \beta \, E^{2} \, \sin\theta \, dE \, d\theta \, d\phi$, it follows that:
\begin{linenowrapper}
\begin{equation*}
1 \equiv \int \, dE \, d\theta \, d\phi \, \beta \, E^{2} \, \sin\theta \, f(E, \theta, \phi) \, W(E|\Ehat) \, \delta(\theta - \thetahat) \cdot \delta(\phi - \phihat) \, ,
\end{equation*}
\end{linenowrapper}
which implies:
\begin{linenowrapper}
\begin{equation}
f(E,\theta,\phi) = \frac{1}{\beta \, E^{2} \, \sin\theta} \, .
\label{eq:TF_f}
\end{equation}
\end{linenowrapper}
Eq.~(\ref{eq:TF_f}) holds for electrons and muons as well as for $\Pbottom$-jets.

\subsection{Relations for mass constraints}
\label{sec:appendix_mass_constraints}

As explained in Section~\ref{sec:mem},
the presence of BW propagators in the ME $\mathcal{M}_{i}(\vecphat)$ renders the numeric integration inefficient,
unless the numeric integration is restricted to those narrow slices in the $6$-particle PS where the mass constraints are satisfied.
We achieve the desired restriction by inserting suitable $\delta$-functions into the integrand on the RHS of Eq.~(\ref{eq:mem3}).
In order to avoid that the insertion of the $\delta$-functions changes the value of the integral,
we formally insert a factor of $1$, which we write as:
\begin{linenowrapper}
\begin{eqnarray}
1 \equiv \textrm{BW} \, \cdot \, \textrm{BW}^{-1} 
 & = & \frac{\pi}{m_{\X} \, \Gamma_{\X}} \, \delta( E_{\X}^{2} - |\vecp_{\X}|^{2} - m_{\X}^{2} ) \cdot 
\left( (E_{\X}^{2} - |\vecp_{\X}|^{2} - m_{\X}^{2})^{2} + (m_{\X} \, \Gamma_{\X})^{2} \right) \nonumber \\
 & = & \pi \, m_{\X} \, \Gamma_{\X} \, \delta( E_{\X}^{2} - |\vecp_{\X}|^{2} - m_{\X}^{2} ) \, ,
\label{eq:deltaFunc}
\end{eqnarray}
\end{linenowrapper}
where we have used the narrow-width approximation to replace the first $\textrm{BW}$ propagator by a $\delta$-function.
The symbol $\X$ in Eq.~(\ref{eq:deltaFunc}) refers to the on-shell particle, of mass $m_{\X}$ and width $\Gamma_{\X}$, which imposes the mass constraint.

We insert Eq.~(\ref{eq:deltaFunc}) into the integrand on the RHS of Eq.~(\ref{eq:mem3})
and then use the $\delta$-function $\delta( E_{\X}^{2} - |\vecp_{\X}|^{2} - m_{\X}^{2} )$ 
to eliminate the integration over $\Ehat$ for one of the daughter particles that the particle $\X$ decays into.
The $\delta$-function rule:
\begin{linenowrapper}
\begin{equation} 
\delta\left( g(x) \right) = \frac{1}{|g^{\prime}(x_{0})|} \, \delta( x - x_{0} ) 
\label{eq:deltaFuncRule}
\end{equation}
\end{linenowrapper}
yields a factor of $|g^{\prime}(x_{0})|^{-1} \defL \left\lvert \frac{\partial g}{\partial x} \right\rvert_{x = x_{0}}$,
which we account for when eliminating the integration over $\Ehat$.
The symbol $x_{0}$ denotes the root of $g(x)$.

\subsubsection{Energy of \texorpdfstring{$\APbottom$}{bbar} produced in \texorpdfstring{$\PHiggs \to \Pbottom\APbottom$}{H->bbar} decay}
\label{sec:appendix_bEn_Hbb}

The condition that the mass of the $2$-particle system of $\Pbottom$ plus $\APbottom$ quark equals $m_{\PHiggs}$ implies that:
\begin{linenowrapper}
\begin{eqnarray}
m_{\PHiggs}^{2} \equiv m_{\Pbottom\APbottom}^{2} 
 & = & ( \Ehat_{\Pbottom} + \Ehat_{\APbottom} )^{2} - ( \vecphat_{\Pbottom} + \vecphat_{\APbottom} )^{2} \nonumber \\
 & = & \Ehat_{\Pbottom}^{2} + \Ehat_{\APbottom}^{2} + 2 \, \Ehat_{\Pbottom} \, \Ehat_{\APbottom} 
- |\vecphat_{\Pbottom}|^{2} - |\vecphat_{\APbottom}|^{2} - 2 \, \vecphat_{\Pbottom} \cdot \vecphat_{\APbottom} \nonumber \\
 & = & \underbrace{\Ehat_{\Pbottom}^{2} - |\vecphat_{\Pbottom}|^{2}}_{= m_{\Pbottom}^{2}} 
+ \underbrace{\Ehat_{\APbottom}^{2} - |\vecphat_{\APbottom}|^{2}}_{= m_{\Pbottom}^{2}} 
+ 2 \, \underbrace{\Ehat_{\Pbottom}}_{\defL a} \, \Ehat_{\APbottom} 
- 2 \, \underbrace{\sqrt{\Ehat_{\Pbottom}^{2} - m_{\Pbottom}^{2}} \, \vecehat_{\Pbottom} \cdot \vecehat_{\APbottom}}_{\defL b} \, 
 \sqrt{\Ehat_{\APbottom}^{2} - m_{\Pbottom}^{2}} \nonumber \\
\Longrightarrow 0 
 & = & \underbrace{\frac{m_{\PHiggs}^{2}}{2} - m_{\Pbottom}^{2}}_{\defL \Delta_{m_{\PHiggs}}} - a \, \Ehat_{\APbottom} + b \, \sqrt{\Ehat_{\APbottom}^{2} - m_{\Pbottom}^{2}} 
  \defR g(\Ehat_{\APbottom}) \, ,
\label{eq:bEn_Hbb1}
\end{eqnarray}
\end{linenowrapper}
where the symbol $\vecehat_{\Pbottom}$ denotes a unit vector in direction of the $\Pbottom$ quark
and the symbol $\APbottom$ a unit vector in direction of the $\APbottom$ quark.
Eq.~(\ref{eq:bEn_Hbb1}) has two solutions:
\begin{equation}
\Ehat_{\APbottom} = \frac{a \, \Delta_{m_{\PHiggs}} \pm |b| \, \sqrt{\Delta_{m_{\PHiggs}}^{2} - (a^{2} - b^{2}) \, m_{\Pbottom}^{2}}}{a^{2} - b^{2}} \, .
\label{eq:bEn_Hbb2}
\end{equation}
We discard the solution of lower energy and consider the solution of higher energy only, 
\ie we take the solution corresponding to the $+$ sign in Eq.~(\ref{eq:bEn_Hbb2}).

The derivative of the RHS of Eq.~(\ref{eq:bEn_Hbb1}) with respect to $\Ehat_{\APbottom}$ amounts to:
\begin{linenowrapper}
\begin{eqnarray}
\frac{1}{|g^{\prime}(\Ehat_{\APbottom})|} 
& = & \frac{1}{\Bigg\lvert a - \frac{b \, \Ehat_{\APbottom}}{\underbrace{\sqrt{\Ehat_{\APbottom}^{2} - m_{\Pbottom}^{2}}}_{= \beta_{\APbottom} \, \Ehat_{\APbottom}}} \Bigg\rvert}
 = \frac{1}{\left\lvert a - \frac{1}{\betahat_{\APbottom}} \, b \right\rvert} \nonumber \\
& = & \frac{1}{\left\lvert \Ehat_{\Pbottom} - \frac{1}{\betahat_{\APbottom}} \,
\smash{\underbrace{\sqrt{\Ehat_{\Pbottom}^{2} - m_{\Pbottom}^{2}}}_{= \betahat_{\Pbottom} \, \Ehat_{\Pbottom}}} \,
\smash{\underbrace{\vecehat_{\Pbottom} \cdot \vecehat_{\APbottom}}_{\defL \cos\sphericalangle(\vecehat_{\Pbottom},\vecehat_{\APbottom})}} \right\rvert}
 = \frac{1}{\left\lvert \Ehat_{\Pbottom} \, \left( 1 - \frac{\betahat_{\Pbottom}}{\betahat_{\APbottom}} \, \cos\sphericalangle(\vecehat_{\Pbottom},\vecehat_{\APbottom}) \right) \right\rvert} \, .
\label{eq:bEn_Hbb3}
\end{eqnarray}
\end{linenowrapper}

\vspace*{1em} 

\subsubsection{Energy of \texorpdfstring{$\Pnu$}{v} produced in \texorpdfstring{$\PW \to \ellnu$}{W->lnu} decay}
\label{sec:appendix_nuEn_Wlnu}

The condition that the mass of the $2$-particle system of $\Plepton$ plus $\Pnu$ equals $m_{\PW}$ implies that:
\begin{linenowrapper}
\begin{eqnarray}
m_{\PW}^{2} \equiv m_{\Plepton\Pnu}^{2} 
 & = & ( \Ehat_{\Plepton} + \Ehat_{\Pnu} )^{2} - ( \vecphat_{\Plepton} + \vecphat_{\Pnu} )^{2} \nonumber \\
 & = & \Ehat_{\Plepton}^{2} + \Ehat_{\Pnu}^{2} + 2 \, \Ehat_{\Plepton} \, \Ehat_{\Pnu} 
- |\vecphat_{\Plepton}|^{2} - |\vecphat_{\Pnu}|^{2} - 2 \, \vecphat_{\Plepton} \cdot \vecphat_{\Pnu} \nonumber \\
 & = & \underbrace{\Ehat_{\Plepton}^{2} - |\vecphat_{\Plepton}|^{2}}_{= m_{\Plepton}^{2} \approx 0} 
+ \underbrace{\Ehat_{\Pnu}^{2} - |\vecphat_{\Pnu}|^{2}}_{= m_{\Pnu}^{2} \approx 0} 
+ 2 \, \Ehat_{\Plepton} \, \Ehat_{\Pnu} 
- 2 \, \underbrace{|\vecphat_{\Plepton}|}_{\approx \Ehat_{\Plepton}} \, \underbrace{|\vecphat_{\Pnu}|}_{\approx \Ehat_{\Pnu}} \, 
 \underbrace{\vecehat_{\Plepton} \cdot \vecehat_{\Pnu}}_{\defL \cos\sphericalangle(\vecehat_{\Plepton},\vecehat_{\Pnu})} \nonumber \\
 & = & 2 \, \Ehat_{\Plepton} \, \Ehat_{\Pnu} \, 
  \underbrace{\left(1 - \cos\sphericalangle(\vecehat_{\Plepton},\vecehat_{\Pnu})\right)}_{= 2 \, \sin^{2}\left(\frac{\sphericalangle(\vecehat_{\Plepton},\vecehat_{\Pnu})}{2}\right)} 
   = 4 \, \Ehat_{\Plepton} \, \Ehat_{\Pnu} \, \sin^{2}\left(\frac{\sphericalangle(\vecehat_{\Plepton},\vecehat_{\Pnu})}{2}\right) \nonumber \\
\Longrightarrow 0 & = & m_{\PW}^{2} - 4 \, \Ehat_{\Plepton} \, \Ehat_{\Pnu} \, \sin^{2}\left(\frac{\sphericalangle(\vecehat_{\Plepton},\vecehat_{\Pnu})}{2}\right)
  \defR g(\Ehat_{\Pnu}) \, ,
\label{eq:nuEn_Wlnu1}
\end{eqnarray}
\end{linenowrapper}
which has the solution:
\begin{equation}
\Ehat_{\Pnu} = \frac{m_{\PW}^{2}}{4 \, \Ehat_{\Plepton} \, \sin^{2}\left(\frac{\sphericalangle(\vecehat_{\Plepton},\vecehat_{\Pnu})}{2}\right)} \, .
\label{eq:nuEn_Wlnu2}
\end{equation}
The symbol $\sphericalangle(\vecehat_{\Plepton},\vecehat_{\Pnu})$ refers to the angle between the directions of the charged lepton and of the neutrino.

The derivative of the RHS of Eq.~(\ref{eq:nuEn_Wlnu1}) with respect to $\Ehat_{\Pnu}$ yields:
\begin{linenowrapper}
\begin{equation}
\frac{1}{|g^{\prime}(\Ehat_{\Pnu})|} 
 = \frac{1}{4 \, \Ehat_{\Plepton} \, \sin^{2}\left(\frac{\sphericalangle(\vecehat_{\Plepton},\vecehat_{\Pnu})}{2}\right)} \, . 
\label{eq:nuEn_Wlnu3}
\end{equation}
\end{linenowrapper}

\subsubsection{Energy of \texorpdfstring{$\Pnu\virt$}{v*} produced in \texorpdfstring{$\PHiggs \to \PW\PW\virt \to \ellnu \, \ellnuStar$}{H->WW*->lvl*v*} decay}
\label{sec:appendix_nuEn_Hww}

As mentioned previously, we denote by $\Plepton\virt$ and $\Pnu\virt$ the charged lepton and the neutrino that originate from the decay of the off-shell $\PW$ boson.
The condition that the mass of the $4$-particle system of $\Plepton$, $\Pnu$, $\Plepton\virt$, and $\Pnu\virt$ equals $m_{\PHiggs}$ implies that:
\begin{linenowrapper}
\begin{eqnarray}
m_{\PHiggs}^{2} \equiv m_{\ellnuellnuStar}^{2} 
 & = & ( \underbrace{\Ehat_{\Plepton} + \Ehat_{\Pnu} + \Ehat_{\ellStar}}_{\defL \Ehat_{\ellnuellStar}} + \Ehat_{\nuStar} )^{2} 
- ( \underbrace{\vecphat_{\Plepton} + \vecphat_{\Pnu} + \vecphat_{\ellStar}}_{\defL \vecphat_{\ellnuellStar}} + \vecphat_{\nuStar} )^{2} \nonumber \\
 & = & \Ehat_{\ellnuellStar}^{2} + \Ehat_{\nuStar}^{2} + 2 \, \Ehat_{\ellnuellStar} \, \Ehat_{\nuStar} 
- |\vecphat_{\ellnuellStar}|^{2} - |\vecphat_{\nuStar}|^{2} - 2 \, \vecphat_{\ellnuellStar} \cdot \vecphat_{\nuStar} \nonumber \\
 & = & \underbrace{\Ehat_{\ellnuellStar}^{2} - |\vecphat_{\ellnuellStar}|^{2}}_{\defL m_{\ellnuellStar}^{2}} 
+ \underbrace{\Ehat_{\nuStar}^{2} - |\vecphat_{\nuStar}|^{2}}_{= m_{\Pnu}^{2} \approx 0} 
+ 2 \, \underbrace{\Ehat_{\ellnuellStar}}_{\defL a} \, \Ehat_{\nuStar} \nonumber \\
 & & \quad - 2 \, \underbrace{\sqrt{\Ehat_{\ellnuellStar}^{2} - m_{\ellnuellStar}^{2}} \, \vecehat_{\ellnuellStar} \cdot \vecehat_{\nuStar}}_{\defL b} \, 
 \underbrace{|\vecphat_{\nuStar}|}_{\approx \Ehat_{\nuStar}} \nonumber \\
 & = & m_{\ellnuellStar}^{2} + 2 \, a \, \Ehat_{\nuStar} - 2 \, b \, \Ehat_{\nuStar} \nonumber \\
\Longrightarrow 0 & = & \underbrace{\frac{m_{\PHiggs}^{2} - m_{\ellnuellStar}^{2}}{2}}_{\defL \Delta_{m_{\PHiggs}}} - a \, \Ehat_{\nuStar} + b \, \Ehat_{\nuStar}
  \defR g(\Ehat_{\nuStar}) \, ,
\label{eq:nuEn_Hww1}
\end{eqnarray}
\end{linenowrapper}
which has the solution:
\begin{linenowrapper}
\begin{eqnarray}
\Ehat_{\nuStar}
 & = & \frac{\Delta_{m_{\PHiggs}}}{a - b} 
 = \frac{m_{\PHiggs}^{2} - m_{\ellnuellStar}^{2}}{2 \, \Big( \Ehat_{\ellnuellStar}
- \underbrace{\sqrt{\Ehat_{\ellnuellStar}^{2} - m_{\ellnuellStar}^{2}}}_{= \betahat_{\ellnuellStar} \, \Ehat_{\ellnuellStar}} \,
 \underbrace{\vecehat_{\ellnuellStar} \cdot \vecehat_{\nuStar}}_{\defL \cos\sphericalangle(\vecehat_{\ellnuellStar},\vecehat_{\nuStar})} \Big)} \nonumber \\
 & = & \frac{m_{\PHiggs}^{2} - m_{\ellnuellStar}^{2}}{2 \, \Ehat_{\ellnuellStar} \, 
\left( 1 - \betahat_{\ellnuellStar} \, \cos\sphericalangle(\vecehat_{\ellnuellStar},\vecehat_{\nuStar}) \right)} \, ,
\label{eq:nuEn_Hww2}
\end{eqnarray}
\end{linenowrapper}
where, for the purpose of shortening the nomenclature, we denote by the symbols $\Ehat_{\ellnuellStar}$ and $\vecphat_{\ellnuellStar}$ the energy and the momentum
of the $3$-particle system comprised of the neutrino originating from the decay of the on-shell $\PW$ boson and of the two charged leptons,
by $\vecehat_{\ellnuellStar}$ a unit vector in direction of $\vecphat_{\ellnuellStar}$,
and by $m_{\ellnuellStar}$ the mass of this $3$-particle system.

The derivative of the RHS of Eq.~(\ref{eq:nuEn_Hww1}) with respect to $\Ehat_{\nuStar}$ amounts to:
\begin{linenowrapper}
\begin{eqnarray}
\frac{1}{|g^{\prime}(\Ehat_{\nuStar})|} 
 & = &\frac{1}{|a - b|} 
  = \frac{1}{\left\lvert\Ehat_{\ellnuellStar} - \smash{\underbrace{\sqrt{\Ehat_{\ellnuellStar}^{2} - m_{\ellnuellStar}^{2}}}_{= \betahat_{\ellnuellStar} \, \Ehat_{\ellnuellStar}}} \, 
\cos\sphericalangle(\vecehat_{\ellnuellStar},\vecehat_{\nuStar})\right\rvert} \nonumber \\[2em]
 & = & \frac{1}{\Ehat_{\ellnuellStar} \left\lvert 1 - \betahat_{\ellnuellStar} \, \cos\sphericalangle(\vecehat_{\ellnuellStar},\vecehat_{\nuStar}) \right\rvert} \, .
\label{eq:nuEn_Hww3}
\end{eqnarray}
\end{linenowrapper}
When inserting Eq.~(\ref{eq:nuEn_Hww3}) into Eq.~(\ref{eq:mem3}) to obtain the expression for the PD $w_{0}(\vecy)$ of the signal hypothesis in Eq.~(\ref{eq:mem_signal}),
we will omit the modulus in the denominator. 
The modulus is redundant, because the argument $1 - \betahat_{\ellnuellStar} \, \cos\sphericalangle(\vecehat_{\ellnuellStar},\vecehat_{\nuStar})$ is never negative.

\subsubsection{Energy of \texorpdfstring{$\Pbottom$}{b}  (\texorpdfstring{$\APbottom$}{bbar}) produced in \texorpdfstring{$\Ptop \to \Pbottom\PW^{+} \to \Pbottom\ellPlusnu$}{t->bW+->bl+v} (\texorpdfstring{$\APtop \to \APbottom\PW^{-} \to \APbottom\ellMinusnu$}{tbar->bbarW- ->bbarl-vbar}) decay}
\label{sec:appendix_bEn_top}

The condition that the mass of of the $3$-particle system comprised of the $\Pbottom$ quark, the charged anti-lepton, and the neutrino equals $m_{\Ptop}$ implies that:
\begin{linenowrapper}
\begin{eqnarray}
m_{\Ptop}^{2} \equiv m_{\Pbottom\ellPlusnu}^{2} 
 & = & ( \Ehat_{\Pbottom} + \underbrace{\Ehat_{\ellPlus} + \Ehat_{\Pnu}}_{\defL \Ehat_{\ellPlusnu}} )^{2} 
- ( \vecphat_{\Pbottom} + \underbrace{\vecphat_{\ellPlus} + \vecphat_{\Pnu}}_{\defL \vecphat_{\ellPlusnu}} )^{2} \nonumber \\
 & = & \Ehat_{\Pbottom}^{2} + \Ehat_{\ellPlusnu}^{2} + 2 \, \Ehat_{\Pbottom} \, \Ehat_{\ellPlusnu} 
- |\vecphat_{\Pbottom}|^{2} - |\vecphat_{\ellPlusnu}|^{2} - 2 \, \vecphat_{\Pbottom} \cdot \vecphat_{\ellPlusnu} \nonumber \\
 & = & \underbrace{\Ehat_{\Pbottom}^{2} - |\vecphat_{\Pbottom}|^{2}}_{= m_{\Pbottom}^{2}} 
+ \underbrace{\Ehat_{\ellPlusnu}^{2} - |\vecphat_{\ellPlusnu}|^{2}}_{= m_{\PW}^{2}} 
+ 2 \, \underbrace{\Ehat_{\ellPlusnu}}_{\defL a} \, \Ehat_{\Pbottom} \nonumber \\
 & & \quad - 2 \, \underbrace{\sqrt{\Ehat_{\ellPlusnu}^{2} - m_{\PW}^{2}} \, \vecehat_{\ellPlusnu} \cdot \vecehat_{\Pbottom}}_{\defL b} \, 
 \sqrt{\Ehat_{\Pbottom}^{2} - m_{\Pbottom}^{2}} \nonumber \\
\Longrightarrow 0 & = & \underbrace{\frac{m_{\Ptop}^{2} - m_{\Pbottom}^{2} - m_{\PW}^{2}}{2}}_{\defL \Delta_{m_{\Ptop}}} - a \, \Ehat_{\Pbottom} + b \, \sqrt{\Ehat_{\Pbottom}^{2} - m_{\Pbottom}^{2}} 
  \defR g(\Ehat_{\Pbottom}) \, ,
\label{eq:bEn_top1}
\end{eqnarray}
\end{linenowrapper}
where we denote the energy of the system of the charged anti-lepton and of the neutrino by the symbol $\Ehat_{\ellPlusnu}$ 
and the momentum of this system by the symbol $\vecphat_{\ellPlusnu}$.
The symbol $\vecehat_{\ellPlusnu}$ denotes a unit vector in direction of $\vecphat_{\ellPlusnu}$.
The mass of this system equals $m_{\PW}$, as the $\PW$ boson produced in the decay $\Ptop \to \Pbottom\PW^{+}$ is on-shell.
Eq.~(\ref{eq:bEn_top1}) has two solutions:
\begin{linenowrapper}
\begin{equation}
\Ehat_{\Pbottom} = \frac{a \, \Delta_{m_{\Ptop}} \pm |b| \, \sqrt{\Delta_{m_{\Ptop}}^{2} - (a^{2} - b^{2}) \, m_{\Pbottom}^{2}}}{a^{2} - b^{2}} \, .
\label{eq:bEn_top2}
\end{equation}
\end{linenowrapper}
We discard the solution of lower energy and consider the solution of higher energy only,
\ie we take the solution corresponding to the $+$ sign in Eq.~(\ref{eq:bEn_top2}).

The derivative of the RHS of Eq.~(\ref{eq:bEn_top1}) with respect to $\Ehat_{\Pbottom}$ yields:
\begin{linenowrapper}
\begin{eqnarray}
\frac{1}{|g^{\prime}(\Ehat_{\Pbottom})|} 
 & = & \frac{1}{\Bigg\lvert a - \frac{b \, \Ehat_{\Pbottom}}{\underbrace{\sqrt{\Ehat_{\Pbottom}^{2} - m_{\Pbottom}^{2}}}_{= \beta_{\Pbottom} \, \Ehat_{\Pbottom}}} \Bigg\rvert}
  = \frac{1}{\left\lvert a - \frac{1}{\betahat_{\Pbottom}} \, b \right\rvert}
  = \frac{1}{\left\lvert \Ehat_{\ellPlusnu} - \frac{1}{\betahat_{\Pbottom}} \,
\smash{\underbrace{\sqrt{\Ehat_{\ellPlusnu}^{2} - m_{\PW}^{2}}}_{= \betahat_{\ellPlusnu} \, \Ehat_{\ellPlusnu}}} \,
\smash{\underbrace{\vecehat_{\ellPlusnu} \cdot \vecehat_{\Pbottom}}_{\defL \cos\sphericalangle(\vecehat_{\ellPlusnu},\vecehat_{\Pbottom})}} \right\rvert} \nonumber \\
 & = & \frac{1}{\left\lvert \Ehat_{\ellPlusnu} \, \left( 1 - \frac{\betahat_{\ellPlusnu}}{\betahat_{\Pbottom}} \, \cos\sphericalangle(\vecehat_{\ellPlusnu},\vecehat_{\Pbottom}) \right) \right\rvert} \, .
\label{eq:bEn_top3}
\end{eqnarray}
\end{linenowrapper}

The corresponding expressions for the case of $\APbottom$ quark, charged lepton, and anti-neutrino are identical to Eqs.~(\ref{eq:bEn_top2}) and~(\ref{eq:bEn_top3}),
except that the symbol $\Pbottom$ is replaced by $\APbottom$ and the symbols $\ellPlus$ and $\Pnu$ are replaced by $\ellMinus$ and $\APnu$.

\bibliography{bbwwMEM}

\begin{thebibliography}{100}
\expandafter\ifx\csname url\endcsname\relax
  \def\url#1{\texttt{#1}}\fi
\expandafter\ifx\csname urlprefix\endcsname\relax\def\urlprefix{URL }\fi
\expandafter\ifx\csname href\endcsname\relax
  \def\href#1#2{#2} \def\path#1{#1}\fi

\bibitem{Higgs-Discovery_ATLAS}
G.~Aad, et~al., {Observation of a new particle in the search for the Standard
  Model Higgs boson with the ATLAS detector at the LHC}, Phys. Lett. B 716
  (2012) 1.
\newblock \href {http://arxiv.org/abs/1207.7214} {\path{arXiv:1207.7214}},
  \href {https://doi.org/10.1016/j.physletb.2012.08.020}
  {\path{doi:10.1016/j.physletb.2012.08.020}}.

\bibitem{Higgs-Discovery_CMS}
S.~Chatrchyan, et~al., {Observation of a new boson at a mass of $125$~\GeV with
  the CMS experiment at the LHC}, Phys. Lett. B 716 (2012) 30.
\newblock \href {http://arxiv.org/abs/1207.7235} {\path{arXiv:1207.7235}},
  \href {https://doi.org/10.1016/j.physletb.2012.08.021}
  {\path{doi:10.1016/j.physletb.2012.08.021}}.

\bibitem{HIG-14-042}
G.~Aad, et~al., {Combined measurement of the Higgs boson mass in
  $\textrm{p}\textrm{p}$ collisions at $\sqrt{s}=7$ and $8$~TeV with the ATLAS
  and CMS experiments}, Phys. Rev. Lett. 114 (2015) 191803.
\newblock \href {http://arxiv.org/abs/1503.07589} {\path{arXiv:1503.07589}},
  \href {https://doi.org/10.1103/PhysRevLett.114.191803}
  {\path{doi:10.1103/PhysRevLett.114.191803}}.

\bibitem{ATLAS:2020coj}
{{ATLAS} Collaboration},
  \href{https://cdsweb.cern.ch/record/2714883}{{Measurement of the Higgs boson
  mass in the $\PHiggs \to \PZ\PZ^{*} \to 4 \Plepton$ decay channel with
  $\sqrt{s}=13$~\TeV $\Pp\Pp$ collisions using the ATLAS detector at the LHC}},
  Tech. rep. (2020).
\newline\urlprefix\url{https://cdsweb.cern.ch/record/2714883}

\bibitem{CMS:2020xrn}
A.~M. Sirunyan, et~al., {A measurement of the Higgs boson mass in the diphoton
  decay channel}, Phys. Lett. B 805 (2020) 135425.
\newblock \href {http://arxiv.org/abs/2002.06398} {\path{arXiv:2002.06398}},
  \href {https://doi.org/10.1016/j.physletb.2020.135425}
  {\path{doi:10.1016/j.physletb.2020.135425}}.

\bibitem{HIG-14-018}
V.~Khachatryan, et~al., {Constraints on the spin-parity and anomalous
  $\PHiggs\textrm{V}\textrm{V}$ couplings of the Higgs boson in proton
  collisions at $7$ and $8$~\TeV}, Phys. Rev. D 92~(1) (2015) 012004.
\newblock \href {http://arxiv.org/abs/1411.3441} {\path{arXiv:1411.3441}},
  \href {https://doi.org/10.1103/PhysRevD.92.012004}
  {\path{doi:10.1103/PhysRevD.92.012004}}.

\bibitem{Aad:2015mxa}
G.~Aad, et~al., {Study of the spin and parity of the Higgs boson in diboson
  decays with the ATLAS detector}, Eur. Phys. J. C 75~(10) (2015) 476,
  [Erratum: Eur.~Phys.~J.~C~76,~no.3,~152~(2016)].
\newblock \href {http://arxiv.org/abs/1506.05669} {\path{arXiv:1506.05669}},
  \href {https://doi.org/10.1140/epjc/s10052-015-3685-1,
  10.1140/epjc/s10052-016-3934-y} {\path{doi:10.1140/epjc/s10052-015-3685-1,
  10.1140/epjc/s10052-016-3934-y}}.

\bibitem{ATLAS:2020evk}
G.~Aad, et~al., {Test of CP invariance in vector-boson fusion production of the
  Higgs boson in the $\PHiggs \to \Pgt\Pgt$ channel in proton-proton collisions
  at $\sqrt{s} = 13$~\TeV with the ATLAS detector}, Phys. Lett. B 805 (2020)
  135426.
\newblock \href {http://arxiv.org/abs/2002.05315} {\path{arXiv:2002.05315}},
  \href {https://doi.org/10.1016/j.physletb.2020.135426}
  {\path{doi:10.1016/j.physletb.2020.135426}}.

\bibitem{CMS:2021nnc}
A.~M. Sirunyan, et~al., {Constraints on anomalous Higgs boson couplings to
  vector bosons and fermions in its production and decay using the four-lepton
  final state}, Phys. Rev. D 104~(5) (2021) 052004.
\newblock \href {http://arxiv.org/abs/2104.12152} {\path{arXiv:2104.12152}},
  \href {https://doi.org/10.1103/PhysRevD.104.052004}
  {\path{doi:10.1103/PhysRevD.104.052004}}.

\bibitem{HIG-15-002}
G.~Aad, et~al., {Measurements of the Higgs boson production and decay rates and
  constraints on its couplings from a combined ATLAS and CMS analysis of the
  LHC $\Pp\Pp$ collision data at $\sqrt{s}=7$ and $8$~\TeV}, JHEP 08 (2016)
  045.
\newblock \href {http://arxiv.org/abs/1606.02266} {\path{arXiv:1606.02266}},
  \href {https://doi.org/10.1007/JHEP08(2016)045}
  {\path{doi:10.1007/JHEP08(2016)045}}.

\bibitem{CMS:2018yfx}
A.~M. Sirunyan, et~al., {Search for invisible decays of a Higgs boson produced
  through vector boson fusion in proton-proton collisions at $\sqrt{s} =
  13$~\TeV}, Phys. Lett. B 793 (2019) 520.
\newblock \href {http://arxiv.org/abs/1809.05937} {\path{arXiv:1809.05937}},
  \href {https://doi.org/10.1016/j.physletb.2019.04.025}
  {\path{doi:10.1016/j.physletb.2019.04.025}}.

\bibitem{CMS:2019ekd}
A.~M. Sirunyan, et~al., {Measurements of the Higgs boson width and anomalous
  $\PHiggs\textrm{V}\textrm{V}$ couplings from on-shell and off-shell
  production in the four-lepton final state}, Phys. Rev. D 99~(11) (2019)
  112003.
\newblock \href {http://arxiv.org/abs/1901.00174} {\path{arXiv:1901.00174}},
  \href {https://doi.org/10.1103/PhysRevD.99.112003}
  {\path{doi:10.1103/PhysRevD.99.112003}}.

\bibitem{Aad:2015pla}
G.~Aad, et~al., {Constraints on new phenomena via Higgs boson couplings and
  invisible decays with the ATLAS detector}, JHEP 11 (2015) 206.
\newblock \href {http://arxiv.org/abs/1509.00672} {\path{arXiv:1509.00672}},
  \href {https://doi.org/10.1007/JHEP11(2015)206}
  {\path{doi:10.1007/JHEP11(2015)206}}.

\bibitem{ATLAS:2018jym}
M.~Aaboud, et~al., {Constraints on off-shell Higgs boson production and the
  Higgs boson total width in $\PZ\PZ \to 4\Plepton$ and $\PZ\PZ \to 2\Plepton
  2\Pnu$ final states with the ATLAS detector}, Phys. Lett. B 786 (2018) 223.
\newblock \href {http://arxiv.org/abs/1808.01191} {\path{arXiv:1808.01191}},
  \href {https://doi.org/10.1016/j.physletb.2018.09.048}
  {\path{doi:10.1016/j.physletb.2018.09.048}}.

\bibitem{Aaboud:2018urx}
M.~Aaboud, et~al., {Observation of Higgs boson production in association with a
  top quark pair at the LHC with the ATLAS detector}, Phys. Lett. B 784 (2018)
  173.
\newblock \href {http://arxiv.org/abs/1806.00425} {\path{arXiv:1806.00425}},
  \href {https://doi.org/10.1016/j.physletb.2018.07.035}
  {\path{doi:10.1016/j.physletb.2018.07.035}}.

\bibitem{HIG-17-035}
A.~M. Sirunyan, et~al., {Observation of $\Ptop\APtop\PHiggs$ production}, Phys.
  Rev. Lett. 120~(23) (2018) 231801.
\newblock \href {http://arxiv.org/abs/1804.02610} {\path{arXiv:1804.02610}},
  \href {https://doi.org/10.1103/PhysRevLett.120.231801}
  {\path{doi:10.1103/PhysRevLett.120.231801}}.

\bibitem{deFlorian:2019app}
D.~de~Florian, I.~Fabre, J.~Mazzitelli, {Triple Higgs production at hadron
  colliders at NNLO in QCD}, JHEP 03 (2020) 155.
\newblock \href {http://arxiv.org/abs/1912.02760} {\path{arXiv:1912.02760}},
  \href {https://doi.org/10.1007/JHEP03(2020)155}
  {\path{doi:10.1007/JHEP03(2020)155}}.

\bibitem{HL-LHC-TDR}
I.~Zurbano~Fernandez, et~al., {High-Luminosity Large Hadron Collider (HL-LHC):
  Technical design report} 10/2020 (12 2020).
\newblock \href {https://doi.org/10.23731/CYRM-2020-0010}
  {\path{doi:10.23731/CYRM-2020-0010}}.

\bibitem{Plehn:2005nk}
T.~Plehn, M.~Rauch, {The quartic Higgs coupling at hadron colliders}, Phys.
  Rev. D 72 (2005) 053008.
\newblock \href {http://arxiv.org/abs/hep-ph/0507321}
  {\path{arXiv:hep-ph/0507321}}, \href
  {https://doi.org/10.1103/PhysRevD.72.053008}
  {\path{doi:10.1103/PhysRevD.72.053008}}.

\bibitem{Binoth:2006ym}
T.~Binoth, S.~Karg, N.~Kauer, R.~{R\"{u}ckl}, {Multi-Higgs boson production in
  the Standard Model and beyond}, Phys. Rev. D 74 (2006) 113008.
\newblock \href {http://arxiv.org/abs/hep-ph/0608057}
  {\path{arXiv:hep-ph/0608057}}, \href
  {https://doi.org/10.1103/PhysRevD.74.113008}
  {\path{doi:10.1103/PhysRevD.74.113008}}.

\bibitem{Grazzini:2018hh}
M.~Grazzini, et~al., Higgs boson pair production at {NNLO} with top quark mass
  effects, JHEP 05 (2018) 059.
\newblock \href {http://arxiv.org/abs/1803.02463} {\path{arXiv:1803.02463}},
  \href {https://doi.org/10.1007/JHEP05(2018)059}
  {\path{doi:10.1007/JHEP05(2018)059}}.

\bibitem{Baglio:2012np}
J.~Baglio, A.~Djouadi, R.~{Gr\"{o}ber}, M.~M. {M\"{u}hlleitner}, J.~Quevillon,
  M.~Spira, {The measurement of the Higgs self coupling at the LHC: theoretical
  status}, JHEP 04 (2013) 151.
\newblock \href {http://arxiv.org/abs/1212.5581} {\path{arXiv:1212.5581}},
  \href {https://doi.org/10.1007/JHEP04(2013)151}
  {\path{doi:10.1007/JHEP04(2013)151}}.

\bibitem{Craig:2013hca}
N.~Craig, J.~Galloway, S.~Thomas, {Searching for signs of the second Higgs
  doublet} (2013).
\newblock \href {http://arxiv.org/abs/1305.2424} {\path{arXiv:1305.2424}}.

\bibitem{Nhung:2013lpa}
D.~T. Nhung, M.~M{\"u}hlleitner, J.~Streicher, K.~Walz, {Higher order
  corrections to the trilinear Higgs self-couplings in the real NMSSM}, JHEP
  1311 (2013) 181.
\newblock \href {http://arxiv.org/abs/1306.3926} {\path{arXiv:1306.3926}},
  \href {https://doi.org/10.1007/JHEP11(2013)181}
  {\path{doi:10.1007/JHEP11(2013)181}}.

\bibitem{Grober:2010yv}
R.~Grober, M.~M{\"u}hlleitner, {Composite Higgs boson pair production at the
  LHC}, JHEP 1106 (2011) 020.
\newblock \href {http://arxiv.org/abs/1012.1562} {\path{arXiv:1012.1562}},
  \href {https://doi.org/10.1007/JHEP06(2011)020}
  {\path{doi:10.1007/JHEP06(2011)020}}.

\bibitem{Contino:2010mh}
R.~Contino, C.~Grojean, M.~Moretti, F.~Piccinini, R.~Rattazzi, {Strong double
  Higgs production at the LHC}, JHEP 1005 (2010) 089.
\newblock \href {http://arxiv.org/abs/1002.1011} {\path{arXiv:1002.1011}},
  \href {https://doi.org/10.1007/JHEP05(2010)089}
  {\path{doi:10.1007/JHEP05(2010)089}}.

\bibitem{Englert:2011yb}
C.~Englert, T.~Plehn, D.~Zerwas, P.~M. Zerwas, {Exploring the Higgs portal},
  Phys. Lett. B 703 (2011) 298.
\newblock \href {http://arxiv.org/abs/1106.3097} {\path{arXiv:1106.3097}},
  \href {https://doi.org/10.1016/j.physletb.2011.08.002}
  {\path{doi:10.1016/j.physletb.2011.08.002}}.

\bibitem{No:2013wsa}
J.~M. No, M.~Ramsey-Musolf, {Probing the Higgs portal at the LHC through
  resonant di-Higgs production}, Phys. Rev. D 89 (2014) 095031.
\newblock \href {http://arxiv.org/abs/1310.6035} {\path{arXiv:1310.6035}},
  \href {https://doi.org/10.1103/PhysRevD.89.095031}
  {\path{doi:10.1103/PhysRevD.89.095031}}.

\bibitem{Randall:1999ee}
L.~Randall, R.~Sundrum, {A Large mass hierarchy from a small extra dimension},
  Phys. Rev. Lett. 83 (1999) 3370.
\newblock \href {http://arxiv.org/abs/hep-ph/9905221}
  {\path{arXiv:hep-ph/9905221}}, \href
  {https://doi.org/10.1103/PhysRevLett.83.3370}
  {\path{doi:10.1103/PhysRevLett.83.3370}}.

\bibitem{Buchalla:2015wfa}
G.~Buchalla, O.~Cata, A.~Celis, C.~Krause, {Note on Anomalous Higgs-Boson
  Couplings in Effective Field Theory}, Phys. Lett. B 750 (2015) 298.
\newblock \href {http://arxiv.org/abs/1504.01707} {\path{arXiv:1504.01707}},
  \href {https://doi.org/10.1016/j.physletb.2015.09.027}
  {\path{doi:10.1016/j.physletb.2015.09.027}}.

\bibitem{Goertz:2014qta}
F.~Goertz, A.~Papaefstathiou, L.~L. Yang, J.~Zurita, {Higgs boson pair
  production in the D=6 extension of the SM}, JHEP 04 (2015) 167.
\newblock \href {http://arxiv.org/abs/1410.3471} {\path{arXiv:1410.3471}},
  \href {https://doi.org/10.1007/JHEP04(2015)167}
  {\path{doi:10.1007/JHEP04(2015)167}}.

\bibitem{HIG-13-032}
V.~Khachatryan, et~al., {Search for two Higgs bosons in final states containing
  two photons and two bottom quarks in proton-proton collisions at $8$~\TeV},
  Phys. Rev. D 94~(5) (2016) 052012.
\newblock \href {http://arxiv.org/abs/1603.06896} {\path{arXiv:1603.06896}},
  \href {https://doi.org/10.1103/PhysRevD.94.052012}
  {\path{doi:10.1103/PhysRevD.94.052012}}.

\bibitem{HIG-15-013}
A.~M. Sirunyan, et~al., {Search for Higgs boson pair production in the
  $\Pbottom\Pbottom\Pgt\Pgt$ final state in proton-proton collisions at
  $\sqrt{s}=8$~\TeV}, Phys. Rev. D 96~(7) (2017) 072004.
\newblock \href {http://arxiv.org/abs/1707.00350} {\path{arXiv:1707.00350}},
  \href {https://doi.org/10.1103/PhysRevD.96.072004}
  {\path{doi:10.1103/PhysRevD.96.072004}}.

\bibitem{HIG-17-006}
A.~M. Sirunyan, et~al., {Search for resonant and nonresonant Higgs boson pair
  production in the $\Pbottom\APbottom\Plepton\Pnu\Plepton\Pnu$ final state in
  proton-proton collisions at $\sqrt{s}=13$~\TeV}, JHEP 01 (2018) 054.
\newblock \href {http://arxiv.org/abs/1708.04188} {\path{arXiv:1708.04188}},
  \href {https://doi.org/10.1007/JHEP01(2018)054}
  {\path{doi:10.1007/JHEP01(2018)054}}.

\bibitem{HIG-17-030}
A.~M. Sirunyan, et~al., {Combination of searches for Higgs boson pair
  production in proton-proton collisions at $\sqrt{s} = 13$~\TeV}, Phys. Rev.
  Lett. 122~(12) (2019) 121803.
\newblock \href {http://arxiv.org/abs/1811.09689} {\path{arXiv:1811.09689}},
  \href {https://doi.org/10.1103/PhysRevLett.122.121803}
  {\path{doi:10.1103/PhysRevLett.122.121803}}.

\bibitem{Sirunyan:2020xok}
A.~M. Sirunyan, et~al., {Search for nonresonant Higgs boson pair production in
  final states with two bottom quarks and two photons in proton-proton
  collisions at $ \sqrt{s} $ = 13 TeV}, JHEP 03 (2021) 257.
\newblock \href {http://arxiv.org/abs/2011.12373} {\path{arXiv:2011.12373}},
  \href {https://doi.org/10.1007/JHEP03(2021)257}
  {\path{doi:10.1007/JHEP03(2021)257}}.

\bibitem{Aad:2015xja}
G.~Aad, et~al., {Searches for Higgs boson pair production in the
  $\PHiggs\PHiggs \to \Pbottom\Pbottom\Pgt\Pgt, \Pgamma\Pgamma\PW\PW^{*},
  \Pgamma\Pgamma\Pbottom\Pbottom, \Pbottom\Pbottom\Pbottom\Pbottom$ channels
  with the ATLAS detector}, Phys. Rev. D 92 (2015) 092004.
\newblock \href {http://arxiv.org/abs/1509.04670} {\path{arXiv:1509.04670}},
  \href {https://doi.org/10.1103/PhysRevD.92.092004}
  {\path{doi:10.1103/PhysRevD.92.092004}}.

\bibitem{Aaboud:2018knk}
M.~Aaboud, et~al., {Search for pair production of Higgs bosons in the
  $\Pbottom\APbottom\Pbottom\APbottom$ final state using proton-proton
  collisions at $\sqrt{s} = 13$~\TeV with the ATLAS detector}, JHEP 01 (2019)
  030.
\newblock \href {http://arxiv.org/abs/1804.06174} {\path{arXiv:1804.06174}},
  \href {https://doi.org/10.1007/JHEP01(2019)030}
  {\path{doi:10.1007/JHEP01(2019)030}}.

\bibitem{Aaboud:2018ftw}
M.~Aaboud, et~al., {Search for Higgs boson pair production in the
  $\Pgamma\Pgamma\Pbottom\APbottom$ final state with 13 TeV $\Pp\Pp$ collision
  data collected by the ATLAS experiment}, JHEP 11 (2018) 040.
\newblock \href {http://arxiv.org/abs/1807.04873} {\path{arXiv:1807.04873}},
  \href {https://doi.org/10.1007/JHEP11(2018)040}
  {\path{doi:10.1007/JHEP11(2018)040}}.

\bibitem{Aaboud:2018sfw}
M.~Aaboud, et~al., {Search for resonant and non-resonant Higgs boson pair
  production in the $\Pbottom\APbottom\Pgt^{+}\Pgt^{-}$ decay channel in
  $\Pp\Pp$ collisions at $\sqrt{s}=13$ TeV with the ATLAS detector}, Phys. Rev.
  Lett. 121~(19) (2018) 191801.
\newblock \href {http://arxiv.org/abs/1808.00336} {\path{arXiv:1808.00336}},
  \href {https://doi.org/10.1103/PhysRevLett.121.191801}
  {\path{doi:10.1103/PhysRevLett.121.191801}}.

\bibitem{Aaboud:2018zhh}
M.~Aaboud, et~al., {Search for Higgs boson pair production in the
  $\Pbottom\APbottom\PW\PW^{*}$ decay mode at $\sqrt{s}=13$~\TeV with the ATLAS
  detector}, JHEP 04 (2019) 092.
\newblock \href {http://arxiv.org/abs/1811.04671} {\path{arXiv:1811.04671}},
  \href {https://doi.org/10.1007/JHEP04(2019)092}
  {\path{doi:10.1007/JHEP04(2019)092}}.

\bibitem{Aaboud:2018ewm}
M.~Aaboud, et~al., {Search for Higgs boson pair production in the
  $\Pgamma\Pgamma\PW\PW^{*}$ channel using $\Pp\Pp$ collision data recorded at
  $\sqrt{s} = 13$~\TeV with the ATLAS detector}, Eur. Phys. J. C 78~(12) (2018)
  1007.
\newblock \href {http://arxiv.org/abs/1807.08567} {\path{arXiv:1807.08567}},
  \href {https://doi.org/10.1140/epjc/s10052-018-6457-x}
  {\path{doi:10.1140/epjc/s10052-018-6457-x}}.

\bibitem{Aaboud:2018ksn}
M.~Aaboud, et~al., {Search for Higgs boson pair production in the
  $\PW\PW^{*}\PW\PW^{*}$ decay channel using ATLAS data recorded at
  $\sqrt{s}=13$~\TeV}, JHEP 05 (2019) 124.
\newblock \href {http://arxiv.org/abs/1811.11028} {\path{arXiv:1811.11028}},
  \href {https://doi.org/10.1007/JHEP05(2019)124}
  {\path{doi:10.1007/JHEP05(2019)124}}.

\bibitem{Baur:2002rb}
U.~Baur, T.~Plehn, D.~L. Rainwater, {Measuring the Higgs boson self coupling at
  the LHC and finite top mass matrix elements}, Phys. Rev. Lett. 89 (2002)
  151801.
\newblock \href {http://arxiv.org/abs/hep-ph/0206024}
  {\path{arXiv:hep-ph/0206024}}, \href
  {https://doi.org/10.1103/PhysRevLett.89.151801}
  {\path{doi:10.1103/PhysRevLett.89.151801}}.

\bibitem{Baur:2002qd}
U.~Baur, T.~Plehn, D.~L. Rainwater, {Determining the Higgs boson self coupling
  at hadron colliders}, Phys. Rev. D 67 (2003) 033003.
\newblock \href {http://arxiv.org/abs/hep-ph/0211224}
  {\path{arXiv:hep-ph/0211224}}, \href
  {https://doi.org/10.1103/PhysRevD.67.033003}
  {\path{doi:10.1103/PhysRevD.67.033003}}.

\bibitem{Baur:2003gpa}
U.~Baur, T.~Plehn, D.~L. Rainwater, {Examining the Higgs boson potential at
  lepton and hadron colliders: a comparative analysis}, Phys. Rev. D 68 (2003)
  033001.
\newblock \href {http://arxiv.org/abs/hep-ph/0304015}
  {\path{arXiv:hep-ph/0304015}}, \href
  {https://doi.org/10.1103/PhysRevD.68.033001}
  {\path{doi:10.1103/PhysRevD.68.033001}}.

\bibitem{Baur:2003gp}
U.~Baur, T.~Plehn, D.~L. Rainwater, {Probing the Higgs self coupling at hadron
  colliders using rare decays}, Phys. Rev. D 69 (2004) 053004.
\newblock \href {http://arxiv.org/abs/hep-ph/0310056}
  {\path{arXiv:hep-ph/0310056}}, \href
  {https://doi.org/10.1103/PhysRevD.69.053004}
  {\path{doi:10.1103/PhysRevD.69.053004}}.

\bibitem{Dolan:2012rv}
M.~J. Dolan, C.~Englert, M.~Spannowsky, {Higgs self-coupling measurements at
  the LHC}, JHEP 10 (2012) 112.
\newblock \href {http://arxiv.org/abs/1206.5001} {\path{arXiv:1206.5001}},
  \href {https://doi.org/10.1007/JHEP10(2012)112}
  {\path{doi:10.1007/JHEP10(2012)112}}.

\bibitem{Papaefstathiou:2012qe}
A.~Papaefstathiou, L.~L. Yang, J.~Zurita, {Higgs boson pair production at the
  LHC in the $\Pbottom\APbottom\PW^{+}\PW^{-}$ channel}, Phys. Rev. D 87~(1)
  (2013) 011301.
\newblock \href {http://arxiv.org/abs/1209.1489} {\path{arXiv:1209.1489}},
  \href {https://doi.org/10.1103/PhysRevD.87.011301}
  {\path{doi:10.1103/PhysRevD.87.011301}}.

\bibitem{deLima:2014dta}
D.~E. Ferreira~de Lima, A.~Papaefstathiou, M.~Spannowsky, {Standard model Higgs
  boson pair production in the $\Pbottom\APbottom\Pbottom\APbottom$ final
  state}, JHEP 08 (2014) 030.
\newblock \href {http://arxiv.org/abs/1404.7139} {\path{arXiv:1404.7139}},
  \href {https://doi.org/10.1007/JHEP08(2014)030}
  {\path{doi:10.1007/JHEP08(2014)030}}.

\bibitem{Wardrope:2014kya}
D.~Wardrope, E.~Jansen, N.~Konstantinidis, B.~Cooper, R.~Falla,
  N.~Norjoharuddeen, {Non-resonant Higgs pair production in the
  $\Pbottom\APbottom\Pbottom\APbottom$ final state at the LHC}, Eur. Phys. J. C
  75~(5) (2015) 219.
\newblock \href {http://arxiv.org/abs/1410.2794} {\path{arXiv:1410.2794}},
  \href {https://doi.org/10.1140/epjc/s10052-015-3439-0}
  {\path{doi:10.1140/epjc/s10052-015-3439-0}}.

\bibitem{Behr:2015oqq}
J.~K. Behr, D.~Bortoletto, J.~A. Frost, N.~P. Hartland, C.~Issever, J.~Rojo,
  {Boosting Higgs pair production in the $\Pbottom\APbottom\Pbottom\APbottom$
  final state with multivariate techniques}, Eur. Phys. J. C 76~(7) (2016) 386.
\newblock \href {http://arxiv.org/abs/1512.08928} {\path{arXiv:1512.08928}},
  \href {https://doi.org/10.1140/epjc/s10052-016-4215-5}
  {\path{doi:10.1140/epjc/s10052-016-4215-5}}.

\bibitem{Li:2015yia}
Q.~Li, Z.~Li, Q.-S. Yan, X.~Zhao, {Probe Higgs boson pair production via the
  $3\Plepton 2\textrm{j} + \MET$ mode}, Phys. Rev. D 92~(1) (2015) 014015.
\newblock \href {http://arxiv.org/abs/1503.07611} {\path{arXiv:1503.07611}},
  \href {https://doi.org/10.1103/PhysRevD.92.014015}
  {\path{doi:10.1103/PhysRevD.92.014015}}.

\bibitem{Adhikary:2017jtu}
A.~Adhikary, S.~Banerjee, R.~K. Barman, B.~Bhattacherjee, S.~Niyogi,
  {Revisiting the non-resonant Higgs pair production at the HL-LHC}, JHEP 07
  (2018) 116.
\newblock \href {http://arxiv.org/abs/1712.05346} {\path{arXiv:1712.05346}},
  \href {https://doi.org/10.1007/JHEP07(2018)116}
  {\path{doi:10.1007/JHEP07(2018)116}}.

\bibitem{PDG}
P.~A. Zyla, et~al., {Review of particle physics}, PTEP 2020~(8) (2020) 083C01.
\newblock \href {https://doi.org/10.1093/ptep/ptaa104}
  {\path{doi:10.1093/ptep/ptaa104}}.

\bibitem{Czakon:2011xx}
M.~Czakon, A.~Mitov, {Top++: A program for the calculation of the top-pair
  cross-section at hadron colliders}, Comput. Phys. Commun. 185 (2014) 2930.
\newblock \href {http://arxiv.org/abs/1112.5675} {\path{arXiv:1112.5675}},
  \href {https://doi.org/10.1016/j.cpc.2014.06.021}
  {\path{doi:10.1016/j.cpc.2014.06.021}}.

\bibitem{ANN}
A.~K. Jain, J.~Mao, K.~M. Mohiuddin, {Artificial neural networks: a tutorial},
  Computer 29~(3) (1996) 31.
\newblock \href {https://doi.org/10.1109/2.485891}
  {\path{doi:10.1109/2.485891}}.

\bibitem{chollet2015keras}
F.~Chollet, et~al., Keras, \url{https://github.com/fchollet/keras} (2015).

\bibitem{Kondo:1988yd}
K.~Kondo, {Dynamical likelihood method for reconstruction of events with
  missing momentum. 1: method and toy models}, J. Phys. Soc. Jap. 57 (1988)
  4126.
\newblock \href {https://doi.org/10.1143/JPSJ.57.4126}
  {\path{doi:10.1143/JPSJ.57.4126}}.

\bibitem{Kondo:1991dw}
K.~Kondo, {Dynamical likelihood method for reconstruction of events with
  missing momentum. 2: mass spectra for $2 \to 2$ processes}, J. Phys. Soc.
  Jap. 60 (1991) 836.
\newblock \href {https://doi.org/10.1143/JPSJ.60.836}
  {\path{doi:10.1143/JPSJ.60.836}}.

\bibitem{deFavereau:2013fsa}
J.~de~Favereau, C.~Delaere, P.~Demin, A.~Giammanco, V.~Lema\^\i{}tre,
  A.~Mertens, M.~Selvaggi, {DELPHES 3: A modular framework for fast simulation
  of a generic collider experiment}, JHEP 02 (2014) 057.
\newblock \href {http://arxiv.org/abs/1307.6346} {\path{arXiv:1307.6346}},
  \href {https://doi.org/10.1007/JHEP02(2014)057}
  {\path{doi:10.1007/JHEP02(2014)057}}.

\bibitem{Neyman:1933wgr}
J.~Neyman, E.~S. Pearson, {On the Problem of the Most Efficient Tests of
  Statistical Hypotheses}, Phil. Trans. Roy. Soc. Lond. A 231~(694-706) (1933)
  289.
\newblock \href {https://doi.org/10.1098/rsta.1933.0009}
  {\path{doi:10.1098/rsta.1933.0009}}.

\bibitem{Volobouev:2011vb}
I.~Volobouev, {Matrix element method in HEP: transfer functions, efficiencies,
  and likelihood normalization}. (2011).
\newblock \href {http://arxiv.org/abs/1101.2259} {\path{arXiv:1101.2259}}.

\bibitem{MadGraph_aMCatNLO}
J.~Alwall, R.~Frederix, S.~Frixione, V.~Hirschi, F.~Maltoni, O.~Mattelaer,
  H.~S. Shao, T.~Stelzer, P.~Torrielli, M.~Zaro, {The automated computation of
  tree-level and next-to-leading order differential cross sections, and their
  matching to parton shower simulations}, JHEP 07 (2014) 079.
\newblock \href {http://arxiv.org/abs/1405.0301} {\path{arXiv:1405.0301}},
  \href {https://doi.org/10.1007/JHEP07(2014)079}
  {\path{doi:10.1007/JHEP07(2014)079}}.

\bibitem{Grinstein:2007iv}
B.~Grinstein, M.~Trott, {A Higgs-Higgs bound state due to new physics at a
  \TeV}, Phys. Rev. D 76 (2007) 073002.
\newblock \href {http://arxiv.org/abs/0704.1505} {\path{arXiv:0704.1505}},
  \href {https://doi.org/10.1103/PhysRevD.76.073002}
  {\path{doi:10.1103/PhysRevD.76.073002}}.

\bibitem{Degrande:2011ua}
C.~Degrande, C.~Duhr, B.~Fuks, D.~Grellscheid, O.~Mattelaer, T.~Reiter, {UFO:
  The universal FeynRules output}, Comput. Phys. Commun. 183 (2012) 1201.
\newblock \href {http://arxiv.org/abs/1108.2040} {\path{arXiv:1108.2040}},
  \href {https://doi.org/10.1016/j.cpc.2012.01.022}
  {\path{doi:10.1016/j.cpc.2012.01.022}}.

\bibitem{Hespel:2014sla}
B.~Hespel, D.~Lopez-Val, E.~Vryonidou, {Higgs pair production via gluon fusion
  in the Two-Higgs-Doublet model}, JHEP 09 (2014) 124.
\newblock \href {http://arxiv.org/abs/1407.0281} {\path{arXiv:1407.0281}},
  \href {https://doi.org/10.1007/JHEP09(2014)124}
  {\path{doi:10.1007/JHEP09(2014)124}}.

\bibitem{Bjorkenx}
J.~D. Bjorken, {Asymptotic sum rules at infinite momentum}, Phys. Rev. 179
  (1969) 1547.
\newblock \href {https://doi.org/10.1103/PhysRev.179.1547}
  {\path{doi:10.1103/PhysRev.179.1547}}.

\bibitem{LHAPDF}
A.~Buckley, J.~Ferrando, S.~Lloyd, K.~Nordstr{\"o}m, B.~Page, M.~R{\"u}fenacht,
  M.~Sch{\"o}nherr, G.~Watt,
  \href{https://doi.org/10.1140/epjc/s10052-015-3318-8}{{LHAPDF6: Parton
  density access in the LHC precision era}}, The European Physical Journal C
  75~(3) (2015) 132.
\newblock \href {https://doi.org/10.1140/epjc/s10052-015-3318-8}
  {\path{doi:10.1140/epjc/s10052-015-3318-8}}.
\newline\urlprefix\url{https://doi.org/10.1140/epjc/s10052-015-3318-8}

\bibitem{MSTW}
A.~D. Martin, W.~J. Stirling, R.~S. Thorne, G.~Watt, {Parton distributions for
  the LHC}, Eur. Phys. J. C 63 (2009) 189.
\newblock \href {http://arxiv.org/abs/0901.0002} {\path{arXiv:0901.0002}},
  \href {https://doi.org/10.1140/epjc/s10052-009-1072-5}
  {\path{doi:10.1140/epjc/s10052-009-1072-5}}.

\bibitem{Fiedler:2010sg}
F.~Fiedler, A.~Grohsjean, P.~Haefner, P.~Schieferdecker, {The matrix element
  method and its application in measurements of the top quark mass}, Nucl.
  Instrum. Meth. A 624 (2010) 203.
\newblock \href {http://arxiv.org/abs/1003.1316} {\path{arXiv:1003.1316}},
  \href {https://doi.org/10.1016/j.nima.2010.09.024}
  {\path{doi:10.1016/j.nima.2010.09.024}}.

\bibitem{SVfitMEM}
L.~Bianchini, B.~Calpas, J.~Conway, A.~Fowlie, L.~Marzola, C.~Veelken,
  L.~Perrini, {Reconstruction of the Higgs mass in events with Higgs bosons
  decaying into a pair of $\Pgt$ leptons using matrix element techniques},
  Nucl. Instrum. Meth. A 862 (2017) 54.
\newblock \href {http://arxiv.org/abs/1603.05910} {\path{arXiv:1603.05910}},
  \href {https://doi.org/10.1016/j.nima.2017.05.001}
  {\path{doi:10.1016/j.nima.2017.05.001}}.

\bibitem{Alwall:2010cq}
J.~Alwall, A.~Freitas, O.~Mattelaer, {The matrix element method and QCD
  radiation}, Phys. Rev. D 83 (2011) 074010.
\newblock \href {http://arxiv.org/abs/1010.2263} {\path{arXiv:1010.2263}},
  \href {https://doi.org/10.1103/PhysRevD.83.074010}
  {\path{doi:10.1103/PhysRevD.83.074010}}.

\bibitem{JME-10-009}
S.~Chatrchyan, et~al., {Missing transverse energy performance of the CMS
  detector}, JINST 6 (2011) P09001.
\newblock \href {http://arxiv.org/abs/1106.5048} {\path{arXiv:1106.5048}},
  \href {https://doi.org/10.1088/1748-0221/6/09/P09001}
  {\path{doi:10.1088/1748-0221/6/09/P09001}}.

\bibitem{NWA}
D.~Berdine, N.~Kauer, D.~Rainwater, {Breakdown of the narrow width
  approximation for new physics}, Phys. Rev. Lett. 99 (2007) 111601.
\newblock \href {http://arxiv.org/abs/hep-ph/0703058}
  {\path{arXiv:hep-ph/0703058}}, \href
  {https://doi.org/10.1103/PhysRevLett.99.111601}
  {\path{doi:10.1103/PhysRevLett.99.111601}}.

\bibitem{VAMP}
T.~Ohl, {Vegas revisited: Adaptive Monte Carlo integration beyond
  factorization}, Comput. Phys. Commun. 120 (1999) 13.
\newblock \href {http://arxiv.org/abs/hep-ph/9806432}
  {\path{arXiv:hep-ph/9806432}}, \href
  {https://doi.org/10.1016/S0010-4655(99)00209-X}
  {\path{doi:10.1016/S0010-4655(99)00209-X}}.

\bibitem{VEGAS}
G.~P. Lepage, {A new algorithm for adaptive multidimensional integration}, J.
  Comput. Phys. 27 (1978) 192.
\newblock \href {https://doi.org/10.1016/0021-9991(78)90004-9}
  {\path{doi:10.1016/0021-9991(78)90004-9}}.

\bibitem{POWHEG1}
P.~Nason, {A new method for combining NLO QCD with shower Monte Carlo
  algorithms}, JHEP 11 (2004) 040.
\newblock \href {http://arxiv.org/abs/hep-ph/0409146}
  {\path{arXiv:hep-ph/0409146}}, \href
  {https://doi.org/10.1088/1126-6708/2004/11/040}
  {\path{doi:10.1088/1126-6708/2004/11/040}}.

\bibitem{POWHEG2}
S.~Frixione, P.~Nason, C.~Oleari, {Matching NLO QCD computations with parton
  shower simulations: the POWHEG method}, JHEP 11 (2007) 070.
\newblock \href {http://arxiv.org/abs/0709.2092} {\path{arXiv:0709.2092}},
  \href {https://doi.org/10.1088/1126-6708/2007/11/070}
  {\path{doi:10.1088/1126-6708/2007/11/070}}.

\bibitem{POWHEG3}
S.~Alioli, P.~Nason, C.~Oleari, E.~Re, {A general framework for implementing
  NLO calculations in shower Monte Carlo programs: the POWHEG BOX}, JHEP 06
  (2010) 043.
\newblock \href {http://arxiv.org/abs/1002.2581} {\path{arXiv:1002.2581}},
  \href {https://doi.org/10.1007/JHEP06(2010)043}
  {\path{doi:10.1007/JHEP06(2010)043}}.

\bibitem{POWHEGTTBAR1}
J.~M. Campbell, R.~K. Ellis, P.~Nason, E.~Re, {Top-pair production and decay at
  NLO matched with parton showers}, JHEP 04 (2015) 114.
\newblock \href {http://arxiv.org/abs/1412.1828} {\path{arXiv:1412.1828}},
  \href {https://doi.org/10.1007/JHEP04(2015)114}
  {\path{doi:10.1007/JHEP04(2015)114}}.

\bibitem{POWHEGTTBAR2}
S.~Alioli, S.-O. Moch, P.~Uwer, {Hadronic top-quark pair-production with one
  jet and parton showering}, JHEP 01 (2012) 137.
\newblock \href {http://arxiv.org/abs/1110.5251} {\path{arXiv:1110.5251}},
  \href {https://doi.org/10.1007/JHEP01(2012)137}
  {\path{doi:10.1007/JHEP01(2012)137}}.

\bibitem{POWHEGHH1}
G.~Heinrich, S.~P. Jones, M.~Kerner, G.~Luisoni, E.~Vryonidou, {NLO predictions
  for Higgs boson pair production with full top quark mass dependence matched
  to parton showers}, JHEP 08 (2017) 088.
\newblock \href {http://arxiv.org/abs/1703.09252} {\path{arXiv:1703.09252}},
  \href {https://doi.org/10.1007/JHEP08(2017)088}
  {\path{doi:10.1007/JHEP08(2017)088}}.

\bibitem{POWHEGHH2}
G.~Heinrich, S.~P. Jones, M.~Kerner, G.~Luisoni, L.~Scyboz, {Probing the
  trilinear Higgs boson coupling in di-Higgs production at NLO QCD including
  parton shower effects}, JHEP 06 (2019) 066.
\newblock \href {http://arxiv.org/abs/1903.08137} {\path{arXiv:1903.08137}},
  \href {https://doi.org/10.1007/JHEP06(2019)066}
  {\path{doi:10.1007/JHEP06(2019)066}}.

\bibitem{NNPDF1}
R.~D. Ball, V.~Bertone, S.~Carrazza, L.~Del~Debbio, S.~Forte, A.~Guffanti,
  N.~P. Hartland, J.~Rojo, {Parton distributions with QED corrections}, Nucl.
  Phys. B 877 (2013) 290.
\newblock \href {http://arxiv.org/abs/1308.0598} {\path{arXiv:1308.0598}},
  \href {https://doi.org/10.1016/j.nuclphysb.2013.10.010}
  {\path{doi:10.1016/j.nuclphysb.2013.10.010}}.

\bibitem{NNPDF2}
R.~D. Ball, V.~Bertone, F.~Cerutti, L.~Del~Debbio, S.~Forte, A.~Guffanti, J.~I.
  Latorre, J.~Rojo, M.~Ubiali, {Unbiased global determination of parton
  distributions and their uncertainties at NNLO and at LO}, Nucl. Phys. B 855
  (2012) 153.
\newblock \href {http://arxiv.org/abs/1107.2652} {\path{arXiv:1107.2652}},
  \href {https://doi.org/10.1016/j.nuclphysb.2011.09.024}
  {\path{doi:10.1016/j.nuclphysb.2011.09.024}}.

\bibitem{NNPDF3}
R.~D. Ball, et~al., {Parton distributions for the LHC Run II}, JHEP 04 (2015)
  040.
\newblock \href {http://arxiv.org/abs/1410.8849} {\path{arXiv:1410.8849}},
  \href {https://doi.org/10.1007/JHEP04(2015)040}
  {\path{doi:10.1007/JHEP04(2015)040}}.

\bibitem{Sjostrand:2014zea}
T.~Sj{\"o}strand, S.~Ask, J.~R. Christiansen, R.~Corke, N.~Desai, P.~Ilten,
  S.~Mrenna, S.~Prestel, C.~O. Rasmussen, P.~Z. Skands, An introduction to
  {\PYTHIA} $8.2$, Comput. Phys. Commun. 191 (2015) 159.
\newblock \href {http://arxiv.org/abs/1410.3012} {\path{arXiv:1410.3012}},
  \href {https://doi.org/10.1016/j.cpc.2015.01.024}
  {\path{doi:10.1016/j.cpc.2015.01.024}}.

\bibitem{Sirunyan:2019dfx}
A.~M. Sirunyan, et~al., Extraction and validation of a new set of {CMS}
  {\PYTHIA} $8$ tunes from underlying-event measurements, Eur. Phys. J. C 80
  (2020) 4.
\newblock \href {http://arxiv.org/abs/1903.12179} {\path{arXiv:1903.12179}},
  \href {https://doi.org/10.1140/epjc/s10052-019-7499-4}
  {\path{doi:10.1140/epjc/s10052-019-7499-4}}.

\bibitem{Cacciari:2008gp}
M.~Cacciari, G.~P. Salam, G.~Soyez, {The anti-$\kt$ jet clustering algorithm},
  JHEP 04 (2008) 063.
\newblock \href {http://arxiv.org/abs/0802.1189} {\path{arXiv:0802.1189}},
  \href {https://doi.org/10.1088/1126-6708/2008/04/063}
  {\path{doi:10.1088/1126-6708/2008/04/063}}.

\bibitem{Cacciari:2011ma}
M.~Cacciari, G.~P. Salam, G.~Soyez, {FastJet user manual}, Eur. Phys. J. C 72
  (2012) 1896.
\newblock \href {http://arxiv.org/abs/1111.6097} {\path{arXiv:1111.6097}},
  \href {https://doi.org/10.1140/epjc/s10052-012-1896-2}
  {\path{doi:10.1140/epjc/s10052-012-1896-2}}.

\bibitem{Cacciari:2008gn}
M.~Cacciari, G.~P. Salam, G.~Soyez, The catchment area of jets, JHEP 04 (2008)
  005.
\newblock \href {http://arxiv.org/abs/0802.1188} {\path{arXiv:0802.1188}},
  \href {https://doi.org/10.1088/1126-6708/2008/04/005}
  {\path{doi:10.1088/1126-6708/2008/04/005}}.

\bibitem{Cacciari:2007fd}
M.~Cacciari, G.~P. Salam, Pileup subtraction using jet areas, Phys. Lett. B 659
  (2008) 119.
\newblock \href {http://arxiv.org/abs/0707.1378} {\path{arXiv:0707.1378}},
  \href {https://doi.org/10.1016/j.physletb.2007.09.077}
  {\path{doi:10.1016/j.physletb.2007.09.077}}.

\bibitem{CMS:2012feb}
S.~Chatrchyan, et~al., {Identification of $\Pbottom$-quark jets with the CMS
  experiment}, JINST 8 (2013) P04013.
\newblock \href {http://arxiv.org/abs/1211.4462} {\path{arXiv:1211.4462}},
  \href {https://doi.org/10.1088/1748-0221/8/04/P04013}
  {\path{doi:10.1088/1748-0221/8/04/P04013}}.

\bibitem{JME-18-001}
A.~M. Sirunyan, et~al., {Pileup mitigation at CMS in $13$~\TeV data}, JINST
  15~(09) (2020) P09018.
\newblock \href {http://arxiv.org/abs/2003.00503} {\path{arXiv:2003.00503}},
  \href {https://doi.org/10.1088/1748-0221/15/09/P09018}
  {\path{doi:10.1088/1748-0221/15/09/P09018}}.

\bibitem{ATLAS:2018txj}
M.~Aaboud, et~al., {Performance of missing transverse momentum reconstruction
  with the ATLAS detector using proton-proton collisions at $\sqrt{s} =
  13$~\TeV}, Eur. Phys. J. C 78~(11) (2018) 903.
\newblock \href {http://arxiv.org/abs/1802.08168} {\path{arXiv:1802.08168}},
  \href {https://doi.org/10.1140/epjc/s10052-018-6288-9}
  {\path{doi:10.1140/epjc/s10052-018-6288-9}}.

\bibitem{ROCcurve}
D.~M. Green, J.~A. Swets, {Signal detection theory and psychophysics}, {Wiley},
  {New York}, 1966.

\bibitem{CUBA}
T.~Hahn, {CUBA: A library for multidimensional numerical integration}, Comput.
  Phys. Commun. 168 (2005) 78.
\newblock \href {http://arxiv.org/abs/hep-ph/0404043}
  {\path{arXiv:hep-ph/0404043}}, \href
  {https://doi.org/10.1016/j.cpc.2005.01.010}
  {\path{doi:10.1016/j.cpc.2005.01.010}}.

\bibitem{Hagiwara:2009aq}
K.~Hagiwara, J.~Kanzaki, N.~Okamura, D.~Rainwater, T.~Stelzer, {Fast
  calculation of HELAS amplitudes using graphics processing unit (GPU)}, Eur.
  Phys. J. C 66 (2010) 477.
\newblock \href {http://arxiv.org/abs/0908.4403} {\path{arXiv:0908.4403}},
  \href {https://doi.org/10.1140/epjc/s10052-010-1276-8}
  {\path{doi:10.1140/epjc/s10052-010-1276-8}}.

\bibitem{Hagiwara:2009cy}
K.~Hagiwara, J.~Kanzaki, N.~Okamura, D.~Rainwater, T.~Stelzer, {Calculation of
  HELAS amplitudes for QCD processes using graphics processing unit (GPU)},
  Eur. Phys. J. C 70 (2010) 513.
\newblock \href {http://arxiv.org/abs/0909.5257} {\path{arXiv:0909.5257}},
  \href {https://doi.org/10.1140/epjc/s10052-010-1465-5}
  {\path{doi:10.1140/epjc/s10052-010-1465-5}}.

\bibitem{Kanzaki:2010ym}
J.~Kanzaki, {Monte Carlo integration on GPU}, Eur. Phys. J. C 71 (2011) 1559.
\newblock \href {http://arxiv.org/abs/1010.2107} {\path{arXiv:1010.2107}},
  \href {https://doi.org/10.1140/epjc/s10052-011-1559-8}
  {\path{doi:10.1140/epjc/s10052-011-1559-8}}.

\bibitem{Hagiwara:2013oka}
K.~Hagiwara, J.~Kanzaki, Q.~Li, N.~Okamura, T.~Stelzer, {Fast computation of
  MadGraph amplitudes on graphics processing unit (GPU)}, Eur. Phys. J. C 73
  (2013) 2608.
\newblock \href {http://arxiv.org/abs/1305.0708} {\path{arXiv:1305.0708}},
  \href {https://doi.org/10.1140/epjc/s10052-013-2608-2}
  {\path{doi:10.1140/epjc/s10052-013-2608-2}}.

\bibitem{Schouten:2014yza}
D.~Schouten, A.~DeAbreu, B.~Stelzer, {Accelerated matrix element method with
  parallel computing}, Comput. Phys. Commun. 192 (2015) 54.
\newblock \href {http://arxiv.org/abs/1407.7595} {\path{arXiv:1407.7595}},
  \href {https://doi.org/10.1016/j.cpc.2015.02.020}
  {\path{doi:10.1016/j.cpc.2015.02.020}}.

\bibitem{Grasseau:2015vfa}
G.~Grasseau, S.~Lisniak, D.~Chamont, {Hybrid implementation of the VEGAS Monte
  Carlo algorithm}, in: {Proceedings, GPU Computing in High Energy Physics
  (GPUHEP2014): Pisa, Italy, September 10-12, 2014}, 2015, p. 103.
\newblock \href {https://doi.org/10.3204/DESY-PROC-2014-05/19}
  {\path{doi:10.3204/DESY-PROC-2014-05/19}}.

\end{thebibliography}

\end{document}